\documentstyle[12pt,bezier]{report}
\newcounter{strand}

\def\comm#1#2{
        \left[#1,#2\right]}
\def\real{\mbox{\rm I}\!\mbox{\rm R}}
\def\ie{\hbox{\it i.e.}}
\def\inprod#1#2{
        \left\langle #1, #2\right\rangle}
\def\C{{\cal C}}
\def\idA{1_{\cal A}}
\def\idU{1_{\cal U}}
\def\Ups{\Upsilon} 
\def\uqg{\mbox{$U_{q}{\/\mbox{\bf g}}$}}
\def\fun{\mbox{Fun$(G_{q})$}}
\newcommand{\tr}{\triangleright}
\def\cross{\mbox{$\times \!\rule{0.3pt}{1.1ex}\,$}}
\def\smash{{\A \cross \U}}
\def\R{\mbox{$\cal R$}}
\newcommand{\Y}{\mbox{$\cal Y$}}
\newcommand{\Z}{\mbox{$\cal Z$}}
\def\A{\mbox{$\cal A$}}
\def\U{\mbox{$\cal U$}}
\newcommand{\DA}{\Delta _{\cal A}}
\newcommand{\AD}{{}_{\cal A}\Delta }
\newcommand{\UD}{{}_{\cal U}\Delta }
\newcommand{\der}{\mbox{d}}
\def\z{\hspace*{9mm}}
\def\x{\hspace{3mm}}
\newcommand{\ad}{\stackrel{\mbox{\scriptsize ad}}{\triangleright}}
\newcommand{\Deutsch}{\cal}
\newcommand{\Schreib}{\cal}
\newcommand{\schreib}{\cal}
\newcommand{\kreuz}{\bf}
\newcommand{\RRI}{\mbox{\mbox{\boldmath $R$}$_{I,I\!I}\,$}}
\newcommand{\RR}{\mbox{\boldmath $R$}}
\def\bigR{\mbox{$\mbox{\bf\rm I}\!\mbox{\bf\rm R}$}}
\newcommand{\bigA}{\mbox{\large A\hspace{-1.6ex}A\hspace{-1.7ex}A}}
\newcommand{\YY}{\mbox{\boldmath $Y$}}
\newcommand{\Aa}{\mbox{\boldmath $A$}}
\newcommand{\mb}{\overline{m}}
\def\I{\mbox{\boldmath $i$}}
\def\Ix#1{\mbox{\boldmath $i$}_{\chi_#1}}
\def\Li{\hbox{\large\it \pounds}}
\def\Lix#1{\hbox{\large\it \pounds}_{\chi_#1}}
\def\Lio#1#2{\hbox{\large\it \pounds}_{O_#1{}^#2}}
\def\dl{\mbox{\bf d}}
\newcommand{\DD}{\mbox{\bf D}}
\newcommand{\om}{\mbox{$\omega $}}
\newcommand{\al}{\mbox{$\alpha $}}
\newcommand{\fum}{\mbox{Fun({\bf M}$_{q}$)}}
\newcommand{\tqm}{\mbox{${\cal T}(\mbox{\bf M}_q)$}}

\def\tq{\mbox{${\cal T}_q$}}
\def\dg{\mbox{\boldmath$\delta $}}
\def\ut{\tilde{U}}
\def\Ht{\tilde{H}}
\newcommand{\da}{\stackrel{\mbox{\scriptsize ad}}{\triangleleft}}
\newcommand{\op}{\stackrel{\mbox{\scriptsize op}}{\triangleright}}
\def\B{\mbox{$\cal B$}}
\def\Il{\mbox{\bf i}}
\begin{document}
\begin{titlepage}
\begin{center}
December 9, 1993    \hfill    LBL-34942\\
                     \hfill    UCB-PTH-93/35\\
                     \hfill    hep-th/9312075\\

\vskip .5in

{\large \bf Quantum Groups, Non-Commutative Differential Geometry and
Applications}\footnote{This
work was supported in part by the Director, Office of Energy Research,
Office of High Energy and Nuclear Physics, Division of High Energy
Physics of the U.S. Department of Energy under Contract
DE-AC03-76SF00098 and in part by the National Science Foundation under
grant PHY-90-21139.}

\vskip .5in

Peter Schupp\footnote{schupp@physics.berkeley.edu}

\vskip .5in

{\em Department of Physics\\
University of California\\
and\\
Theoretical Physics Group\\
Lawrence Berkeley Laboratory\\
University of California\\
Berkeley, California 94720}

\vskip .5in

A dissertation submitted in partial satisfaction of the requirements \\
for the degree of Doctor of Philosophy in Physics.

\vskip .5in

Committee in charge:\\[.2in]
\parbox{6cm}{Professor Bruno Zumino, Chair\\
Professor Stanley Mandelstam\\
Professor Nicolai Y. Reshetikhin}

\end{center}

\pagebreak

\begin{abstract}
The topic of this thesis is the development of a versatile and
geometrically motivated differential calculus on non-commutative or
quantum spaces, providing powerful but easy-to-use mathematical tools for
applications in physics and related sciences. A generalization of
unitary time evolution is proposed and studied for a simple
2-level system, leading to non-conservation of microscopic entropy, a
phenomenon new to quantum mechanics. A Cartan calculus that combines
functions, forms, Lie derivatives and inner derivations along general
vector fields into one big algebra is constructed for quantum groups
and then extended to quantum planes. The construction of a tangent
bundle on a quantum group manifold and an BRST type approach to quantum
group gauge theory are given as further examples of applications.

The material is organized in two parts: Part I studies vector fields on
quantum groups, emphasizing Hopf algebraic structures, but
also introducing a `quantum geometric' construction.
Using a generalized semi-direct product construction we combine
the dual Hopf algebras \A\ of functions and \U\ of left-invariant vector
fields into one fully bicovariant algebra of differential operators. The
pure braid group is introduced as the commutant of $\Delta (\U)$. It provides
invariant maps $\A \to  \U$ and thereby  bicovariant vector fields,
casimirs and metrics. This construction allows the translation of
undeformed matrix expressions into their less obvious quantum algebraic
counter parts. We study this in detail for quasitriangular Hopf
algebras, giving the determinant and
orthogonality relation for the `reflection' matrix. Part II considers
the additional structures of differential forms and finitely generated
quantum Lie algebras --- it is devoted to the construction of the Cartan
calculus, based on an undeformed Cartan identity. We attempt a
classification of various types of quantum Lie algebras and present a
fairly general example for their
construction, utilizing pure braid methods, proving orthogonality
of the adjoint representation and giving a (Killing) metric and the
quadratic casimir.
A reformulation of the Cartan calculus as a braided algebra and its
extension to quantum planes, directly and induced from the group calculus,
are provided.
\end{abstract}
\end{titlepage}

\renewcommand{\thepage}{\roman{page}}
\setcounter{page}{3}
\mbox{ }
\vfill

\begin{center}
{\bf Disclaimer}
\end{center}

\vskip .2in

\begin{scriptsize}
\begin{quotation}
This document was prepared as an account of work sponsored by the United
States Government.  Neither the United States Government nor any agency
thereof, nor The Regents of the University of California, nor any of their
employees, makes any warranty, express or implied, or assumes any legal
liability or responsibility for the accuracy, completeness, or usefulness
of any information, apparatus, product, or process disclosed, or represents
that its use would not infringe privately owned rights.  Reference herein
to any specific commercial products process, or service by its trade name,
trademark, manufacturer, or otherwise, does not necessarily constitute or
imply its endorsement, recommendation, or favoring by the United States
Government or any agency thereof, or The Regents of the University of
California.  The views and opinions of authors expressed herein do not
necessarily state or reflect those of the United States Government or any
agency thereof of The Regents of the University of California and shall
not be used for advertising or product endorsement purposes.
\end{quotation}
\end{scriptsize}

\vskip 2in

\begin{center}
\begin{small}
{\it Lawrence Berkeley Laboratory is an equal opportunity employer.}
\vfill
\mbox{}
\end{small}
\end{center}

\newpage
\renewcommand{\thepage}{\arabic{page}}
\setcounter{page}{1}

\section*{Acknowledgements}

It is a pleasure to thank my advisor, Professor Bruno Zumino, who
introduced me to the topics of this thesis, for many things ---
his support and care,
for sharing his amazing intuition and being patient with his
young student's ignorance --- and especially for creating a pleasant,
cooperative and productive environment.

In this context I also would like to thank my colleagues Chryss
Chryssomalakos and Paul Watts for cooperation rather than competition,
for constructive criticism, ongoing discussions and
generally for a fun time. The same goes to my fellow graduate
students and friends at LBL, on campus, at Caltech and in Heidelberg.
Thanks also to the professors, postdocs and staff at LBL and on
campus, and especially to Sandy Ewing, Betty Moura, Luanne Neumann,
Anne Takizawa and Donna Sakima. We had many interesting conversations,
not only about physics, and their help in bureaucratic matters
made life at Berkeley easier
and quite enjoyable.

Special thanks to Professor Nicolai Yu.\ Reshetikhin for always
having time to help me with my many questions
and for sharing
his elegant approach to Quantum Groups in general and Ribbon Tangles
in particular
and to Professor Stanley Mandelstam
for helpful
discussion,
for being on my committee and for reading my thesis.

Helpful conversations about aspects of the subject matter of this thesis
with Paolo Aschieri, Scott Hotes, John Madore,
Peter Neu, Marc Rosso, Michael Schlieker, Alexander Sevrin, Leehwa
Yeh and participants of the
XXII-th DGM conference are gratefully acknowledged.
I would also like to thank Professor W.\ J.\ Freeman, Dirk Graudenz, Andreas
Herz and David MacKay for initiating me into the interesting field of
neural systems  and Professor L.\ Falicov for introducing me to condensed
matter theory. I am indebted to Professor J\"orn
Knoll for early teaching and guidance
and to Dr.\ Heister and Renker GmbH.\ for their support and
faith in my abilities
as a student.

This thesis would not have been possible without the ongoing
support and encouragement of my parents Hans and Marlis Schupp.
Finally, no thanks could be enough for
Claudia Herold, for her love, encouragement,
welcome distraction
and her patience with the INS, physics
and me.

This work was supported in part by the Director, Office of Energy
Research, Office of High Energy and Nuclear Physics, Division of High
Energy Physics of the U.S. Department of Energy under Contract
DE-AC03-76SF00098 and in part by the National Science Foundation under
grants PHY-90-21139 and PHY-89-04035.

\tableofcontents

\pagebreak

\chapter*{Introduction}
\addcontentsline{toc}{part}{Introduction}

The topic of this thesis is non-commutative geometry in general and
the development of powerful and easy to use differential calculi
on quantum spaces and some examples of their application in particular.
I will try to give an as geometric picture as possible while including
all necessary mathematical tools. The emphasis will be on the formation
of concepts (Begriffsbildung).

In classical differential geometry we have a choice between two dual and
equivalent descriptions: we can either work with points on a manifold
$\cal M$ or with the algebra {\boldmath$C$}$({\cal M})$ of functions on
$\cal M$.
Non-commutative geometry is based on the idea that the algebra
{\boldmath$C$}$({\cal M})$ need not be commutative. Such a space is
called a quantum space --- in analogy to the quantization
of the commutative algebra of functions on phase-space that yields the
non-commutative operator algebra of quantum mechanics. More general,
a non-commutative algebra, viewed
as if it was a function algebra on a (possibly non-existing) topological
space, is called a quantum or pseudo space. One could call it a ``theory
of shadows'' --- shadows of classical concepts and objects.

The poor understanding of physics at very short distances indicates that
the small scale structure of space-time might not be adequately
described by classical continuum geometry. At the Planck scale one
expects that the notion of classical geometry has to be generalized to
incorporate quantum effects. No convincing alternative is presently
known, but several possibilities have been proposed; one of them is
the introduction into physics of non-commutative geometry.
Such new physical theories would allow, roughly speaking, the necessary
fuzziness for a successful description of the space-time ``foam'' expected
at tiny distances. See for instance the interesting gedanken experiment
\cite{Ma} concerning generalized uncertainty relations.

This certainly was one of the motivations behind the work on quantum
deformations of the Lorentz and Poincare groups \cite{OSWZ2,OSWZ1,LuRu}
and of Minkowski space in
terms of a parameter $q$ and of course behind Connes program \cite{Co1} of
non-commutative geometry, but there are also many other possible applications
of non-commutative calculi in physics
like generalized symmetries  ({\it e.g.} quantum group gauge
theory) and stochastics (master equations, random walks, \ldots), to
mention a few.
Continuous deformations of symmetry groups in physical theories
have historically
been proven to be rather successful in enlarging the class of phenomena
that these theories describe well; one of the most famous examples is
special relativity. For this reason it would be very interesting
in elementary particle physics to study deformations of semi-simple
Lie groups. Unfortunately these groups allow only trivial deformations
as long as one stays within the category of Lie groups, hence giving
another motivation for the study of the less rigid quantum groups.

Such generalizations of physical theories might have welcome and also
unexpected side effects: One of them is the possibility that some
$q$-deformed quantum field theories might be naturally finite.
This is expected if the deformation parameter has dimensions of length,
in analogy to amplitudes in string theory which were proven to be
finite to all orders by S. Mandelstam \cite{Mm2}. Even if $q$
turns out not to be a physical parameter, such a theory might still be
interesting as a new way to regularize infinities \cite{pZ,Md5},
using $q$-identities, known from the study of $q$-functions, which were
first introduced in the context of combinatorics nearly a century ago.
Here we should also mention a quick and easy approach, due to
\cite{DMS}, to lattice gauge theory based on a minimal non-commutative
calculus.
In chapter~\ref{C:AQMM} we will show at the example of a simple toy model
that modified time evolution equations, that could be motivated from
deformed space time symmetries, lead to non-conservation of entropy.
This might be of interest in connection with the black hole evaporation
paradox.
Connes \cite{Co} and Connes \& Lott \cite{CoL}
consider  a minimal generalization of classical gauge theory
and study a Kaluza-Klein theory
with a 2-point internal space and use non-commutative geometric methods
to define metric properties; note that it is also
possible with these methods to
gauge {\em discrete} spaces. This lead to a new approach to the standard
model. Fr\"ohlich and collaborators  \cite{CFF} introduced gravity in this
context.
As an example of new symmetries in ``old'' theories we would like to
mention the work of
the Hamburg Group of Mack and collaborators \cite{Mac}: They  showed that the
internal symmetries of (low-dimensional) quantum field theories with
braid group statistics form a larger class than groups and were able to
motivate from basic axioms of such field theories that elements of weak
quasitriangular quasi Hopf algebras with $*$-structures should act as
symmetry operators in the Hilbert space of physical states.
Particle physics {\em phenomenology} from $q$-deformed Poincare algebra is
for example considered in \cite{DC},
where evidence of $q$-deformed space time
is sought in the observed spectrum of $\rho -a, \omega -f, K^{0}$ mesons and
remarkably
good agreement of theory and experiment, similar to, if not better, than
Regge pole theory is found.

The theory of non-commutative spaces is quite old, going back to early
work of Kac \cite{Ka}, Taksaki \cite{Ta} and Schwarz \& Enock \cite{SEn}.
Recently, the interest got revived by the discovery of non-trivial
examples. Quantum groups, which are a content rich example of quantum
spaces, arise naturally in several different branches
of physics and mathematics: in the context of integrable models, quantum
inverse scattering method, Yang-Baxter-equations and their solutions,
the so called $R$-matrices,
Knizhnik-Zamolodchikov equations, rational
conformal field theory and in the theory of knot and
ribbon invariants. Concerning knot theory
we should in particular mention the discovery
of the Jones polynomial \cite{Jo} and its generalizations, which were
then reconstructed from quantum $R$-matrices in the work of
Reshetikhin \& Turaev \cite{ReTu}
and later related to the topological Chern-Simons action by Witten
\cite{Wi}.
It was pointed out by Drinfeld that these examples
find an adequate description in the language of Hopf algebras.

There are at least three major approaches to the construction of quantum
deformations of Lie groups: Drinfeld and Jimbo introduce a deformation
parameter on the Lie algebra level and provided us with consistent
deformations for all semi-simple Lie groups. The St Petersburg Group
impose $q$-dependent commutation relations in terms of numerical
$R$-matrices among the matrix elements of
a matrix representation. Manin finally identifies quantum groups with
endomorphisms of quantum planes.

A large part of this thesis is devoted to the study of differential
calculi on quantum groups rather than quantum planes (these will
be considered in the second part of this thesis).
This path was in part taken because
quantum groups have more structure than quantum planes
and hence provide more guidance in the search for the correct axioms.
Apart from this purely practical reason,
the importance of differential geometry in the theory of (quantum)
Lie groups and vice versa
should, however, not be underestimated. Lie groups make their appearance
in differential geometry, {\it e.g.} in principal and associated fiber
bundles and in the infinite graded Lie algebra of the Cartan generators
($\Li, \I, \dl$).
Differential geometry
on group manifolds on the other hand
gives rise to the concepts of tangent Lie algebra
and infinitesimal representation --- and infinitesimal group generators,
like {\it e.g.} the angular momentum operator
play obviously a very important role in physics.
Covariant differential calculi on  quantum  {\em groups} were first introduced
by S. Woronowicz \cite{W2}; differential calculi on linear quantum {\em planes}
were constructed by J. Wess \& B. Zumino \cite{WZ}.
Since then much effort \cite{RTF,SWZ,B,CC,CW} has been devoted to the
construction of differential
geometry on  quantum groups.
Most approaches
are unfortunately rather specific: many papers deal with the subject by
considering the quantum group in question as defined by its R-matrix,
and others limit themselves to particular cases.
In this thesis we will develop a more abstract formulation which depends
primarily on the underlying Hopf algebraic structure of a quantum group;
it will therefore  be a generalization of many previously obtained
results, and the task of constructing specific examples of differential
calculi is greatly simplified. We have to stop short of giving a ``cook book
recipe'', however, because of case specific problems in the identification of
finite bases of generators.

The thesis is divided into two parts: Part I studies vector fields on
quantum groups; an algebraic and a geometric construction of a
bicovariant quantum algebra of differential operators is given.
Here we are mainly interested in the underlying Hopf algebra and
bicovariance considerations, introducing the pure braid group and
the canonical element in this context. Part II introduces additional
structure in form of  a Cartan calculus of
differential forms, Lie derivatives and
inner derivations; it is devoted to
differential calculi on quantum groups and quantum planes
and examples of their application.

\part{Bicovariant Quantum Algebras}

\chapter{Quantum Algebras and Quantum Groups}

\section{Introduction}

There are two dual approaches to the quantization of Lie groups.
Drinfeld \cite{Df} and Jimbo \cite{Ji} have given quantum deformations of
all simple Lie algebra in terms of a numerical parameter $q$. For the
case of SL${}_q(2)$ one has for instance
\begin{equation}
[H,X_{\pm }] = \pm  2 X_{\pm }, \z [X_{+},X_{-}] = \frac {q^{H} - q^{-H}}{q -
q^{-1}}
\end{equation}
and consistent rules for taking tensor product representations, given in
terms of coproducts, that we
will come back to later.
The second approach is due to the Russian school of Faddeev, Reshetikhin
and Takhtadzhyan. Consider again SL${}_q(2)$ which can be defined in terms of
a two by two matrix
\begin{equation}
T = \left(\begin{array}{cc} a & b\\c & d\end{array}\right),
\end{equation}
its fundamental representation. But instead of behaving like {\boldmath
$C$}-numbers, the group parameters $a,b,c,d$ now obey non-trivial
commutation relations
\begin{equation}
\begin{array}{c}
ab=qba,\z ac=qca,\z bc=cb,\\
bd=qdb,\z cd=qdc,\z ad-da=\lambda bc
\end{array}
\end{equation}
where $\lambda =(q-q^{-1})$, and
\begin{equation}
\mbox{det}_{q}(T)=ad-qbc=1.
\end{equation}
The remarkable property of such quantum matrices is that, given two
identical but mutually commuting copies of these matrices, their matrix
product is again a quantum matrix who's elements satisfy the same
commutation relations, as given above. Later we will express this
property in terms of the coproduct of $T$, which is an algebra
homomorphism.

In the following we will give a more formal introduction to quantum
groups.

\subsection{Quasitriangular Hopf Algebras}

A Hopf algebra \A\ is an algebra $(\A\,\cdot ,+,k)$ over a field $k$,
equipped with a coproduct $\Delta :\A \rightarrow \A \otimes \A$,
an antipode $S:\A \rightarrow \A$,
and a counit $\epsilon :\A \rightarrow k$, satisfying
\begin{eqnarray}
(\Delta \otimes i\!d)\Delta (a) & = & (i\!d \otimes \Delta )\Delta (a), \z
\mbox{(coassociativity),}\\
\cdot (\epsilon \otimes i\!d)\Delta (a) & = & \cdot (i\!d \otimes
\epsilon )\Delta (a) = a, \z
\mbox{(counit),}\\
\cdot (S \otimes i\!d)\Delta (a) & = & \cdot (i\!d \otimes S)\Delta (a)
= 1 \epsilon (a), \z
\mbox{(coinverse),}
\label{coalgebra}
\end{eqnarray}
for all $a \in \A$. These axioms are dual to the axioms of an algebra.
There are also a number of consistency conditions between the algebra
and the coalgebra structure,
\begin{eqnarray}
\Delta (a b) & = & \Delta (a) \Delta (b),\\
\epsilon (a b) & = & \epsilon (a) \epsilon (b),\\
S(a b) & = & S(b) S(a), \z \mbox{(antihomomorphism)},\\
\Delta (S(a)) & = & \tau (S \otimes S)\Delta (a), \z
\mbox{with}\x\tau (a \otimes b) \equiv b \otimes a,\\
\epsilon (S(a)) & = & \epsilon (a),\z\mbox{and}\\
\Delta (1) & = & 1 \otimes 1,\z S(1) = 1,\z \epsilon (1) = 1_{k},
\end{eqnarray}
for all $a,b \in \A$. We will often use Sweedler's \cite{SW} notation for the
coproduct:
\begin{equation}
\Delta (a) \equiv a_{(1)} \otimes a_{(2)}\z \mbox{\em (summation
is understood)\/.}
\label{sweedler}
\end{equation}
Note that a Hopf algebra is in general
non-cocommutative, i.e. $\tau \circ \Delta \neq \Delta $.

A quasitriangular Hopf algebra \U\ \cite{Df} is a Hopf algebra with a
{\em universal\/} $\R \in$ $\U \hat{\otimes} \U$ \x that keeps the
non-cocommutativity under control,
\begin{equation}
\tau (\Delta (a)) = \R \Delta (a) \R^{-1},
\label{quasi}
\end{equation}
and satisfies,
\begin{eqnarray}
(\Delta \otimes i\!d) \R & = & \R^{13} \R^{23},\x \mbox{and}\nonumber \\
(i\!d \otimes \Delta ) \R & = & \R^{13} \R^{12},
\label{coprodR}
\end{eqnarray}
where {\em upper\/} indices denote the position of the components of \R\
in the tensor product {\em algebra\/}\x $\U\hat{\otimes}
\U\hat{\otimes} \U$ :
if \x $\R \equiv \alpha _{i} \otimes \beta _{i}$ \x {\em(summation is
understood),\/} then {\it e.g.}  \x $\R^{13}\equiv \alpha _{i} \otimes 1
\otimes\beta _{i}$ . Equation (\ref{coprodR}) states that \R\ generates
an algebra map \x $\langle \R , .
\otimes i\!d \rangle \!\! :$ $\U^{*} \rightarrow
\U$ \x and an antialgebra map \x $\langle \R , i\!d \otimes . \rangle \!\!:$
$\U^{*} \rightarrow \U$.\footnote{Notation: ``.'' denotes an argument
to be inserted and ``$i\!d$'' is the
identity map, {\it e.g.} $\langle \R, i\!d \otimes f \rangle $ $\equiv
\alpha _{i} \langle \beta _{i} , f \rangle $; $\R \equiv \alpha _{i} \otimes
\beta _{i} \in \U \hat{\otimes} \U,
$ $f \in \U^{*}$.}
The following equalities are consequences of the
axioms:
\begin{eqnarray}
\R^{12}\R^{13}\R^{23} & = & \R^{23}\R^{13}\R^{12},\x \mbox{(quantum
Yang-Baxter equation),}\\
(S \otimes i\!d)\R & = & \R^{-1},\\
(i\!d \otimes S)\R^{-1} & = & \R,\x\mbox{and}\\
(\epsilon \otimes i\!d)\R & = & (i\!d \otimes \epsilon )\R = 1.
\end{eqnarray}
An example of a quasitriangular Hopf algebra that is of particular
interest here is the deformed universal enveloping algebra \uqg\ of a
Lie algebra {\Deutsch g}. Dual to \uqg\ is the Hopf algebra of\ ``functions
on the quantum group'' \mbox{\fun\ ;} in fact, \uqg\ and \fun\ are
{\em dually paired}. We call two Hopf algebras \U\ and \A\ dually paired
if there exists a non-degenerate inner product $<\;,\;>:$
$\U \otimes \A \rightarrow k$, such that:
\begin{eqnarray}
<x y,a> & = & <x \otimes y, \Delta (a)> \equiv <x,a_{(1)}><y,a_{(2)}>,
\label{multinduced}\\
<x,a b> & = & <\Delta (x),a \otimes b> \equiv <x_{(1)},a><x_{(2)},b>,
\label{inducedmult}\\
<S(x),a> & = & <x,S(a)>,\\
<x,1> & = & \epsilon (x),\z \mbox{and}\z <1,a>=\epsilon (a),
\end{eqnarray}
for all  $x,y \in \U$ and $a,b \in \A$. In the following we will assume
that \U\ (quasitriangular) and \A\ are dually paired Hopf algebras,
always keeping \uqg\ and \fun\ as concrete realizations in mind.

In the next subsection we will sketch how to obtain \fun\ as a matrix
representation of \uqg.

\subsection{Dual Quantum Groups}
\label{Dual}

We cannot speak about a quantum group $\mbox{\Deutsch G}_{q}$ directly,
just as ``phase
space'' loses its meaning in quantum mechanics, but in the spirit of
geometry on non-commuting spaces the (deformed) functions on the quantum
group \fun\ still make sense.
This can be made concrete, if we write $\fun$ as a pseudo matrix group
\cite{W1}, generated by the elements of an $N \times N$
matrix $A \equiv ({A^{i}}_{j})_{i,j = 1...N} \in M_{N}(\fun)$%
\footnote{We are automatically dealing with
$GL_{q}(N)$ unless there are explicit
or implicit restrictions on the matrix elements of $A$.}%
. We require that ${\rho ^{i}}_{j} \equiv < .\; , {A^{i}}_{j}>$ be a
matrix representation of $\uqg$, i.e.
\begin{equation}
\begin{array}{lr}
{\rho ^{i}}_{j} : \uqg \rightarrow k, & \\
{\rho ^{i}}_{j}(xy) = \sum _{k}^{} {\rho ^{i}}_{k}(x) {\rho ^{k}}_{j}(y), &\x
\mbox{for }
\forall x,y \in \uqg,
\label{rep}
\end{array}
\end{equation}
just like in the classical case%
\footnote{The quintessence of this construction is that the coalgebra
of \fun\ is undeformed i.e. we keep the familiar
matrix group expressions of the classical theory.}%
. The universal $\R  \in \uqg \hat{\otimes}
\uqg$ coincides in this representation with the numerical $R$-matrix:
\begin{equation}
<\R,{A^{i}}_{k} \otimes {A^{j}}_{l}> = {R^{ij}}_{kl}.
\end{equation}
It immediately follows from (\ref{multinduced}) and (\ref{rep}) that the
coproduct of $A$ is given by matrix multiplication \cite{W1,RTF},
\begin{equation}
\Delta A = A \dot{\otimes} A, \z\mbox{i.e.}\x\Delta ({A^{i}}_{j})=
{A^{i}}_{k}\otimes{A^{k}}_{j}. \label{coofA}
\end{equation}
Equations (\ref{quasi}), (\ref{inducedmult}),
and (\ref{rep})  imply \cite{Df,RTF},
\begin{equation}
\begin{array}{rcl}
<x , {A^{j}}_{s} {A^{i}}_{r}>& = & <\Delta x ,{A^{j}}_{s} \otimes
{A^{i}}_{r}>\\
&=&< \tau \circ \Delta x,{A^{i}}_{r} \otimes {A^{j}}_{s}>\\
&=&<\R (\Delta x) \R^{-1}, {A^{i}}_{r} \otimes {A^{j}}_{s}>\\
&=&{R^{ij}}_{kl}<\Delta x , {A^{k}}_{m} \otimes
{A^{l}}_{n}>{(R^{-1})^{mn}}_{rs}\\
&=&< x , {R^{ij}}_{kl} {A^{k}}_{m} {A^{l}}_{n} {(R^{-1})^{mn}}_{rs}>,
\end{array}
\end{equation}
i.e. the matrix elements of $A$
satisfy the following commutation relations,
\begin{equation}
{R^{ij}}_{kl} {A^{k}}_{m} {A^{l}}_{n}
                             =   {A^{j}}_{s} {A^{i}}_{r} {R^{rs}}_{mn},
\end{equation}
which can be written more compactly in tensor product notation as:
\begin{eqnarray}
R_{12} A_{1} A_{2} & = &   A_{2} A_{1} R_{12};\label{AA} \\
R_{12} = (\rho _{1} \otimes \rho _{2}) (\R)&  \equiv
&  <\R , A_{1} \otimes A_{2}>.
\end{eqnarray}
{\em Lower\/} numerical indices shall denote here the position of the
respective matrices in the tensor product of {\em representation spaces
(modules)\/}.
The contragredient representation \cite{Rn}\x
$\rho ^{-1} = < .\; ,SA>$\x gives the antipode
of \fun\ in matrix form:\x$S({A^{i}}_{j}) = {(A^{-1})^{i}}_{j}$.
The counit is:\x$\epsilon ({A^{i}}_{j}) = <1,{A^{i}}_{j}> = {\delta ^{i}}_{j}$.

Higher (tensor product) representations can be
constructed from $A$:\\
$A_{1} A_{2}$, $A_{1} A_{2} A_{3}$, \ldots , $A_{1} A_{2} \cdots A_{m}$.
We find numerical \RR-matrices \cite{Md1} for any pair of such
representations:
\begin{equation}
  \begin{array}{rcl}
  \hspace{-5mm}\RR_{\underbrace{(1',2',\ldots,n')}_{I},
\underbrace{(1,2,\ldots,m)}_{I\!I}}
  &\equiv & <\R,A_{1'} A_{2'}\cdots A_{n'} \otimes A_{1} A_{2}\cdots A_{m}>\\
  &   =  &
           \begin{array}[t]{cccc}
           R_{1'm}&\cdot \; R_{1'(m-1)}&\cdot \;\ldots\;\cdot & R_{1'1}\\
  \cdot \; R_{2'm}&\cdot \; R_{2'(m-1)}&\cdot \;\ldots\;\cdot & R_{2'1}\\
        \vdots    &    \vdots          &                    &\vdots  \\
  \cdot \; R_{n'm}&\cdot \; R_{n'(m-1)}&\cdot \;\ldots\;\cdot & R_{n'1}
           \end{array}
  \end{array}
\label{bigR}
\end{equation}
Let \x$\Aa_{I} \equiv A_{1'} A_{2'}\cdots A_{n'}$\x and \x$\Aa_{I\!I} \equiv
A_{1} A_{2}\cdots A_{m}$,\x then:
\begin{equation}
\RRI \Aa_{I} \Aa_{I\!I} = \Aa_{I\!I} \Aa_{I} \RRI.
\label{RRA}
\end{equation}
\RRI is the ``partition function''
of exactly solvable models. We will need it in section~\ref{S:MdotP}.

We can also write \uqg\ in matrix form \cite{RTF,Rn} by taking representations
$\varrho$ --- {\it e.g.} $\varrho = <.\,,\Aa>$ --- of \R\ in its first or
second tensor product space,
\begin{eqnarray}
L^{+}_{\varrho} & \equiv & (i\!d \otimes \varrho)(\R ) , \z
        L^{+} \; \equiv \; <\R^{21},A \otimes i\!d>,\\
SL^{-}_{\varrho} & \equiv & (\varrho \otimes i\!d) (\R ), \z
        SL^{-} \; \equiv \; <\R,A \otimes i\!d>,\\
L^{-}_{\varrho} & \equiv & (\varrho \otimes i\!d) (\R^{-1}) , \z
        L^{-} \; \equiv \; <\R,S A \otimes i\!d>.
\end{eqnarray}
The commutation relations for all these matrices follow directly from
the quantum Yang-Baxter equation, {\it e.g.}
\begin{equation}
\begin{array}{rcrcccl}
0 & = &  <&\!\!\!\R^{23}\R^{13}\R^{12}& -& \R^{12}\R^{13}\R^{23}\;,&
                                         i\!d \otimes A_{1} \otimes A_{2}>\\
  & = && \!\!\! R_{12} L^{+}_{2} L^{+}_{1}& -& L^{+}_{1} L^{+}_{2} R_{12}\;,&
\end{array}
\end{equation}
where upper ``algebra'' indices should not be confused with lower ``matrix''
indices.
Similarly one finds:
\begin{eqnarray}
R_{12} L^{-}_{2} L^{-}_{1}&=& L^{-}_{1} L^{-}_{2} R_{12},\\
R_{12} L^{+}_{2} L^{-}_{1}&=& L^{-}_{1} L^{+}_{2} R_{12}.
\end{eqnarray}

\section{Quantized Algebra of Differential Operators}
\setcounter{footnote}{0}

Here we would like to show how two dually paired Hopf algebras can be combined
using a Hopf algebra analog of a semi-direct product construction.
We obtain an algebra of differential operators consisting of
elements of \uqg\ with function coefficients from \fun. Both the
inner product with and the action on elements of \fun\ by elements of
\uqg\ will be encoded in the product of the new combined algebra.
Using this construction we can avoid having to work with
convolution products and similar abstract and sometimes clumsy
constructions. In fact we will be able to extend the $R$-matrix approach
of \cite{RTF} so that all (Hopf algebra) relations can be written
in terms of simple commutation relations of operator-valued
matrices; see for example \cite{SWZ}.

\subsection{Actions and Coactions}

\paragraph{Actions.}
A {\em left action} of an algebra $A$ on a vector space $V$ is
a bilinear map,
\begin{equation}
\tr :\; {A} \otimes {V}
\rightarrow {V}:\; x \otimes v \mapsto x \tr v,
\end{equation}
such that:
\begin{equation}
(x y) \tr v = x \tr (y \tr v),\z 1 \tr v = v.
\end{equation}
$V$ is called a left $A$-module.
In the case of the left action of a Hopf algebra $H$ on an algebra $A'$
we can in addition ask that this action preserve the algebra structure
of $A'$, i.e. \x $x \tr (a b) =
(x_{(1)} \tr a)\,(x_{(2)} \tr b)$\footnote{$x\tr$ is called a
{\em generalized derivation}.} \x and
\x $x \tr 1 = 1\,\epsilon (x)$, \x for all \x $x \in H,\: a,b \in A'$.
$A'$ is then called a left $H$-module algebra.
Right actions and modules are defined in complete analogy.
A left action of an algebra on a (finite dimensional) vector space
induces a right action of the same algebra on the dual vector space
and vice versa, via pullback. Of particular interest to us is the
left action of \U\ on \A\ induced by the right multiplication in \U:
\begin{equation}
\begin{array}{l}
<y , x \tr a > := <y x , a> = <y \otimes x,\Delta a> =
<y ,a_{(1)} < x , a_{(2)} >\!> ,\\
\Rightarrow \x x \tr a = a_{(1)} < x , a_{(2)} >,
\z \mbox{for } \forall \;x,y \in \U,\,a \in \A,
\end{array}
\label{UonA}
\end{equation}
where again $\Delta a \equiv a_{(1)} \otimes a_{(2)}$.
This action of \U\ on \A\ respects the algebra structure of \A, as
can easily be checked. The action of \U\ on itself given by right
or left multiplication does {\em not} respect the algebra structure of \U;
see however (\ref{adjoint}) as an example of an algebra-respecting
``inner'' action.
\paragraph{Coaction.}
In the same sense as comultiplication is the dual operation to
multiplication, {\em right {\rm or} left coactions} are dual to left or
right actions respectively. One therefore defines a right coaction
of a coalgebra $C$ on a vector space $V$ to
be a linear map,
\begin{equation}
\Delta _{C} : \; V \rightarrow V \otimes C:\;
v \mapsto \Delta _{C}(v) \equiv v^{(1)} \otimes v^{(2)'},
\end{equation}
such that,
\begin{equation}
(\Delta _{C} \otimes i\!d)\Delta _{C} = (i\!d \otimes \Delta )\Delta _{C},\z
(i\!d \otimes \epsilon )\Delta _{C} = i\!d.
\end{equation}
Following \cite{Md1} we have introduced here a notation for the
coaction that resembles Sweedler's notation (\ref{sweedler}) of the coproduct.
The prime on the second factor marks a right coaction.
If we are dealing with the right coaction of a Hopf algebra $H$ on
an algebra $A$, we say that the coaction respects the algebra structure
and $A$ is a right $H$-comodule algebra, if \x $\Delta _{H}(a\cdot b) =
\Delta _{H}(a) \cdot \Delta _{H}(b)$ and $\Delta _{H}(1) = 1 \otimes 1$,
\x for all \x$a,b \in A$. In the case of a coaction on a Hopf algebra,
there might be additional compatability relations between its coproduct
and antipode and the coaction.
\paragraph{Duality of Actions and Coactions.}
If the coalgebra $C$ is dual to an algebra $A$ in the sense of
(\ref{multinduced}), then a {\em right} coaction of $C$ on $V$ will
induce a {\em left} action of $A$ on $V$ and vice versa, via
\begin{equation}
x \tr v = v^{(1)}<x,v^{(2)'}>,\z {\em (general)},
\end{equation}
for all \x$x \in A,\;v \in V$.
Applying this general formula to the specific case
of our dually paired Hopf algebras \U\ and \A, we see that the right
coaction $\DA$ of \A\ on itself, corresponding to the left action
of \U\ on \A, as given by (\ref{UonA}), is just the coproduct
$\Delta $ in \A, i.e. we pick:
\begin{equation}
\DA (a) \equiv a^{(1)} \otimes a^{(2)'}
= a_{(1)} \otimes a_{(2)},\z \mbox{for }
\forall a \in \A.
\end{equation}

To get an intuitive picture we may think of the left action (\ref{UonA})
as being a generalized {\em specific left translation} generated by a
left invariant ``tangent vector'' $x \in \U$ of the quantum group.
The coaction $\DA$ is then the generalization of an {\em unspecified
translation}.
If we supply for instance a vector $x \in \U$ as transformation parameter, we
recover the generalized specific transformation (\ref{UonA});
if we use $1 \in \U$, i.e. evaluate at the ``identity of the quantum
group'', we get the identity transformation; but the quantum analog
to a classical finite translation through left or right multiplication by
a {\em specific} group element does not exist.
In section~\ref{genervf} we will give a much more detailed and
geometric discussion of these matter.

\paragraph{Quantum Matrix Formulation.}
The dual quantum group in its matrix form stays very close to the classical
formulation and we want to use it to illustrate some of the above
equations. For the matrix $A \in M_{N}(\fun)$ and $x \in \uqg$ we find,
\begin{equation}
\begin{array}{l}
\fun     \rightarrow \fun \otimes \fun : \\
\DA \, A      =      A A', \z \mbox{(right coaction)},
\end{array}
\end{equation}
\begin{equation}
\begin{array}{l}
\fun     \rightarrow \fun \otimes \fun :  \\
\AD \, A      =      A' A, \z \mbox{(left coaction)},
\end{array}
\end{equation}
\begin{equation}
\begin{array}{l}
\uqg \otimes \fun \rightarrow \fun : \\
x \tr A  =  A <x,A>, \z  \mbox{(left action)},
\label{xonA}
\end{array}
\end{equation}
where matrix multiplication is implied. Following common custom we have
used a prime to distinguish copies of the matrix $A$ in different
tensor product spaces. We see that in complete analogy to the
classical theory of Lie algebras, we first evaluate $x \in \uqg$, interpreted
as a left invariant vector field,  on $A \in M_{n}(\fun)$ at the
``identity of $\mbox{\Deutsch G}_{q}$'', giving a numerical matrix
$< x , A > \in M_{n}(k)$,
and then shift the result by left matrix multiplication with $A$ to an
unspecified ``point" on the quantum group.
Unlike a Lie group, a quantum group is not a manifold in the classical
sense and we hence cannot talk about its elements, except for the identity
(which is also the counit of \fun).
For $L^{+} \in M_{N}(\uqg)$ equation (\ref{xonA}) becomes,
\begin{equation}
 L^{+}_{2} \tr A_{1} \x = \x A_{1} < L^{+}_{2} , A_{1} > \x = \x A_{1} R_{12},
\end{equation}
and similarly for $L^{-} \in M_{N}(\uqg)$:
\begin{equation}
 L^{-}_{2} \tr A_{1} \x = \x A_{1} < L^{-}_{2} , A_{1} > \x = \x A_{1}
R^{-1}_{21}.
\end{equation}

\subsection{Commutation Relations}
\label{ComRel}

The left action of $x \in \U$ on products in \A\ , say $b f$, is given
via the coproduct in \U\ ,
\begin{equation}
\begin{array}{rcl}
x \tr b f & = & (b f)_{(1)} <x,(b f)_{(2)}> \\
& = & b_{(1)} f_{(1)} <\Delta (x),b_{(2)} \otimes f_{(2)}> \\
& = & \cdot \Delta x \tr (b \otimes f)
= b_{(1)} <x_{(1)},b_{(2)}> \: x_{(2)} \tr f.
\end{array}
\label{prod}
\end{equation}
Dropping the ``$\tr$'' we can write this for
arbitrary functions $f$ in the form of commutation
relations,
\begin{equation}
x\;b = \Delta x \tr (b \otimes i\!d) = b_{(1)} <x_{(1)},b_{(2)}> \: x_{(2)} .
\label{commrel}
\end{equation}
This commutation relation provides \x $\A \otimes \U$\x
with an algebra structure via the {\em cross product},
\begin{equation}
\begin{array}{l}
\cdot :\;(\A \otimes \U) \otimes (\A \otimes \U) \rightarrow
\A \otimes \U: \\
a x \otimes b y \mapsto a x\cdot b y
= a \: b_{(1)} <x_{(1)},b_{(2)}> \, x_{(2)} \: y.
\end{array}
\label{crossprod}
\end{equation}
That $\A \otimes \U$ is indeed an associative algebra with this multiplication
follows from the Hopf algebra axioms;
it is denoted \A \cross \U\ and we call it the {\em quantized algebra
of differential
operators}. The commutation relation (\ref{commrel})
should be interpreted as a product in \A \cross \U\ . (Note that we omit
$\otimes$-signs wherever they are obvious, but we sometimes
insert a product sign ``$\cdot $'' for clarification of the formulas.)
Right actions and the corresponding commutation relations are also possible:
\newcommand{\links}[1]{\stackrel{\leftarrow}{#1}}
$ b \triangleleft \links{x}  =   <\links{x},b_{(1)}> b_{(2)}$  and
$b \links{x}  =   \links{x}_{(1)} <\links{x}_{(2)},b_{(1)}> b_{(2)}$.

Equation (\ref{commrel}) can be used to calculate arbitrary inner products of
\U\ with \A\ , if we define \cite{Z2}  a {\em right vacuum} ``$>$" to act like
the
counit in \U\ and a {\em left vacuum} ``$<$" to act like the counit in \A\ ,
\begin{equation} \begin{array}{rcl}
<x\: b> & = & <b_{(1)} <x_{(1)},b_{(2)}> \: x_{(2)}> \\
      & = &\epsilon (b_{(1)}) <x_{(1)},b_{(2)}> \: \epsilon (x_{(2)}) \\
      & = & <\cdot (i\!d\otimes\epsilon )\Delta (x),\:
            \cdot (\epsilon \otimes i\!d)\Delta (b)> \\
      & = & <x,b>,\z \mbox{for } \forall \x x \in \U ,\, b \in \A .
\end{array} \end{equation}
Using only the right vacuum we recover formula (\ref{UonA}) for
left actions,
\begin{equation}
\begin{array}{rcl}
x\: b > & = & b_{(1)} <x_{(1)},b_{(2)}>  x_{(2)}> \\
      & = & b_{(1)} <x_{(1)},b_{(2)}>  \epsilon (x_{(2)}) \\
      & = & b_{(1)} < x , b_{(2)}> \\
      & = & x \tr b,\z \mbox{for } \forall \x x \in \U ,\, b \in \A .
\end{array}
\label{rightvac}
\end{equation}
As an example we will write the preceding equations for $A$, $L^{+}$, and
$L^{-}$:
\begin{eqnarray}
L^{+}_{2} A_{1} & = & A_{1} R_{12} L^{+}_{2},\z \mbox{(commutation relation for
$L^{+}$ with $A$),}\label{LPA}\\
L^{-}_{2} A_{1} & = & A_{1} R^{-1}_{21} L^{-}_{2},\z \mbox{(commutation
relation for
$L^{-}$ with $A$),}\label{LMA}\\
< A & = & I <,\z \mbox{(left vacuum for $A$),}\\
L^{+}> & = & L^{-}> \x = \x > I,\z \mbox{(right vacua for $L^{+}$ and
$L^{-}$).}
\end{eqnarray}

Equation (\ref{rightvac}) is not the only way to
define left actions of \U\ on \A\ in terms of the product in \A \cross \U\ .
An alternate definition utilizing the coproduct and antipode in \U\ ,
\begin{equation}
\begin{array}{rcl}
x_{(1)}\: b\, S(x_{(2)}) & =
& b_{(1)} <x_{(1)},b_{(2)}>  x_{(2)}\, S(x_{(3)})%
\footnotemark \\
        & = & b_{(1)} <x_{(1)},b_{(2)}>  \epsilon (x_{(2)})\\
        & = & b_{(1)} < x , b_{(2)}> \\
        & = & x \tr b,\z \mbox{for }
              \forall \x x \in \U ,\, b \in \A ,
\end{array}
\end{equation}
is in a sense more satisfactory because it readily generalizes
to left actions of \U\ on \A \cross \U\ ,
\footnotetext{Notation:\z $
     \begin{array}[t]{l}
     (\Delta \otimes i\!d)\Delta (x) = (i\!d \otimes \Delta )\Delta
     (x) =  x_{(1)} \otimes x_{(2)} \otimes x_{(3)} = \Delta ^{2}(x),\\
     x_{(1)} \otimes x_{(2)} \otimes x_{(3)} \otimes x_{(4)}= \Delta ^{3}(x),\x
     \mbox{etc., see \cite{Md1}.}
     \end{array}$} \addtocounter{footnote}{-1}
\begin{equation}
\begin{array}{rcl}
x \tr b y & := & x_{(1)}\: b y\, S(x_{(2)}) \\
        & = & x_{(1)}\: b\, S(x_{(2)})\;x_{(3)}\: y\, S(x_{(4)})%
\footnotemark \vspace{2mm}\\
        & = & (x_{(1)} \tr b)\,( x_{(2)} \ad y),
            \z \mbox{for } \forall \x x,y \in \U ,\, b \in \A ,
\end{array}
\label{leftad}
\end{equation}
where we have introduced the left adjoint (inner) action in \U\ :
\begin{equation}
x \ad y \x = \x x_{(1)} y \,S(x_{(2)}), \z \mbox{for } \forall \x x,y \in \U .
\label{adjoint}
\end{equation}

\subsection{Complex Structure}

In the previous section we constructed a generalized semi-direct product
algebra \A \cross \U\ using commutation relations
\begin{equation}
x\, a = a_{(1)}<x_{(1)},a_{(2)}> x_{(2)}
\end{equation}
that allow ordering of all
elements of \A \cross \U\ in the form $\A\otimes\U$. After some easy
manipulations we derive an alternative commutation relation
\begin{equation}
a\, x = x_{(2)}<S^{-1} x_{(1)},a_{(2)}> a_{(1)},\label{opcomm}
\end{equation}
good for ordering in the form $\U\otimes\A$.
We can now introduce complex conjugation on \A \cross \U\ as an
antimultiplicative involution, {\em i.e.}
\begin{equation}
\overline{a}\, \overline{x} = \overline{x a} =
\overline{x_{(2)}}<x_{(1)},a_{(2)}>^{*}
\overline{a_{(1)}}.
\end{equation}
Comparing this equation to equation (\ref{opcomm}) gives the following
natural choices:
\begin{eqnarray}
<x,a>^{*} & = & <S^{-1} \overline{x},\overline{a}>,\\
\Delta (\overline{a}) & = & \overline{a_{(1)}}\otimes \overline{a_{(2)}},
\end{eqnarray}
and hence
\begin{equation}
S^{-1} \overline{x} = \overline{S x}.
\end{equation}
In this context let us also define a {\em unitary representation}: A
unitary representation $T \in M_{n}(\A)$ satisfies $T^{\dagger } \equiv
\overline{T}^{t} = S T$ so that
\begin{equation}
<\overline{x}, T> = <x,\overline{S T}>^{*} = <x,T>^{\dagger },
\end{equation}
{\em i.e.} the matrix representing the complex conjugate of an element
in \U\
is equal to the adjoint of the matrix representing the original element.
In the next section we would like to give two examples to
illustrate the material presented so far. The first one, SU${}_{q}$(2), is
by now the standard example for a quantum group; it is due to \cite{W0}.
We pick it as a
representative for the $R$-matrix approach to quantum groups. Dropping
the reality and the unit determinant conditions one can obtain the
further examples of SL${}_{q}$(2) and GL${}_{q}$(2) respectively. The
second example is  the Quantum Euclidean Group ---
we show how one can obtain it
via a contraction procedure from SU${}_{q}$(2);
a more complete treatment of this original
work can be found in \cite{SWZ2}.

\section{SU${}_{q}$(2) and E${}_{q}$(2)}

In this section we will present  SU${}_{q}(2)$  and show how the deformed
Euclidean
group E${}_{q}(2)$ and its dual, the deformed Lie algebra $U_{q}$su(2), can
be obtained from it by contraction.
The Euclidean group E(2) is a simple example of an inhomogeneous group.
Deformations of such groups in general have been studied in
\cite{SWW}.  Celeghini {\it et al.} \cite{CGST} found a deformation of $Ue(2)$
by
contracting $U_{q}$su(2) and simultaneously letting the deformation
parameter $h \equiv \ln q$\, go to zero. Here we are interested in the
case where $q$ is left untouched.

\subsection{E${}_{q}(2)$ by contraction of SU${}_{q}(2)$}

The commutation relations for SU${}_{q}(2)$
\cite{RTF,Z2},
may be written in compact matrix notation as
\begin{eqnarray}
\label{RTT}
R_{12}T_{1}T_{2} = T_{2}T_{1}R_{12}, & det_{q}T = 1, & T^{\dagger } = T^{-1},
\nonumber \\
\Delta (T) = T \dot{\otimes} T, & \epsilon (T) = I, & S(T)=T^{-1},
\end{eqnarray}
where
\begin{eqnarray}
T =  \left( \begin{array}{lr}
\alpha & -q \bar{\gamma } \\ \gamma & \bar{\alpha }
\end{array} \right), & R = q^{-1/2} \left( \begin{array}{cccc}
q & 0 & 0 & 0 \\
0 & 1 & 0 & 0 \\
0 & \lambda & 1 & 0 \\
0 & 0 & 0 & q
\end{array} \right),
\end{eqnarray}
$\lambda = q-q^{-1}$ and $\bar{q}=q$.  Now set
$$\alpha \equiv v,\x \bar{\alpha } \equiv  \bar{v},\x \gamma \equiv \ell
\bar{n}\x \mbox{and}\x\bar{\gamma } \equiv \ell n,$$
where
$\ell \in \real - \{0\}$ is a contraction parameter.  Written in terms of
$v$, $\bar{v}$, $n$ and $\bar{n}$, relations (\ref{RTT}) become
\begin{eqnarray}
&det_{q}T=v \bar{v} + q^{2} \ell ^{2} n \bar{n} = \bar{v}v + \ell ^{2} \bar{n}
n
=1,& \nonumber \\
&n \bar{n} = \bar{n} n,\,\, v n = q n v, \,\, v \bar{n} = q \bar{n} v,\,\,
\mbox{\it etc.} & \nonumber
\end{eqnarray}
and give E${}_{q}(2)$ in agreement with  \cite{W3}
as a contraction of SU${}_{q}(2)$ in the limit
$\ell \rightarrow 0$:
\begin{eqnarray}
\label{Eq2}
v\bar{v} = \bar{v} v =1, & n \bar{n} = \bar{n} n, & vn=qnv,\nonumber \\
n \bar{v} = q \bar{v} n, & v \bar{n}=q \bar{n} v, & \bar{n} \bar{v} = q
\bar{v} \bar{n},\nonumber \\
\Delta (n)=n \otimes \bar{v} + v \otimes n, & \Delta (v) = v \otimes v,
\nonumber \\
\Delta (\bar{n}) = \bar{n} \otimes v + \bar{v} \otimes \bar{n}, & \Delta
(\bar{v}) = \bar{v} \otimes \bar{v},\\
\epsilon (n) = \epsilon (\bar{n}) = 0, & \epsilon (v) = \epsilon (\bar{v}) =1,
\nonumber \\
S(n)=-q^{-1}n,& S(v)=\bar{v},\nonumber \\
S(\bar{n}) = -q \bar{n}, & S(\bar{v})=v. \nonumber
\end{eqnarray}
It is convenient to introduce the operators
$\theta $, $\bar{\theta }$, $m$, and $\bar{m}$, defined by
\begin{eqnarray}
& v=e^{\frac {i}{2} \theta },\:\: \bar{\theta }=\theta ,\: \: m=nv, \:\:
\bar{m} = \bar{v} \bar{n}. &
\end{eqnarray}
In this basis, the coproducts take on the particularly nice form
\begin{eqnarray}
\Delta (m) =& m \otimes 1 + e^{i \theta } \otimes m, & \Delta ( \bar{m}) =
\bar{m} \otimes 1 + e^{-i \theta } \otimes \bar{m},\nonumber \\
\Delta (\theta )  =& \theta \otimes 1 + 1 \otimes \theta .
\end{eqnarray}
The matrix $E$ given by
\begin{equation}
E = \left( \begin{array}{ll}
e^{i \theta } & m \\ 0 & 1
\end{array} \right)
\end{equation}
satisfies the relations
\begin{eqnarray}
\Delta (E) = E \dot{\otimes} E, & S(E) = E^{-1}, & \epsilon (E) = I.
\end{eqnarray}
These are exactly the relations one would expect for an element of a quantum
matrix group.  Notice that the action of $E$ on the column vector $\left(
\begin{array}{c}
z \\ 1
\end{array} \right)$, where $z$ is a complex coordinate, is given by
\begin{eqnarray}
z \mapsto e^{i \theta } z + m, & \bar{z} \mapsto e^{-i \theta } \bar{z}+
\bar{m}.
\end{eqnarray}
We may therefore identify $E$ as an element of the deformed 2-dimensional
Euclidean group E${}_{q}(2)$.
Fun(E${}_{q}(2))$ is the algebra of all $C^{\infty }$ functions in the group
parameters of E${}_{q}(2)$, $i.e.$ the algebra spanned by ordered monomials in
$\theta $, $m$, and $\bar{m}$.  Thus, Fun(E${}_{q}(2))$ is taken to be $span \{
\theta ^{a} m^{b} \bar{m}^{c} \mid a,b,c =0,1,\ldots \}$.

\subsection{$U_{q}$e(2) by contraction of $U_{q}$su(2)}
\label{Peter2}

The deformed universal enveloping algebra $U_{q}$su(2), dual to
Fun(SU${}_{q}(2))$,
is generated by hermitian operators $H$, $X_{+}$, $X_{-}$ satisfying
\begin{eqnarray}
\label{HXX}
\comm{H}{X_{\pm }}= \pm 2 X_{\pm }, & \comm{X_{+}}{X_{-}} = \frac
{q^{H}-q^{-H}}{ q-q^{-1}},
\nonumber \\
\Delta (H) = H \otimes 1 + 1 \otimes H, & \Delta (X_{\pm }) =
X_{\pm } \otimes q^{H/2} + q^{-H/2} \otimes X_{\pm }, \nonumber \\
\epsilon (H)= \epsilon (X_{\pm })=0, \\
S(H) = -H, & S(X_{\pm }) = -q^{\pm 1} X_{\pm }. \nonumber
\end{eqnarray}
Following \cite{RTF} these relations can be rewritten as
\begin{eqnarray}
\label{RLL}
R_{12}L^{\pm }_{2}L^{\pm }_{1}=L^{\pm }_{1}L^{\pm }_{2}R_{12},&
R_{12}L^{+}_{2}L^{-}_{1}=L^{-}_{1}L^{+}_{2}R_{12},\nonumber \\
\Delta (L^{\pm }) = L^{\pm } \dot{\otimes} L^{\pm }, & \epsilon (L^{\pm })=I,
\\
S(L^{\pm })=(L^{\pm })^{-1}, \nonumber
\end{eqnarray}
where $L^{\pm }$ are given by
\begin{eqnarray}
L^{+} =  \left( \begin{array}{lr}
q^{-H/2} & q^{-1/2} \lambda X_{+} \\ 0 & q^{H/2}
\end{array} \right),&  L^{-} = \left( \begin{array}{lr}
q^{H/2} & 0 \\ -q^{1/2} \lambda X_{-} & q^{-H/2}
\end{array} \right).
\end{eqnarray}
Using this matrix notation, we can state the duality
between the group and the algebra by means of commutation relations
\begin{eqnarray}
\label{TRL}
L^{+}_{1} T_{2} = T_{2} R_{21} L^{+}_{1}, & L^{-}_{1} T_{2} = T_{2} R^{-1}_{12}
L^{-}_{1},
\end{eqnarray}
as explained in section~\ref{ComRel}. Equations (\ref{TRL}) are not only
consistent with the inner products
\begin{eqnarray}
<{L^{+}_{1}},{T_{2}}>=R_{21}, & <{L^{-}_{1}},{T_{2}}>=R^{-1}_{12},
\end{eqnarray}
given in \cite{RTF} but also contain information about the coproducts
of $L^{+}$, $L^{-}$ and $T$ so that equations (\ref{RLL}) can actually be
derived as
consistency conditions to (\ref{RTT}) and (\ref{TRL}).  Complex
conjugation can be defined as an involution
on the extended algebra generated by products of
$T$ and $L^{\pm }$. This agrees with
\begin{eqnarray}
\Delta (\bar{h}) = \overline{\Delta (h)}, & \overline{S(\bar{h})}=S^{-1}(h)
\end{eqnarray}
and
\begin{equation}
\label{dual}
<{\bar{\chi }},{h}>=<{\chi },{S^{-1}(\bar{h})}>^{*}.
\end{equation}
Unitarity of $T$ then
implies $(L^{+})^{\dagger }= (L^{-})^{-1}$, $i.e.$ $\bar{H} = H$,
$\overline{X_{\pm }}=X_{\mp }$.
In the present case equations (\ref{TRL}) become
\begin{eqnarray}
\label{ga}
Hv=vH-v, & X_{+}v=q^{1/2}vX_{+}-\ell qnq^{H/2}, & X_{-}v=q^{1/2}vX_{-},
\nonumber \\
\ell H \bar{n}=\ell (\bar{n}H-\bar{n}), & \ell X_{+} \bar{n} = q^{1/2} \bar{n}
\ell X_{+} + \bar{v}q^{H/2}, & \ell X_{-} \bar{n}= \ell q^{1/2}\bar{n}X_{-},
\end{eqnarray}
plus the complex conjugate relations.

The way that the deformation parameter $\ell $ appears in these relations
suggests
the definition of new operators
$$P_{+} \equiv \ell X_{+},\x P_{-} \equiv  \overline{P_{+}} =\ell
X_{-}\x\mbox{and}\x J \equiv H/2,$$
so that we will retain
non-trivial commutation relations for $P_{\pm }$ and $J$ with $v$, $\bar{v}$,
$n$ and $\bar{n}$ in the limit $\ell \rightarrow 0$.
Inserting $P_{\pm }$ and $J$ into equation (\ref{HXX})  we
obtain $U_{q}$e(2)
as a contraction of $U_{q}$su(2) in this limit:
$\bar{J}=J$, $\overline{P_{\pm }}=P_{\mp }$, and
\begin{eqnarray}
\label{eq2}
\comm{J}{P_{\pm }}=\pm P_{\pm }, & \comm{P_{+}}{P_{-}}=0, \nonumber \\
\Delta (P_{\pm })=P_{\pm } \otimes q^{J} + q^{-J} \otimes P_{\pm }, & \Delta
(J) =
J \otimes 1 + 1 \otimes J, \\
\epsilon (P_{\pm }) = \epsilon (J) =0, \nonumber \\
S(J) = -J, & S(P_{\pm })=-q^{\pm 1} P_{\pm }. \nonumber
\end{eqnarray}
Note that the {\em algebra} obtained in (\ref{eq2}) is the same as the
classical
2-dimensional Euclidean algebra $e(2)$ (with $P_{\pm }=P_{x}\pm i P_{y}$ and
$J$ as hermitian generators) \cite{CGST}.  Note, however, as a {\em Hopf
algebra} it is still deformed; the deformation parameter $q$ remains unchanged.

It was shown by Paul Watts \cite{SWZ2} that this Hopf algebra is
identical to the one obtained by directly constructing the dual
Hopf algebra of $Fun(E_{q}(2))$ using methods similar to \cite{R}.
The result was
\begin{eqnarray}
<\nu ^{k} \mu ^{l} \xi ^{n},\theta ^{a} m^{b} \bar{m}^{c}>=[k]_{q}!
[l]_{q^{-1}}! n! \delta _{na} \delta _{lb} \delta _{kc}, & [x]_{q}! \equiv
{\displaystyle \prod _{y=1}^{x} \frac {q^{2y}-1}{q^{2}-1}},
\end{eqnarray}
where $\{ \nu ^{k} \mu ^{l} \xi ^{n} \mid k,l,n=0,1,\ldots \}$ is a basis for
$U_{q}e(2)$ which is  related to    our
operators $J$, $P_{+}$, and $P_{-}$ via
\begin{eqnarray}
J \equiv i\xi , & P_{+} \equiv q q^{-i \xi }\nu , & P_{-} \equiv -q^{-1} \mu
q^{-i \xi }.
\end{eqnarray}
These two constructions  are
summarized in the following (commutative) diagram:

\begin{picture}(300,70)
\put(90,50){$SU_{q}(2)$}
\put(210,50){$E_{q}(2)$}
\put(90,0){$U_{q}su(2)$}
\put(210,0){$U_{q}e(2)$}
\put(105,45){\vector(0,-1){30}}
\put(225,45){\vector(0,-1){30}}
\put(130,54){\vector(1,0){80}}
\put(132,4){\vector(1,0){76}}
\put(155,56){\tiny contraction}
\put(162,48){\tiny $\ell \rightarrow 0$}
\put(155,6){\tiny contraction}
\put(162,-2){\tiny $\ell \rightarrow 0$}
\put(106,30){\tiny dual}
\put(226,30){\tiny dual}
\end{picture}
\vspace{2 mm}

\chapter{Bicovariant Calculus} \label{C:BC}

Having extended the left \U-module \A\ to \A \cross \U through the
construction of the cross product algebra, we would now like
to also extend the definition of the coaction of \A\
to \A \cross \U, making the quantized algebra of differential operators an
\A-bicomodule.

\section{Left and Right Covariance}
\label{Bico}
In this section we would like to study the transformation properties
of the differential operators in \A \cross \U\
under left and right translations, i.e. the coactions $\AD$ and $\DA$
respectively.
We will require,
\begin{eqnarray}
\AD(b y) & = & \AD(b) \AD(y) = \Delta (b) \AD(y)\x\in \A\otimes\A\cross\U,\\
\DA(b y) & = & \DA(b) \DA(y) = \Delta (b) \DA(y)\x\in \A\cross\U\otimes\A,
\label{DAonby}
\end{eqnarray}
for all \x $b\in\A,\,y\in\U,$\x
so that we are left only to define $\AD$ and $\DA$ on elements of \U.
We already mentioned that we would like to interpret
\U\ as the algebra of {\em left invariant} vector fields; consequently
we will try
\begin{equation}
\AD(y) \x = \x 1 \otimes y \z \in \A \otimes \U,
\label{ady}
\end{equation}
as a left coaction. It is easy to see that this coaction respects
not only the left action (\ref{UonA}) of \U\ on \A,
\begin{equation}
\begin{array}{rcl}
\AD (x \tr b) & = & \AD(b_{(1)})<x,b_{(2)}>\\
              & = & 1 \, b_{(1)} \otimes b_{(2)}<x,b_{(3)}>\\
              & = & x^{(1)'} b_{(1)} \otimes (x^{(2)} \tr b_{(2)})\\
              & =: & \AD(x) \tr \AD(b),
\end{array}
\end{equation}
but also the algebra structure (\ref{commrel}) of \A \cross \U,
\begin{equation}
\begin{array}{rcl}
\AD (x \cdot b) & = & \AD(b_{(1)})<x_{(1)},b_{(2)}>\AD(x_{(2)})\\
              & = & b_{(1)}\,1 \otimes b_{(2)}<x_{(1)},b_{(3)}>x_{(2)}\\
              & = & 1 \, b_{(1)}\otimes b_{(2)}<x_{(1)},b_{(3)}>x_{(2)}\\
              & = & x^{(1)'}  b_{(1)} \otimes (x^{(2)} \cdot b_{(2)})\\
              & =: & \AD(x) \cdot \AD(b).
\end{array}
\end{equation}

The right coaction, $\DA :$ $\U \rightarrow \U\otimes\A$,
is considerably harder to find. We will approach this problem by extending the
commutation relation (\ref{commrel}) for elements of \U\ with elements of \A\
to a generalized commutation relation for elements of \U\ with elements of
\A\cross\U,
\begin{equation}
x \cdot b y =: (b y)^{(1)} <x_{(1)}\,,(b y)^{(2)'}>x_{(2)},
\label{gencom}
\end{equation}
for all$\x x,y\in\U,\;b\in\A$. In the special case $b=1$ this states,
\begin{equation}
x \cdot y = y^{(1)} <x_{(1)}\,,y^{(2)'}>x_{(2)},\z x,y \in \U,
\label{altxy}
\end{equation}
and gives an implicit definition of the right coaction
\x$\DA(y) \equiv y^{(1)} \otimes y^{(2)'}$\x of \A\ on \U.
Let us check whether $\DA$ defined in this way respects the left action
(\ref{UonA}) of \U\ on \A:
\begin{equation}
\begin{array}{rcl}
<z \otimes y,\DA(x \tr b)> & = & <z y\:,\: x \tr b> \\
& = & <z y\:,\: b_{(1)}><x\:,\: b_{(2)}>\\
& = & <z y x\:,\: b> \\
& = & <z (x^{(1)} < y_{(1)}, x^{(2)'}> y_{(2)})\:,\: b>\\
& = & <z x^{(1)} \otimes y_{(1)} \otimes y_{(2)}\:,\: b_{(1)} \otimes
x^{(2)'} \otimes b_{(2)}> \\
& = & <z x^{(1)} \otimes y\:,\:b_{(1)} \otimes  x^{(2)'} b_{(2)}>\\
& = & <z \otimes y\:,\: (x^{(1)} \tr b_{(1)}) \otimes x^{(2)'} b_{(2)}> \\
& =: & <z \otimes y\:,\: \DA(x) \tr \DA(b)>,
\end{array}
\end{equation}
for all $x, y, z \in \U,\;b \in \A$, q.e.d. .\\
Given a linear basis $\{e_{i}\}$ of \U\ and
the dual basis $\{f^{j}\}$ of $\A = \U^{*}$, $<e_{i} , f^{j}> = \delta
_{i}^{j}$,
we can derive an explicit expression \cite{pW} for $\DA$ from (\ref{altxy}):
\begin{equation}
\DA(e_{i}) = e_{j} \ad e_{i} \otimes f^{j},
\end{equation}
or equivalently, by linearity of $\DA$:
\begin{equation}
\DA(y) =  e_{j} \ad y \otimes f^{j},\z y \in \U.
\label{gencoact}
\end{equation}
It is then easy to show that,
\begin{eqnarray}
(\DA \otimes i\!d)\DA(y) & = & (i\!d \otimes \Delta )\DA(y),\\
(i\!d \otimes \epsilon )\DA(y) & = & y,
\end{eqnarray}
proving that $\DA$ satisfies the requirements of a coaction on \U, and,
\begin{equation}
\DA(x y) =\DA(x) \DA(y),
\end{equation}
showing that $\DA$ is an \U-algebra homomorphism; $\DA$
is however in general not a \U-Hopf algebra homomorphism.
Using the explicit expression for $\DA$ we can now prove that it
respects the algebra structure of \A\cross\U:
\begin{equation}
\begin{array}{rcl}
\DA(x a) & = & \DA(a_{(1)}<x_{(1)},a_{(2)}> x_{(2)})\\
&=& \Delta (a_{(1)}) <x_{(1)},a_{(2)}> \DA(x_{(2)})\\
&=& (a_{(1)} \otimes a_{(2)})(<x_{(1)},a_{(3)}> x_{(2)}{}^{(1)} \otimes
x_{(2)}{}^{(2)'})\\
&=& (a_{(1)} \otimes a_{(2)})(<x_{(1)},a_{(3)}>e_{i_{(1)}} x_{(2)}
Se_{i_{(2)}}\otimes f^{i})\\
&=& a_{(1)}<x_{(1)},a_{(3)}>e_{i_{(1)}} x_{(2)} Se_{i_{(2)}} \otimes a_{(2)}
f^{i} Sa_{(4)} a_{(5)}\\
&=& a_{(1)}<e_{k} \otimes x_{(1)} \otimes Se_{l},a_{(2)}\otimes a_{(3)}\otimes
a_{(4)}>e_{i}\ad
x_{(2)}\otimes f^{k} f^{i} f^{l} a_{(5)}\\
&=& a_{(1)}<e_{i_{(1)}} x_{(1)} Se_{i_{(3)}},a_{(2)}>e_{i_{(2)}}\ad
x_{(2)}\otimes f^{i} a_{(3)}\\
&=& a_{(1)}<e_{i_{(1)}} x_{(1)} Se_{i_{(4)}},a_{(2)}>e_{i_{(2)}} x_{(2)}
Se_{i_{(3)}}\otimes f^{i} a_{(3)}\\
&=& e_{i_{(1)}} x Se_{i_{(2)}} a_{(1)} \otimes f^{i} a_{(2)}\\
&=& (e_{i}\ad x \otimes f^{i})(a_{(1)}\otimes a_{(2)})\\
&=& \DA(x) \DA(a).\z\Box
\end{array}
\end{equation}
This not only proofs that $\DA$ is a $\smash$-algebra homomorphism but
also that the algebra structure of $\smash$ is compatible with
$\DA$\footnote{In more mathematical terms: The two-sided ideal
$I := x a - a_{(1)}<x_{(1)},a_{(2)}> x_{(2)}$
that we factored out of $U(\A\otimes\U)$ to obtain $\smash$ is invariant
under $\DA$ in the sense $\DA(I) \subset I \otimes \A$.}.
Clearly a less complicated way to see this would be quite welcome. In
the next section we will see that $\DA$ can be obtained for all elements
of \A\cross\U\ via conjugation by the canonical element $C \in
\U\otimes\A$ so that the \A\cross\U-homomorphism property of $\DA$ is
then obvious.

\section{The Canonical Element}
So far we have  shown how the two dual
Hopf algebras $\A$ ``functions
on the quantum group'' and $\U$ ``deformed universal enveloping algebra''
can be combined into a new algebra,
the cross product or generalized semi-direct
product algebra $\smash$, and that this algebra may be
viewed as consisting of bicovariant
differential operators and the functions they act on.
This algebra is not a Hopf algebra but it has $\A$ and $\U$ as Hopf subalgebras
and can in principle be reconstructed from either one of them.
As we shall show, the transformation properties of the elements of $\smash$
are simply given
through conjugation by the {\em canonical element} $C$  of $\U \otimes \A$
--- furthermore, we can recover many of the familiar relations for quantum
groups from the consistency relations which $C$ satisfies in the case where
$\U$
is quasitriangular \cite{RTF,Z2}.
One could even take an extreme point of view and base everything on the
canonical element $C$ in $\smash$ and its commutation relations,
making any explicit reference to the coalgebra structures ($\Delta , S,
\epsilon $) of $\A$ and $\U$ superfluous. \\
The expression of the coaction in terms of the canonical element was
found in collaboration with Paul Watts \cite{CSW}.

\subsubsection{Defininition and Relations}

So let us now introduce the canonical element $C$ in $\U \otimes \A$
\begin{equation}
C\equiv e_{i}\otimes f^{i}.
\end{equation}
$C$ satisfies several relations; for instance, note that
\begin{eqnarray}
((S\otimes id)(C))\, C&=& S(e_{i})e_{j}\otimes f^{i}f^{j}\nonumber \\
&=& D_{k}^{ij}S(e_{i})e_{j}\otimes f^{k}\nonumber \\
&=& (m \circ (S\otimes id)\circ \Delta )(e_{k})\otimes f^{k}\nonumber \\
&=& \idU \epsilon (e_{k})\otimes f^{k}\nonumber \\
&=& \idU \otimes E_{k}f^{k}\nonumber \\
&=&\idU \otimes \idA,
\end{eqnarray}
where $m$ is the multiplication map, $D_{k}^{ij}$ is the matrix that
describes the coproduct in \U\ and $E_{k}$ is the vector corresponding to
the counit in \U, so
\begin{equation}
(S\otimes id)(C)=C^{-1}.
\end{equation}
Similar
calculations also give
\begin{equation}
(id \otimes S)(C)=C^{-1},
\end{equation}
as well as the following:
\begin{eqnarray}
(\Delta \otimes id)(C)&=&C_{13}C_{23},\\
(id \otimes \Delta )(C)&=&C_{12}C_{13},\\
(\epsilon \otimes id)(C)=(id \otimes \epsilon )(C)&=&\idU \otimes \idA.
\end{eqnarray}
There is more to $C$ than just the above relations; this is seen by computing
the right coaction of a basis vector in $\U$.  Using (\ref{gencoact})
\begin{eqnarray}
\DA(e_{i})&=&(e_{j}\triangleright e_{i})\otimes f^{j}\nonumber \\
&=&(e_{j})_{(1)}e_{i}S((e_{j})_{(2)})\otimes f^{j}\nonumber \\
&=&D_{j}^{mn}e_{m}e_{i}S(e_{n})\otimes f^{j}\nonumber \\
&=&e_{m}e_{i}S(e_{n})\otimes f^{m}f^{n}\nonumber \\
&=&(e_{m}\otimes f^{m})(e_{i}\otimes \idA)(S(e_{n})\otimes f^{n})\nonumber \\
&=&C(e_{i}\otimes \idA)(S \otimes id)(C),
\end{eqnarray}
so for any $x \in \U$,
\begin{equation}
\DA(x)=C(x\otimes 1)C^{-1}.
\end{equation}
However, when we think of $C$ as living in $(\smash)\otimes (\smash)$, with
$e_{i}$ and $f^{i}$ as the bases for the subalgebras $\U$ and $\A$ of $\smash$
respectively, further results follow.  For instance, for $a \in \A$,
\begin{eqnarray}
C(a\otimes 1)C^{-1}&=&e_{i}aS(e_{j})\otimes f^{i}f^{j}\nonumber \\
&=&(a_{(1)}(e_{i})_{(2)}\inprod{(e_{i})_{(1)}}{a_{(2)}})S(e_{j})\otimes
D_{k}^{ij}f^{k}\nonumber \\
&=&a_{(1)}\inprod{(e_{k})_{(1)}}{a_{(2)}}(e_{k})_{(2)}S((e_{k})_{(3)})\otimes
f^{k}\nonumber \\
&=&a_{(1)}\otimes \inprod{e_{k}}{a_{(2)}}f^{k}\nonumber \\
&=&a_{(1)}\otimes a_{(2)},
\end{eqnarray}
(where $1 = 1_{\smash}\equiv \idA \otimes \idU$) so that
\begin{equation}
C(a\otimes 1)C^{-1}=\Delta (a).
\end{equation}
Thus, the right coaction of $\A$ on $\smash$ is obtained through
{\em conjugation} by $C$
\begin{equation}
\DA(\alpha )=C(\alpha \otimes 1)C^{-1} \label{Cgencoact}
\end{equation}
for any $\alpha \in \smash$.  This expression shows explicitly that
$\DA$ is an algebra homomorphism
\begin{equation}
\begin{array}{rcl}
\DA(\alpha \beta )
     & = & C (\alpha \beta \otimes 1) C^{-1}\\
     & = & C (\alpha \otimes 1) C^{-1} C (\beta \otimes 1) C^{-1}\\
     & = & \DA(\alpha ) \DA(\beta )
\end{array}
\end{equation}
for $\alpha ,\beta \in \A\cross\U$, and that it is consistent with the algebra
structure of $\smash$
\begin{equation}
\begin{array}{rcl}
C(x a)C^{-1}
& = & C(a_{(1)}<x_{(1)},a_{(2)}>x_{(2)} \otimes 1)C^{-1}\\
& = & C(a_{(1)} \otimes 1)C^{-1} <x_{(1)},a_{(2)}> C(x_{(2)} \otimes 1)C^{-1}\\
& = & \Delta (a_{(1)})<x_{(1)},a_{(2)}> \DA(x_{(2)})\\
& = & \DA(x a).
\end{array}
\end{equation}
We can continue doing calculations
along these
lines, and we find
\begin{equation}
C^{-1}(1 \otimes x)C=\Delta (x)
\end{equation}
for $x \in \U$.  For elements of the cross product algebra this
gives the left \U-coaction
\begin{equation}
\UD(\alpha ) \equiv \alpha _{1'} \otimes \alpha _{2} = C^{-1} (1 \otimes \alpha
) C,
\end{equation}
that appears in the general commutation relation
\begin{equation}
\alpha \beta = \beta ^{(1)} <\alpha _{1'},\beta ^{(2)'}>\alpha _{2}.
\end{equation}
Using these results, together with the coproduct relations for
$C$, we obtain the equation
\begin{equation}
C_{23}C_{12}=C_{12}C_{13}C_{23}.
\end{equation}
(Interestingly, this equation can be viewed as giving the multiplication on
$\smash$ as defined in (\ref{mult}).)

\subsubsection{Quasitriangular Case}

In the case where $\U$ is a quasitriangular Hopf algebra with universal
R-matrix $\R$, the coproduct relations involving $C$ imply the following
consistency conditions:
\begin{eqnarray}
\R_{12}C_{13}C_{23}&=&C_{23}C_{13}\R_{12},\nonumber \\
\R_{23}C_{12}&=&C_{12}\R_{13}\R_{23}, \nonumber \\
\R_{13}C_{23}&=&C_{23}\R_{13}\R_{12}.\label{rcquasi}
\end{eqnarray}
To see the added significance of these equations, note that
\begin{equation}
\inprod{C}{a \otimes id}=a,
\end{equation}
where $a \in \A$, and we use the notation
\begin{equation}
\inprod{x}{id}=x
\end{equation}
for $x \in \U$.  Let $\rho : \U \rightarrow M_{n}(k)$ be a matrix
representation of $\U$, and define the $n \times n$ matrices
$A^{i}{}_{j}\in {\cal A}$ by
\begin{equation}
\inprod{x}{A^{i}{}_{j}}\equiv \rho ^{i}{}_{j}(x).
\end{equation}
(These $A^{i}{}_{j}$'s are what are usually viewed as the non-commuting
matrix elements of the pseudo-matrix group associated with $\U$ \cite{W1}.)
Given $\rho $, we can define the $\U$-valued matrices
\begin{eqnarray}
L^{+}&\equiv &(id \otimes \rho )(\R),\nonumber \\
L^{-}&\equiv &(\rho \otimes id)(\R^{-1}),
\end{eqnarray}
and the numerical R-matrix
\begin{equation}
R\equiv (\rho \otimes \rho )(\R).
\end{equation}
Furthermore, it is easily seen that $(\rho \otimes id)(C)=A$.  Now let us apply
$(\rho ^{i}{}_{k}\otimes \rho ^{j}{}_{l}\otimes id)$ to the first of equations
(\ref{rcquasi}); the left side gives
\begin{eqnarray}
(\rho ^{i}{}_{k}\otimes \rho ^{j}{}_{l} \otimes id)(\R_{12}C_{13}C_{23})&=&
(\rho ^{i}{}_{m}\otimes \rho ^{j}{}_{n})(\R)(\rho ^{m}{}_{k}\otimes id)(C)
(\rho ^{n}{}_{l}\otimes id)(C)\nonumber \\
&=&R^{ij}{}_{mn}A^{m}{}_{k}A^{n}{}_{l}.
\end{eqnarray}
The right hand side gives $A^{i}{}_{m}A^{j}{}_{n}R^{mn}{}_{kl}$, so using the
usual notation, we obtain
\begin{equation}
RA_{1}A_{2}=A_{2}A_{1}R,
\end{equation}
which gives the commutation relations between the elements of $A$.  Doing
similar gymnastics with the other two equations in (\ref{rcquasi}) gives
\begin{eqnarray}
L^{+}_{1}A_{2}&=&A_{2}R_{21}L^{+}_{1},\nonumber \\
L^{-}_{1}A_{2}&=&A_{2}R^{-1}L^{-}_{1},
\end{eqnarray}
which give the commutation relations between elements of $\U$ and $\A$
within $\smash$.  (Of course, we also have the commutation relations
\begin{eqnarray}
RL_{2}^{\pm }L_{1}^{\pm }&=&L_{1}^{\pm }L_{2}^{\pm }R,\nonumber \\
RL_{2}^{+}L_{1}^{-}&=&L_{1}^{-}L_{2}^{+}R,
\end{eqnarray}
between elements of $\U$, obtained as above from $\R_{12} \R_{13} \R_{23}=
\R_{23} \R_{13} \R_{12}$, the quantum Yang-Baxter equation.)  Thus, we recover
all the commutation relations between $A$ and $L^{\pm }$ given in \cite{Z2}.

\section{Bicovariant Vector Fields}

The appearance of an infinite sum in equation (\ref{gencoact})
or for that matter
(\ref{Cgencoact}) suggests that the elements of $\U$ have in general very
complicated transformation properties. In contrast, the functions in
$\A$, especially those constructed from the matrix elements of $A$,
have very simple transformation properties given by the coproduct in
$\A$ (\ref{coofA}).
We would like to show how to construct vector fields corresponding
to --- and inheriting the simple behavior of --- these functions.
This construction can then be used to find a basis of vector fields that
closes under coaction and hence under (mutual) adjoint actions.
First we need to proof the following lemma.

\noindent {\bf Lemma: } {\em Let $\Ups \equiv \Ups_{i} \otimes \Ups^{i}
\in \U \otimes \U$ such that
$\Ups \Delta (x) = \Delta (x) \Ups$ for all $x \in \U$, then it follows that
$\: \Ups_{i} \otimes (x \ad \Ups^{i}) = (\Ups_{i} \da x) \otimes \Ups^{i}$
with $\Ups_{i} \da x \equiv S(x_{(1)}) \Ups_{i} x_{(2)}$ for all $x \in \U$.}

\noindent {\bf Proof:}
\begin{equation}
\begin{array}{rcl}
\Ups_{i} \otimes (x \ad \Ups^{i}) & \equiv &
             \Ups_{i} \otimes x_{(1)} \Ups^{i} S(x_{(2)})\\
& = & S(x_{(1)}) x_{(2)} \Ups_{i} \otimes x_{(3)} \Ups^{i} S(x_{(4)})\\
& = & S(x_{(1)}) \Ups_{i} x_{(2)} \otimes \Ups^{i} x_{(3)} S(x_{(4)})\\
& = & (\Ups_{i} \da x) \otimes \Ups^{i}.\z\Box
\end{array}
\end{equation}
For any function $b \in \A$, define
\begin{equation}
Y_{b} := \inprod{\Ups}{b \otimes i\!d} \: \in \U .
\end{equation}
{\bf Proposition:} This vector field has the following transformation property:
\begin{equation}
\Delta _{\A}(Y_{b}) = Y_{b(2)} \otimes S(b_{(1)}) b_{(3)}\label{proposition}
\end{equation}
{\bf Proof:}
\begin{equation}
\begin{array}{rcl}
\Delta _{\A}(Y_{b}) & = & \inprod{\Ups_{i}}{b} (e_{k} \triangleright \Ups^{i})
\otimes f^{k}\\
&=& \inprod{\Ups_{i} \triangleleft e_{k}}{b} \Ups^{i} \otimes f^{k}\\
&=& \inprod{\Ups_{i} \otimes e_{k}}{b_{(2)}
    \otimes S(b_{(1)}) b_{(3)}} \Ups^{i} \otimes f^{k}\\
& = & Y_{b(2)} \otimes S(b_{(1)}) b_{(3)}.\z\Box
\end{array}
\end{equation}

\noindent {\bf Example:} {\em Let $\Ups :=$ $ \R_{21} \R_{12}$ and
$b :=$ $ A^{i}{}_{j}$,
then $Y^{i}{}_{j} :=$ $Y_{A^{i}{}_{j}} =$ $\inprod{\R_{21} \R_{12}}{A^{i}{}_{j}
\otimes i\!d}$  is
the well-known matrix of vector fields $L^{+} S(L^{-})$ introduced in
\cite{RSTS} with
coaction:\\ $\Delta _{\A}(Y^{i}{}_{j}) =$ $ Y^{k}{}_{l} \otimes S(A^{i}{}_{k})
A^{l}{}_{j}$.}

This last example may in some cases (when \U\ is factorizable \cite{pR})
provide a way of computing the
canonical element $C$ from $\R_{21} \R_{12}$: Let $\mu $ be the map
\begin{equation}
\mu : \A \rightarrow \U: \:\: b \mapsto \inprod{\R_{21} \R_{12}}{b \otimes
i\!d},
\end{equation}
then $(i\!d \otimes \mu ) (C) = e_{i} \inprod{\R_{21} \R_{12}}{f^{i} \otimes
i\!d} = \R_{21} \R_{12}$ and, in cases where $\mu $ is invertible;
\begin{equation}
C = (i\!d \otimes \mu ^{-1}) (\R_{21} \R_{12}).
\end{equation}

In the next section we will elaborate more on elements like $\Ups$
and their connection to the ``Pure Braid Group''. There we will
also proof the reverse of Proposition~\ref{proposition}.

\section{The Pure Braid Group}

\subsection*{Introduction}

In the classical theory of Lie algebras we start the construction of
a bicovariant calculus by introducing a matrix
$\Omega = A^{-1}\der A \in \Gamma $ of one-forms that is invariant
under left transformations,
\begin{eqnarray}
A \rightarrow A' A:&&\der \rightarrow\der,\x \Omega \rightarrow \Omega ,
\end{eqnarray}
and covariant under right transformations,
\begin{eqnarray}
 A \rightarrow A A':&&\der \rightarrow\der,\x \Omega
 \rightarrow A'^{-1} \Omega A'.
\end{eqnarray}
The dual basis to the entries of this matrix $\Omega $ form a matrix $X$
of vector fields with the same transformation properties as
$\Omega $:
\begin{equation}
\langle {\Omega ^{i}}_{j}, {X^{k}}_{l}\rangle
= {\delta ^{i}}_{l} {\delta ^{k}}_{j}
\z \mbox{\small\em (classical)}.
\end{equation}
We find,
\begin{equation}
 X = (A^{T} \frac {\partial }{\partial A})^{T}
\z \mbox{\small\em (classical)}.
\label{clas-vec}
\end{equation}

Woronowicz \cite{W2} was able to extend the definition of a
bicovariant calculus to
quantum groups. His approach via differential forms has the advantage that
coactions (transformations) $\AD : \Gamma \rightarrow \A \otimes \Gamma $
and $\DA : \Gamma \rightarrow \Gamma \otimes  \A$ can be introduced very
easily through,
\begin{eqnarray}
\AD (\der a) & = & (i\!d \otimes \der )\Delta a,
\label{forms1}\\
\DA (\der a) & = & (\der \otimes i\!d)\Delta a,
\label{forms2}
\end{eqnarray}
where \A\ is the Hopf algebra of `functions on the quantum group',
$a \in \A$ and $\Delta $ is the coproduct in \A\ . Equations
(\ref{forms1},\ref{forms2}) rely on the existence of an invariant map
$\der:\A \rightarrow \Gamma $ provided by the exterior derivative.
A construction of the bicovariant calculus starting directly from the
vector fields is much harder because  simple formulae like
(\ref{forms1},\ref{forms2}) do not seem to exist a priori. The properties of
the element $\Ups$ that we introduced in the previous section however indicates
exceptions:
We will show that for
Hopf algebras that allow ``pure braid elements'' $\Ups$, like e.g.
quasitriangular Hopf algebras, invariant maps from \A\ to
the quantized algebra of differential
operators \A \cross \U\ can indeed be constructed.
Using these maps we will then construct differential
operators with simple transformation properties and in particular a
bicovariant matrix of vector fields roughly corresponding to (\ref{clas-vec}).

In the next subsection we will hence describe a map, $\Phi :$ $\A
\rightarrow \A\cross\U$,
that is invariant under (right) coactions and can be used to find $\DA$
on  specific elements $\Phi (b)\in\U$ in terms of $\DA$ on $b\in\A$:\x
$\DA(\Phi (b))=(\Phi \otimes i\!d)\DA(b)$.

\subsection[Invariant Maps]{Invariant Maps and the Pure Braid Group}
\label{Braid}

A basis of generators for the pure braid group $B_{n}$ on $n$ strands can be
realized in \U, or for that matter \uqg, as follows in terms of the
universal \R:
$$\begin{array}{l}
\R^{21}\R^{12} ,\z \R^{21}\R^{31}\R^{13}\R^{12} \equiv
(i\!d \otimes \Delta )\R^{21}\R^{12},\x \ldots \x,\\
\R^{21}\cdots\R^{n1}\R^{1n}%
\cdots\R^{12} \equiv ({i\!d}^{(n-2)} \otimes \Delta )({i\!d}^{(n-3)}
\otimes \Delta )\cdots(i\!d \otimes \Delta )\R^{21}\R^{12},
\end{array}$$ and their inverses; see figure~\ref{GOPBG} and ref.\cite{Rn}.
All polynomials in these generators are central in $\Delta ^{(n-1)}\U$ $\equiv
\{\Delta ^{(n-1)}(x)\: |\: x \in \U \}$; in fact we can take,
\begin{equation}
\mbox{span}\{B_{n}\} := \{\Z_{n} \in \U^{\hat{\otimes} n} |
\Z_{n} \Delta ^{(n-1)}(x)
= \Delta ^{(n-1)}(x) \Z_{n} , \x\mbox{for} \forall x \in \U \},
\end{equation}
as a definition.\\
{\em Remark:} Elements of span$\{B_{n}\}$ do not have to be written in terms
of the universal $\R$, they also arise from central elements and coproducts
of central elements. This is particularly important in cases where \U\
is not a quasitriangular Hopf algebra.

There is a map,\x$\Phi _{n}:$
$\A \rightarrow \A \otimes \U^{\otimes (n-1)} \hookrightarrow
(\A\cross\U)^{\otimes (n-1)}$,\x associated to each element  of
span$\{B_{n}\}$:
\begin{equation}
\Phi _{n}(a) := \Z_{n} \tr (a \otimes i\!d^{(n-1)}),\z \mbox{with }
\Z_{n} \in \mbox{span}\{B_{n}\},\; a \in \A.
\label{phin}
\end{equation}
\setlength{\unitlength}{2.3pt}
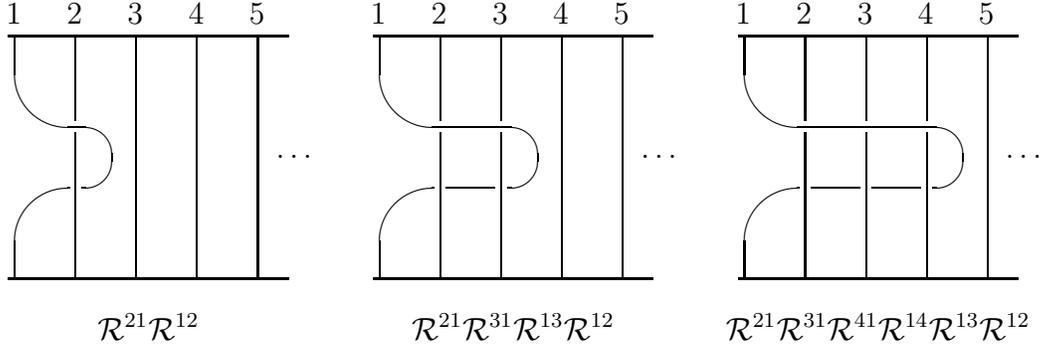
\begin{figure}
\begin{picture}(150,60)
\put(10,10){\line(0,1){24}}
\put(10,36){\line(0,1){14}}
\multiput(20,10)(10,0){3}{\line(0,1){40}}
\put(9,10){\oval(18,30)[tl]}
\put(11,50){\oval(22,30)[bl]}
\put(11,30){\oval(10,10)[r]}
\multiput(-1,52)(10,0){5}{\addtocounter{strand}{1}\makebox(0,0)[b]{
\arabic{strand}}}
\setcounter{strand}{0}
\put(22,0){\makebox(0,0)[b]{$\R^{21}\R^{12}$}}
\put(70,10){\line(0,1){24}}
\put(80,10){\line(0,1){24}}
\put(70,36){\line(0,1){14}}
\put(80,36){\line(0,1){14}}
\multiput(90,10)(10,0){2}{\line(0,1){40}}
\put(69,10){\oval(18,30)[tl]}
\put(81,50){\oval(42,30)[bl]}
\put(81,30){\oval(10,10)[r]}
\put(71,25){\line(1,0){8}}
\multiput(59,52)(10,0){5}{\addtocounter{strand}{1}\makebox(0,0)[b]{
\arabic{strand}}}
\setcounter{strand}{0}
\put(82,0){\makebox(0,0)[b]{$\R^{21}\R^{31}\R^{13}\R^{12}$}}
\put(130,10){\line(0,1){24}}
\put(140,10){\line(0,1){24}}
\put(150,10){\line(0,1){24}}
\put(130,36){\line(0,1){14}}
\put(140,36){\line(0,1){14}}
\put(150,36){\line(0,1){14}}
\put(160,10){\line(0,1){40}}
\put(129,10){\oval(18,30)[tl]}
\put(151,50){\oval(62,30)[bl]}
\put(151,30){\oval(10,10)[r]}
\put(131,25){\line(1,0){8}}
\put(141,25){\line(1,0){8}}
\multiput(119,52)(10,0){5}{\addtocounter{strand}{1}\makebox(0,0)[b]{
\arabic{strand}}}
\put(142,0){\makebox(0,0)[b]{$\R^{21}\R^{31}\R^{41}\R^{14}\R^{13}\R^{12}$}}
\multiput(43,30)(60,0){3}{\makebox(0,0)[bl]{\ldots}}
\thicklines
\multiput(-1,10)(60,0){3}{\line(1,0){46}}
\multiput(-1,50)(60,0){3}{\line(1,0){46}}
\end{picture}
\caption{Generators of the pure braid group.} \label{GOPBG}
\end{figure}
We will first consider the case $n = 2$. Let $\Ups\equiv \Ups_{1_{i}} \otimes
\Ups_{2_{i}}$ be an element of span$\{B_{2}\}$ and\x $\Phi (b) = \Ups \tr (b
\otimes i\!d)
= b_{(1)}<\Ups_{1_{i}},b_{(2)}>\Ups_{2_{i}}$,\x for $b\in\A$. We compute,
\begin{equation}
\begin{array}{rcl}
x \cdot \Phi (b) & = & \Delta (x) \tr \Phi (b) \\
                & = & \Delta (x) \Ups \; \tr (b \otimes i\!d) \\
                & = & \Ups \Delta (x) \; \tr (b \otimes i\!d) \\
                & = & \Ups \tr (x\cdot b) \\
                & = & \Phi (b_{(1)}) <x_{(1)} , b_{(2)}>x_{(2)},
\end{array}
\label{xphin}
\end{equation}
which, when compared to the {\em generalized} commutation relation
(\ref{gencom}), i.e.
\begin{equation}
x \cdot \Phi (b) \x = \x [\Phi (b)]^{(1)} <x_{(1)},[\Phi (b)]^{(2)'}> x_{(2)},
\end{equation}
gives,
\begin{equation}
\begin{array}{l}
\DA(\Phi (b)) \equiv [\Phi (b)]^{(1)} \otimes [\Phi (b)]^{(2)'} = \Phi
(b_{(1)})
\otimes b_{(2)}\\
\Rightarrow  \DA(\Phi (b))=(\Phi \otimes i\!d)\DA(b),
\end{array}
\label{dapy}
\end{equation}
as promised.
However we are especially interested in the transformation properties
of elements of \U, so let us define,
\begin{equation}
\Ups_{b} := <\Ups , b \otimes i\!d> =
<\Ups_{1_{i}},b>\Ups_{1_{i}},
\end{equation}
for $\Ups\in\mbox{span}(B_{2}),\,b\in\A$.
Using (\ref{DAonby},\ref{dapy}) we recover the result of
Proposition~\ref{proposition}
\begin{equation}
\DA (\Ups_{b}) = \Ups_{b_{(2)}} \otimes S(b_{(1)}) b_{(3)}.\\
\label{dayb}
\end{equation}
Let us now proof the reverse statement:\\
{\bf Proposition:} If there is a linear map $\Ups:\A \rightarrow \U$,
realized and labelled by some element $\Ups \in \U \hat{\otimes} \U$ via
$b \mapsto \Ups_{b} \equiv <\Ups,b \otimes i\!d>,\x\forall b\in\A$, such that
the
resulting element in \U\ transforms like
$\DA \Ups_{b} = \Ups_{b_{(2)}} \otimes S b_{(1)} b_{(3)}$; then $\Ups
\in$span$(B_{2})$,
{\em i.e.} $\Ups$ must commute with all coproducts.\\
{\bf Proof:} For all $x \in \U$ and $b \in \A$
\begin{equation}
\begin{array}{rcl}
<\Delta x \Ups,b \otimes i\!d> & = & <\Delta x,b_{(1)}
             \otimes i\!d><\Ups,b_{(2)}\otimes i\!d>\\
        & = &<x_{(1)},b_{(1)}>x_{(2)}\Ups_{b_{(2)}}\\
        & = &<x_{(1)},b_{(1)}>\Ups_{b_{(3)}}<x_{(2)},S b_{(2)} b_{(4)}>
x_{(3)}\\
        & = &\Ups_{b_{(3)}}<x_{(1)},b_{(1)} S b_{(2)} b_{(4)}> x_{(2)}\\
        & = &\Ups_{b_{(1)}}<x_{(1)},b_{(2)}> x_{(2)}\\
        & = &<\Ups \Delta x,b\otimes i\!d>.\z\Box
\end{array}
\end{equation}
{}From this follows an important {\bf Corollary:}\\
If there exists a map $\phi :\A \rightarrow \A\cross\U$
such that $\DA \circ \phi  = (\phi  \otimes i\!d) \circ \Delta $; then it
follows that
$\phi (b) = b_{(1)}<\Ups,b_{(2)}\otimes i\!d>$ with $\Ups \in$span$(B_{2})$ for
all
$b \in \A$ and vice versa.\\

Here are a few important examples for ``pure braid elements'':
For the simplest non-trivial example in the case of a  quasitriangular
Hopf algebra
$\Y \equiv \R^{21}\R^{12}$ and
$b \equiv {A^{i}}_{j}$, we obtain the `reflection-matrix'\cite{Ku}
$Y \in \mbox{M}_{n}(\U)$,
which has been introduced before by other authors \cite{RSTS,J}
in connection with integrable models and the
differential calculus on quantum groups,
\begin{equation}
\begin{array}{rcl}
{Y^{i}}_{j} & := & Y_{{A^{i}}_{j}}\\
         & = & <\R^{21}\R^{12},{A^{i}}_{j} \otimes i\!d>\\
         & = & {(<\R^{31}\R^{23},A \dot{\otimes} A \otimes i\!d>)^{i}}_{j}\\
         & = & {(<\R^{21},A \otimes i\!d><\R^{12}, A \otimes i\!d>)^{i}}_{j}\\
         & =&  {(L^{+} SL^{-})^{i}}_{j},
\end{array}
\label{defY}
\end{equation}
with transformation properties,
\begin{eqnarray}
A & \rightarrow & A A':\z {Y^{i}}_{j} \rightarrow \DA({Y^{i}}_{j})
\begin{array}[t]{l}
= {Y^{k}}_{l} \otimes S({A^{i}}_{k}) {A^{l}}_{j} \\ \equiv {((A')^{-1} Y
A')^{i}}_{j},
\end{array} \\
A & \rightarrow & A' A:\z {Y^{i}}_{j} \rightarrow \AD({Y^{i}}_{j})
= 1 \otimes {Y^{i}}_{j}.
\end{eqnarray}
The commutation relation (\ref{commrel}) becomes in this case,
\begin{equation}
\begin{array}{rcl}
Y_{2} A_{1} & = & L^{+}_{2} SL^{-}_{2} A_{1}\\
        & = & L^{+}_{2} A_{1} SL^{-}_{2} R_{21}\\
        & = & A_{1} R_{12} L^{+}_{2} SL^{-}_{2} R_{21}\\
        & = & A_{1} R_{12} Y_{2} R_{21},
\end{array}
\label{YA}
\end{equation}
where we have used (\ref{LPA}), (\ref{LMA}), and the associativity of the
cross product (\ref{crossprod});
note that we did not have to use any explicit expression for the coproduct
of $Y$.
The matrix $\Phi ({A^{i}}_{j}) = {A^{i}}_{k} {Y^{k}}_{j}$ transforms exactly
like $A$,
as expected, and interestingly even satisfies the same commutation relation
as $A$,
\begin{equation}
R_{12} (A Y)_{1} (A Y)_{2} = (A Y)_{2} (A Y)_{1} R_{12},
\end{equation}
as can be checked by direct computation.
C. Chryssomalakos \cite{pC} found an ``explanation'' for this fact by
expressing $A Y$ in terms of casimirs. We will come back to this in the
next section.

The choice, $\Y \equiv (1 - \R^{21}\R^{12})/\lambda $, where $\lambda
\equiv q - q^{-1}$, and again $b \equiv {A^{i}}_{j}$ gives us a matrix
$X \in \mbox{M}_{n}(\U)$,
\begin{equation}
{X^{i}}_{j} :=<(1 - \R^{21}\R^{12})/\lambda ,{A^{i}}_{j} \otimes i\!d> =
{((I - Y)/\lambda )^{i}}_{j},
\end{equation}
that we will encounter again in section \ref{Lie}.
$X$ has the same transformation
properties as $Y$ and is the quantum analog of the classical matrix
(\ref{clas-vec}) of vector fields.

Finally, the particular choice $b \equiv \det_{q} A$ in conjunction with
$\Y \equiv \R^{21}\R^{12}$ can serve as the definition of the quantum
determinant of $Y$,
\begin{equation}
\mbox{Det} Y := Y_{\det_{q} A} = <\R^{21}\R^{12},{\det}_{q} A \otimes i\!d> ;
\label{dety}
\end{equation}
we will come back to this in the next section, but let us just mention
that this definition of Det$Y$ agrees with,
\begin{equation}
\begin{array}{rcl}
{\det}_{q}(AY) & = & {\det}_{q}(A<\R^{21}\R^{12},A \otimes i\!d>)\\
             & = & {\det}_{q} A <\R^{21}\R^{12},{\det}_{q} A \otimes i\!d>\\
             & = & {\det}_{q} A \;\mbox{Det} Y.
\end{array}
\end{equation}

Before we can consider maps $\Phi _{n}$ for $n > 2$ we need to extend the
algebra and coalgebra structure of \A\cross\U\ to $(\A\cross\U)^{ \otimes
(n-1)}$.
It is sufficient to consider $(\A\cross\U)^{\otimes 2}$; all other cases
follow by analogy. If we let
\begin{equation}
(a \otimes b) (x \otimes y) \x = \x a x \otimes b y,\z\mbox{for }
\forall \: a,b \in \A,\:x,y \in \U,
\end{equation}
then it follows that
\begin{equation}
\begin{array}{rcl}
x\cdot a \otimes y\cdot b
&=& a_{(1)} <x_{(1)}\:,\:a_{(2)}> x_{(2)} \otimes
    b_{(1)} <y_{(1)}\:,\:b_{(2)}> y_{(2)} \\
&=& {(a\otimes b)}_{(1)} <{(x\otimes y)}_{(1)}\:,\:{(a\otimes
b)}_{(2)}> {(x\otimes y)}_{(2)}\\
&=& (x\otimes y)\cdot (a\otimes b),\z\mbox{for }
\forall \: a,b \in \A,\:x,y \in \U,
\end{array}
\end{equation}
as expected from a tensor product algebra.
If we coact with \A\ on $\A\cross\U^{\otimes 2}$, or higher powers, we
simply collect all the contributions of $\DA$ from each
tensor product space in one space on the right:
\begin{equation}
\begin{array}{l}
\DA(a x \otimes b y) \x = \x (a x)^{(1)} \otimes (b y)^{(1)}
\otimes (a x)^{(2)'} (b y)^{(2)'},\\\mbox{for }
\forall \: a,b \in \A,\:x,y \in \U.
\end{array}
\end{equation}

\section{Casimirs}

Casimirs play an important role in the theory of quantum groups, even
more so than in classical group theory. They, or rather characters
related to them, label representations;
casimirs --- in particular tr${}_{q}(Y)$ and Det${}_{q}(Y)$
show up as coefficients in the
characteristic polynomial for the matrix of bicovariant generators $Y$
and finally extra non-classical generators in Quantum Lie Algebras
are given by casimirs. Here we want to collect some formulas for
casimir operators and comment on a few of their uses.

\subsubsection{Casimirs related to $\Delta ^{Ad}$-invariant elements of \A}

Centrality of elements of \U\ is synonymous to their invariance under
the right \A-coaction because of
\begin{equation}
x y = y^{(1)}<x_{(1)},y^{(2)'}> x_{(2)},\z \DA y \equiv y^{(1)} \otimes
y^{(2)'}
\end{equation}
--- if $\DA c = c \otimes 1$ then $x c = c<x_{(1)},1> x_{(2)} = c \epsilon
(x_{(1)}) x_{(2)}
= c x$. In the previous sections we have shown how to construct
elements of \U\ from elements of \A, preserving their transformation
properties under adjoint coaction. The quantum determinant and the
quantum traces are invariant under $\Delta ^{Ad}$, giving our first group of
examples for casimir operators:
\begin{equation}
\mbox{tr}_{q}(Y^{k}) = <\Ups,\mbox{tr}_{q}(A^{k}) \otimes i\!d>,\z
\mbox{Det}_{q}(Y) = <\Ups,\mbox{Det}_{q}(A) \otimes i\!d>,
\end{equation}
where $\Ups$ is an element of the pure braid group, $i.e.$
$\Ups \Delta (y) = \Delta (y) \Ups$ for all $y \in \U$.
In the case of $\Ups = \R^{21} \R^{12}$ the first set of casimirs coincides
with the ones given in \cite{RTF}; there are in fact as many independent
ones as the rank of the corresponding group.

\subsubsection{Casimirs arising from the pure braid group}

Let $\gamma  := \Ups_{i} S \Ups^{i}$, where $\Ups \equiv \Ups_{i} \otimes
\Ups^{i}$ is an
element of the pure braid group. Here is a proof that $\gamma $ is a casimir:
\begin{equation}
\begin{array}{rcl}
\Ups_{i} y_{(1)} \otimes \Ups^{i} y_{(2)} & = & y_{(1)} \Ups_{i} \otimes
y_{(2)} \Ups^{i}\\
\Leftrightarrow \Ups_{i} y_{(1)} S(y_{(2)}) S(\Ups^{i}) & = & y_{(1)} \Ups_{i}
S(\Ups^{i})
S(y_{(2)})\\
\Leftrightarrow \epsilon (y) \gamma  & = & y_{(1)} \gamma  S(y_{(2)}) \\
\Leftrightarrow y \gamma  & = & \gamma  y.\z\Box
\end{array}
\end{equation}
More casimirs like $S(\Ups_{i}) \Ups^{i},\, \epsilon (\Ups_{i}) \Ups^{i},\,
\Ups_{i} \epsilon (\Ups^{i})$
can be obtained in similar ways.
\paragraph{Relation to Drinfeld's casimir $c$.}
Drinfeld \cite{Df,ReTu}
showed that the $S^{2}$ automorphism is realized as conjugation by
an element $u$ in quasitriangular Hopf algebras. Let $\R = \alpha _{i} \otimes
\beta _{i}$, then $u = S(\beta _{i}) \alpha _{i}$, $S(u) = \alpha _{j} S(\beta
_{j})$ and $c = u S(u)$.
If we choose $\Ups = \R^{21} \R^{12}$ as our pure braid element, then
\begin{equation}
\begin{array}{rcl}
\Ups_{i} S(\Ups^{i}) & = & \beta _{i}\alpha _{j}S(\beta _{j})S(\alpha _{j}) \\
        & = & \beta _{i}S(u)S(\alpha _{j})\left(S(u)\right)^{-1} S(u)\\
        & = & \beta _{i} S^{-1}(\alpha _{i}) \alpha _{j} S(\beta _{j})\\
        & = & u S(u)\,=\,c
\end{array}
\end{equation}
and similar $S(\Ups_{i}) \Ups^{i} = S^{-1}(c)$.

\subsubsection{Extra Generators}

Classically the commutator of Lie bracket of a casimir $c$ and some
vector field $y$ vanishes because of the centrality of $c$; so casimirs
do not play a role in classical Lie algebras. In the quantum case the
commutator is replaced by the adjoint action and then
\begin{equation}
c \ad y = c_{(1)} y S(c_{(2)}) \neq 0
\end{equation}
in general. We however still have
\begin{equation}
y \ad c = y_{(1)} c S(y_{(2)}) = \epsilon (y) c,
\end{equation}
which is zero if $\epsilon (y) =0$, as is usually the case for a generator of
a quantum Lie algebra.

\subsubsection{Special properties of $A Y$}

We remarked earlier that $A Y = A L^{+} SL^{-}$ satisfies the same algebra
as $A$ does. C.Chryssomalakos \cite{pC} found that this is
also true for $A Y^{k}$ and gave a nice explanation for this fact that
I would like to quote here:
Using the coproduct of $c$
\begin{equation}
\Delta  c = (\R^{21} \R^{12})^{-2}(c \otimes c)
\end{equation}
one easily derives
\begin{equation}
A Y = \alpha  c^{-1} A c,
\end{equation}
where $\alpha  \delta _{i}^{j} = <c,A^{j}{}_{i}>$. In the case of a Ribbon Hopf
Algebra
\cite{Rn,ReTu}
there is a central element $w$ that implements the square root of $c$;
its coproduct is
\begin{equation}
\Delta  w = (\R^{21} \R^{12})^{-1}(w \otimes w),
\end{equation}
leading to
\begin{equation}
A Y = \alpha ^{-\frac {1}{2}} w^{-1} A w
\end{equation}
and more general
\begin{equation}
A Y^{k} = \alpha ^{-\frac {k}{2}} w^{-k} A w^{k}.\label{crysspeter}
\end{equation}
In the case that we are not dealing with a ribbon Hopf algebra, there
is an alternative expression \cite{pC} based on another algebra
homomorphism $A \mapsto A D^{-1}$, where $D =$ $<u,A>$,
\begin{equation}
A Y = \alpha  u A D^{-1} u^{-1}.\label{cryss}
\end{equation}
{}From the form of these equations it is clear that the
map $Cr:A \mapsto A Y^{k}$ is an algebra homomorphism. It also
follows quite easily that this map is invariant in the
sense $\DA \circ Cr = (Cr \otimes i\!d) \circ \Delta $. This
immediately poses the question of a relation to our theory
of bicovariant generators and pure braid elements.
For the ``Ribbon'' case we find
\begin{equation}
\begin{array}{rcl}
Y^{k}      & = & \alpha ^{-\frac {k}{2}} S(A) w^{-k} A w^{k}\\
        & = & \alpha ^{-\frac {k}{2}} <(w^{-k})_{(1)}, A>(w^{-k})_{(2)} w^{k},
\end{array}
\end{equation}
so that $Y^{k} = <\Ups,A \otimes i\!d>$ with the pure braid element
\begin{equation}
\Ups = \alpha ^{-\frac {k}{2}} \Delta (w^{-k}) (1 \otimes w^{k}).
\end{equation}
The ``Non-Ribbon'' case gives
\begin{equation}
\begin{array}{rcl}
Y       & = & \alpha  S(A) u A D^{-1} u^{-1}\\
        & = & \alpha  <u_{(1)},A> u_{(2)} <u^{-1},A> u^{-1}\\
        & = & \alpha  <u_{(1)} u^{-1},A> u_{(2)} u^{-1},
\end{array}
\end{equation}
such that again $Y = <\Ups',A \otimes i\!d>$ with another pure braid
element
\begin{equation}
\Ups' = \alpha  \Delta (u) (u^{-1} \otimes u^{-1}).
\end{equation}
Both examples are hence as expected
special cases of the pure braid formulation.

\chapter{\mbox{$\Schreib R$} - Gymnastics}
\label{Rgym}

In this chapter we would like to study for the example of $Y \in M_{N}(\U)$ the
matrix form of \U\ as introduced at the end of section \ref{Dual}.
Let us first derive commutation relations for $Y$ from the quantum
Yang-Baxter  equation (QYBE): Combine the following two copies of the
QYBE,
$$ \R^{12}\R^{13}\R^{23} = \R^{23}\R^{13}\R^{12},\mbox{ and }
   \R^{21}\R^{31}\R^{32} = \R^{32}\R^{31}\R^{21},$$
resulting in,
$$\R^{21}\R^{31}\underline{\R^{32}\R^{12}\R^{13}}\R^{23}=
\R^{32}\underline{\R^{31}\R^{21}\R^{23}}\R^{13}\R^{12},$$
and apply the QYBE to the underlined part to find,
$$\R^{21}(\R^{31}\R^{13})\R^{12}(\R^{32}\R^{23})=
(\R^{32}\R^{23})\R^{21}(\R^{31}\R^{13})\R^{12},$$
which, when evaluated on $<\;.\;,\:A_{1} \otimes A_{2} \otimes i\!d>$, gives:
\begin{equation}
R_{21} Y_{1} R_{12} Y_{2} = Y_{2} R_{21} Y_{1} R_{12}.
\label{YY}
\end{equation}

\section{Higher Representations and the $\bullet$-Product}

As was pointed out in section \ref{Dual}, tensor product representations
of \U\ can be constructed by combining $A$-matrices. This product of $A$-%
matrices defines a new product for \U\, which we will denote ``$\bullet$''.
The idea is to combine $Y$-matrices (or $L^{+},L^{-}$ matrices) in the same
way as $A$-matrices to get higher dimensional matrix representations,
\begin{eqnarray}
Y_{1} \bullet Y_{2} & := & <\R^{21} \R^{12} , A_{1}A_{2} \otimes i\!d>,
\label{YdotY}\\
L^{+}_{1} \bullet L^{+}_{2} & := & <\R^{21} , A_{1}A_{2} \otimes i\!d>,\\
SL^{-}_{1} \bullet SL^{-}_{2} & := & <\R^{12} , A_{1}A_{2} \otimes i\!d>.
\end{eqnarray}
Let us evaluate (\ref{YdotY}) in terms of the ordinary product in \U,
\begin{equation}
\begin{array}{rcl}
Y_{1} \bullet Y_{2} & = &<(\Delta \otimes i\!d)\R^{21} \R^{12} , A_{1}
\otimes A_{2} \otimes i\!d>\\
& = & <\R^{32}\R^{31}\,\R^{13}\R^{23},A_{1} \otimes A_{2} \otimes i\!d>\\
& = & <(\R^{-1})^{12}\,\R^{31}\R^{13}\,\R^{12}\,\R^{32}\R^{23},
A_{1} \otimes A_{2} \otimes i\!d>\\
& = & R_{12}^{-1} Y_{1} R_{12} Y_{2},
\end{array}
\label{YcY}
\end{equation}
where we have used,
\begin{eqnarray*}
\R^{32}\R^{31}\R^{13}\R^{23} & = & ((\R^{-1})^{12} \underline{\R^{12})
\R^{32} \R^{31}} \R^{13} \R^{23}\\
& = & (\R^{-1})^{12} \R^{31} \underline{\R^{32} \R^{12} \R^{13}} \R^{23}\\
& = & (\R^{-1})^{12}\R^{31}\R^{13}\R^{12}\R^{32}\R^{23}.
\end{eqnarray*}
Similar expressions for $L^{+}$ and $SL^{-}$ are:
\begin{eqnarray}
L^{+}_{1} \bullet L^{+}_{2} & = & L^{+}_{2}L^{+}_{1},\label{lpclp}\\
SL^{-}_{1} \bullet SL^{-}_{2} & = & SL^{-}_{1} SL^{-}_{2}.\label{lmclm}
\end{eqnarray}
All matrices in $M_{N}(\U)$ satisfy by definition the same commutation
relations (\ref{AA}) as $A$, when written in terms of the $\bullet$-
product,
\begin{eqnarray}
R_{12} L^{+}_{1} \bullet L^{+}_{2}  =   L^{+}_{2}  \bullet L^{+}_{1} R_{12} &
\Leftrightarrow
& R_{12} L^{+}_{2} L^{+}_{1} = L^{+}_{1} L^{+}_{2} R_{12},\\
R_{12} SL^{+}_{1} \bullet SL^{+}_{2}  =  SL^{+}_{2}  \bullet SL^{+}_{1} R_{12}
& \Leftrightarrow
& R_{12} SL^{+}_{1} SL^{+}_{2} = SL^{+}_{2} SL^{+}_{1} R_{12},\\
R_{12} Y_{1} \bullet Y_{2} = Y_{2} \bullet  Y_{1}  R_{12} &
\Leftrightarrow
&\begin{array}[t]{lr}\makebox[11mm][l]{$\!\!R_{12} (R_{12}^{-1} Y_{1}
R_{12} Y_{2})$}&\\
&  =  (R_{21}^{-1} Y_{2} R_{21} Y_{1}) R_{12}
\end{array} \nonumber \\ \label{YYY}
& \Leftrightarrow & \x R_{21} Y_{1} R_{12} Y_{2} = Y_{2} R_{21} Y_{1} R_{12}.
\end{eqnarray}
{\em Remark:} Equations incorporating the $\bullet$-product are
mathematically very similar to the expressions introduced in ref.\cite{Md2}
for braided linear algebras --- our analysis was in fact motivated by
that work --- but on a conceptional level things are quite different:
We are not dealing with a braided algebra with a braided multiplication
but rather with a rule for combining matrix representations that turns out
to be very useful, as we will see, to find conditions on the matrices in
$M_{N}(\U)$ from algebraic relations for matrices in $M_{N}(\A)$.

\subsection{Multiple $\bullet$-Products} \label{S:MdotP}

We can define multiple (associative) $\bullet$-products by,
\begin{equation}
Y_{1} \bullet Y_{2} \bullet \ldots \bullet Y_{k} :=
<\R^{21}\R^{12} , A_{1} A_{2} \cdots A_{k} \otimes i\!d>,
\end{equation}
but this equation is not very useful to evaluate these multiple
$\bullet$-products in practice. However, the ``big'' \RR-matrix of equation
(\ref{bigR}) can be used to calculate multiple
$\bullet$-products recursively: Let \x$\YY_{I} \equiv Y_{1'} \bullet Y_{2'}
\bullet\ldots  \bullet
Y_{n'}$\x and \x$\YY_{I\!I} \equiv Y_{1} \bullet Y_{2} \bullet\ldots \bullet
Y_{m}$,\x then:
\begin{equation}
\YY_{I} \bullet \YY_{I\!I} = \RRI^{-1} \YY_{I} \RRI \YY_{I\!I};
\end{equation}
compare to (\ref{RRA}) and (\ref{YcY}).
The analog of equation (\ref{YYY}) is also true:
\begin{eqnarray}
&&\RRI \YY_{I} \bullet \YY_{I\!I} = \YY_{I\!I} \bullet \YY_{I} \RRI\\
&&\Leftrightarrow \x \RR_{I\!I,I} \YY_{I} \RRI \YY_{I\!I} =
\YY_{I\!I} \RR_{I\!I,I} \YY_{I} \RRI.
\end{eqnarray}
The $\bullet$-product of three $Y$-matrices, for example,
reads in terms of the
ordinary multiplication in \U\ as,
\begin{equation}
\begin{array}{rcl}
Y_{1} \bullet (Y_{2} \bullet Y_{3})
&=& \RR^{-1}_{1,(23)} Y_{1} \RR_{1,(23)} (Y_{2} \bullet Y_{3})\\
&=&(R^{-1}_{12} R^{-1}_{13} Y_{1} R_{13} R_{12}) (R^{-1}_{23} Y_{2} R_{23})
Y_{3}.
\end{array}
\end{equation}
This formula generalizes to higher
$\bullet$-products,\footnote{All products are ordered according to increasing
multiplication parameter, e.g.  $$\displaystyle\prod _{i=1}^{k}
\bullet Y_{i} \equiv
Y_{1} \bullet Y_{2} \bullet \ldots \bullet Y_{k}.$$}
\begin{equation}
\begin{array}{l}
\YY_{(1\ldots 2)} \equiv {\displaystyle \prod _{i=1}^{k} \bullet Y_{i} =
\prod _{i=1}^{k}
Y^{(i)}_{1\ldots k}},
\z \mbox{where:}\vspace{3mm}\\
Y^{(i)}_{1\ldots k} = \left\{ \begin{array}{l}
R^{-1}_{i\:(i+1)} R^{-1}_{i\:(i+2)}\cdots R^{-1}_{i\:k}
Y_{i} R_{i\:k}\cdots R_{i\:(i+1)},\x 1\leq i < k,\\
Y_{k},\x i = k.
\end{array} \right.
\end{array}
\label{mult}
\end{equation}

\section{Quantum Determinants}

Assuming that we have defined the quantum determinant $\det_{q} A$ of $A$
in a suitable way --- e.g. through use of the quantum $\varepsilon_{q}$-tensor,
which  in turn can be derived from the quantum exterior plane --- we can then
use the invariant maps $\Phi _{n}$  for $n = 2$ to find the corresponding
expressions in \U; see (\ref{dety}).   Let us consider a couple of examples:
\begin{eqnarray}
\mbox{Det} Y & := & <\R^{21}\R^{12},{\det}_{q} A \otimes i\!d>,\\
\mbox{Det} L^{+} & := & <\R^{21},{\det}_{q} A \otimes i\!d>,\\
\mbox{Det} SL^{-} & := & <\R^{12},{\det}_{q} A \otimes i\!d>.
\end{eqnarray}
Because of equations (\ref{lpclp}) and (\ref{lmclm}) we can identify,
\begin{equation}
\mbox{Det} L^{+} \equiv {\det}_{q^{-1}} L^{+}, \x \mbox{Det} SL^{-} \equiv
{\det}_{q} SL^{-}.
\end{equation}
Properties of ${\det}_{q} A$, namely:
\begin{eqnarray}
A \;{\det}_{q} A & = & {\det}_{q} A \;A\z\mbox{\em (central),}\label{cent}\\
\Delta ({\det}_{q} A) & = & {\det}_{q} A \otimes {\det}_{q} A
\z\mbox{\em (group-like),}
\label{grpl}
\end{eqnarray}
translate into corresponding properties of ``Det''.
For example, here is a short proof of the centrality of $\mbox{Det} Y$
$\equiv Y_{{\det}_{q} A}$
based on equations (\ref{altxy}) and (\ref{dayb}):\footnote{This
proof easily generalizes to show the centrality of {\em any} (right)
invariant $c \in \U$, $\DA(c) = c \otimes 1$, an example being the invariant
traces tr$(D^{-1} Y^{k})$ \cite{RTF}.}
\begin{equation}
\begin{array}{rcl}
x \: Y_{b} & = & Y_{b_{(2)}} < x_{(1)}\:,\:S(b_{(1)}) b_{(3)}> x_{(2)},\z
\forall x \in \U;\\
\Rightarrow x \: Y_{{\det}_{q} A} & = & Y_{{\det}_{q} A}
<x_{(1)}\:,\:S({\det}_{q} A) {\det}_{q} A> x_{(2)}\\
& = & Y_{{\det}_{q} A} <x_{(1)}\:,\:1> x_{(2)}\\
& = & Y_{{\det}_{q} A} \: x,\z\forall x \in \U.
\end{array}
\end{equation}
The determinant of $Y$ is central in the algebra, so its matrix
representation must be proportional to the identity matrix,
\begin{equation}
<\mbox{Det} Y, A > = \kappa I,
\label{kappa}
\end{equation}
with some proportionality constant $\kappa $ that is equal to one in the case
of special quantum groups; note that (\ref{kappa}) is equivalent to:
\begin{equation}
{\det}_{1} (R_{21} R_{12}) = \kappa I_{12},
\end{equation}
where ${\det}_{1}$ is the ordinary determinant taken in the first
pair of matrix indices.
We can now compute the  commutation relation of Det$ Y$ with $A$ \cite{SWZ},
\begin{equation}
\begin{array}{rcl}
\mbox{Det} Y A & = & A<\mbox{Det} Y, A>\:\mbox{Det} Y\\
               & = & \kappa A \; \mbox{Det} Y,
\end{array}
\end{equation}
showing that in the case of special quantum groups the determinant
of $Y$ is actually
central in $\A\cross\U$.\footnote{The invariant traces are central
only in \U\ because they are not group-like.}

Using (\ref{grpl})
in the definition of Det$Y$,
\begin{equation}
\begin{array}{rcl}
\mbox{Det} Y & = & <\R^{21}\R^{12},{\det}_{q} A \otimes i\!d>\\
& = & <\R^{31}\R^{23},\Delta ({\det}_{q} A) \otimes i\!d>\\
& = & <\R^{31}\R^{23},{\det}_{q} A \otimes {\det}_{q} A \otimes i\!d>\\
& = & {\det}_{q^{-1}} L^{+} \cdot {\det}_{q} SL^{-},
\end{array}
\end{equation}
we see that ``Det$Y$'' coincides with the definition of the determinant
of $Y$ given in \cite{Z2}.

A practical calculation of Det$Y$ in terms of the matrix elements of $Y$
starts from,
\begin{equation}
{\det}_{q} A \x\varepsilon_{q}^{i_{1}\cdots i_{N}} = {\left( \prod _{k=1}^{N}
A_{k}
\right)^{i_{1}\cdots i_{N}}}{}_{j_{1}\cdots j_{N}}
\x\varepsilon_{q}^{j_{1}\cdots j_{N}},
\end{equation}
and uses Det$Y = {\det}_{q} \bullet\!Y$, i.e. the q-determinant with the
$\bullet$-multiplication:
\begin{equation}
\mbox{Det} Y \x\varepsilon_{q}^{i_{1}\cdots i_{N}} = {\left( \prod _{k=1}^{N}
\bullet Y_{k} \right)^{i_{1}\cdots i_{N}}}{}_{j_{1}\cdots j_{N}}
\x\varepsilon_{q}^{j_{1}\cdots j_{N}}.
\end{equation}
Now we use equation (\ref{mult}) and get:
\begin{equation}
\begin{array}{l}
\mbox{Det} Y \x\varepsilon_{q}^{i_{1}\cdots i_{N}} = {\left(
{\displaystyle  \prod _{k=1}^{N}
Y^{(k)}_{1\ldots N}}\right)^{i_{1}\cdots i_{N}}}{}_{j_{1}\cdots j_{N}}
\x\varepsilon_{q}^{j_{1}\cdots j_{N}}, \z \mbox{where:}\vspace{3mm}\\
Y^{(i)}_{1\ldots k} = \left\{ \begin{array}{l}
R^{-1}_{i\:(i+1)} R^{-1}_{i\:(i+2)}\cdots R^{-1}_{i\:k}
Y_{i} R_{i\:k}\cdots R_{i\:(i+1)},\x 1\leq i < k,\\
Y_{k},\x i = k.
\end{array} \right.
\end{array}
\end{equation}

It is interesting to see what happens if we use a matrix $T \in M_{N}(\A)$
with determinant ${det}_{q} T = 1$, e.g. $T := A/({det}_{q} A)^{1/N}$,  to
define a matrix $Z \in M_{N}(\U)$ \cite{SWZ}
in analogy to equation (\ref{defY}),
\begin{equation}
Z := <\R^{21}\R^{12} , T \otimes i\!d>;
\end{equation}
we find that $Z$ is automatically of unit determinant:
\begin{equation}
\begin{array}{rcl}
Det Z & :=& <\R^{21}\R^{12} , {det}_{q} T \otimes i\!d> \\
      & = & <\R^{21}\R^{12} , 1 \otimes i\!d>\\
      & = & (\epsilon \otimes i\!d)(\R^{21}\R^{12})\x =\x 1.
\end{array}
\end{equation}

\section{An Orthogonality Relation for $Y$}
If we want to consider only such transformations
\begin{equation}
x \mapsto \AD(x) = A \dot{\otimes} x,\z x \in \mbox{\kreuz C}_{q}^{N},\:A\in
M_{N}(\A),
\end{equation}
of the quantum plane that leave lengths invariant, we need to impose an
orthogonality condition on $A$; see \cite{RTF}. Let $C \in M_{N}(k)$ be
the appropriate metric and $x^{T} C x$ the length squared of $x$ then we find,
\begin{equation}
A^{T} C A = C \z \mbox{\em (orthogonality)},
\end{equation}
as the condition for an invariant length,
\begin{equation}
x^{T} C x \mapsto \AD(x^{T} C x) = 1 \otimes x^{T} C x.
\end{equation}
If we restrict $A$ --- and thereby \A\ ---  in this way we should also
impose a corresponding orthogonality condition in \U.
Use of the $\bullet$-product makes this, as in the case of the quantum
determinants, an easy task: we can simply copy the orthogonality condition
for $A$ and propose,
\begin{eqnarray}
(L^{+})^{T} \bullet C L^{+} & = & C \x \Rightarrow \x L^{+} C^{T} (L^{+})^{T} =
C^{T},\\
(SL^{-})^{T} \bullet C SL^{-} & = & C \x \Rightarrow \x (SL^{-})^{T} C SL^{-}
=  C,\\
Y^{T} \bullet C Y & = & C,\z \mbox{(matrix multiplication understood)},
\end{eqnarray}
as orthogonality conditions in \U. The first two equations were derived
before in \cite{RTF} in a different way. Let us calculate the condition
on $Y$ in terms of the ordinary multiplication in \U,
\begin{equation}
\begin{array}{rcl}
C_{ij} & = & {Y^{k}}_{i} \bullet C_{kl} {Y^{l}}_{j}\\
       & = & C_{kl} {(Y_{1} \bullet Y_{2})^{kl}}_{ij}\\
       & = & C_{kl} {(R^{-1}_{12} Y_{1} R_{12} Y_{2})^{kl}}_{ij},
\end{array}
\end{equation}
or, using $C_{ij} = q^{(N-1)} {R^{lk}}_{ij} C_{kl}$:
\begin{equation}
C_{ij} = q^{(N-1)} C_{mn} {(Y_{1} R_{12} Y_{2})^{nm}}_{ij}.
\end{equation}
{\em Remark:} Algebraic relations  on the matrix elements of $Y$
like the ones given in the
previous two sections  also give implicit conditions on $\R$;
however we purposely did not specify  $\R$, but rather formally assume
its existence and focus on the numerical R-matrices that appear in
all final expressions. Numerical R-matrices are known for most
deformed Lie algebras of interest \cite{RTF} and many other quantum groups.
One could presumably use some of the techniques outlined in this article
to actually derive  relations for numerical R-matrices or even for the
universal $\R$.

\section{About the Coproduct of $Y$}

It would be nice if we could express the coproduct of $Y$,
\begin{equation}
\Delta (Y) = <(i\!d \otimes \Delta ) \R^{21}\R^{12}, A \otimes i\!d>,
\end{equation}
in terms of the matrix elements of the matrix $Y$ itself, as it is
possible for the coproducts of the matrices $L^{+}$ and $L^{-}$.
Unfortunately, simple expressions have only been found in some
special cases; see e.g. \cite{C1,C2,DJS}. A short calculation gives,
\begin{equation}
\Delta ({Y^{i}}_{j}) = (\R^{-1})^{12} (1 \otimes {Y^{i}}_{k}) \R^{12}
({Y^{k}}_{j}
\otimes 1);
\end{equation}
this could be interpreted as some kind of braided tensor product
\cite{Md2,Md3},
\begin{equation}
\Delta ({Y^{i}}_{j}) =: {Y^{i}}_{k} \tilde{\otimes} {Y^{k}}_{j},
\end{equation}
but for practical purposes one usually introduces a new matrix,
\begin{equation}
{O_{(ij)}}^{(kl)} := {(L^{+})^{i}}_{k} S{(L^{-})^{l}}_{j}\x \in M_{N \times
N}(\U),
\end{equation}
such that,
\begin{equation}
\Delta (Y_{A}) = {O_{A}}^{B} \otimes Y_{B},
\end{equation}
where capital letters stand for pairs of indices. The coproduct of
${X^{i}}_{j} = {(I - Y)^{i}}_{j}/\lambda $ is in this notation:
\begin{equation}
\Delta (X_{A}) = X_{A} \otimes 1 + {O_{A}}^{B} \otimes X_{B}.
\label{DeltaX}
\end{equation}

We will only use ${O_{A}}^{B}$ in formal expressions involving the
coproduct of $Y$. It will usually not show up in any practical
calculation, because commutation relation (\ref{YA}) already
implicitly contains $\Delta (Y)$ and all inner products of
$Y$ with strings of $A$-matrices following from it.

\chapter{Vectorfields on Quantum Groups}

\label{C:VoQG}
In this chapter we are trying to find quantum analogs of two important
and closely related concepts in the classical theory of Lie groups:
Lie algebras of left-invariant vector fields and general vector fields over
the group manifold. We will come back to both subjects in part~2, after
developing the additional structure of an exterior differential
calculus.
Our approach will be heuristic in nature;
stress is on  formation of concepts (Begriffsbildung).
The concept of vector fields can also be approached from differential
forms, see \cite{Asc}.

\section{Quantum Lie Algebras}
\label{Lie}
\subsection{Adjoint Action and Jacobi Identities}

Classically the (left) adjoint actions of the generators $\chi _{i}$ of a
Lie algebra {\Deutsch g} on each other are given by the commutators,
\begin{equation}
\chi _{i} \ad \chi _{j} = [ \chi _{i} , \chi _{j} ] = \chi _{k}
f_{i}{}^{k}{}_{j},
\label{commutator}
\end{equation}
expressible in terms of the structure constants $f_{i}{}^{k}{}_{j}$,
whereas the (left) adjoint action of elements of the corresponding
Lie group \mbox{\Deutsch G} is given by conjugation,
\begin{equation}
h \ad g = h g h^{-1}, \z h,g \in \mbox{\Deutsch G}.
\end{equation}
Both formulas generalize in Hopf algebra language to the same
expression,
\begin{eqnarray}
&&\chi _{i} \ad \chi _{j}  =  \chi _{i_{(1)}} \chi _{j} S(\chi
_{i_{(2)}}),\z\mbox{with:}
\x S(\chi ) = - \chi ,\nonumber \\
&& \Delta (\chi ) \equiv \chi _{(1)} \otimes \chi _{(2)} = \chi \otimes
1 + 1 \otimes \chi ,\z \mbox{for } \forall
\chi \in \mbox{\Deutsch g},\vspace{5mm}\label{adcom}\\
&&h \ad g = h_{(1)} g S(h_{(2)}),\z\mbox{with:}
\x S(h) = h^{-1},\nonumber\\
&&\Delta (h) \equiv h_{(1)} \otimes h_{(2)} = h \otimes h,\z \mbox{for }
\forall
h \in \mbox{\Deutsch G},
\end{eqnarray}
and agree with our formula (\ref{adjoint}) for the (left) adjoint action
in \U.
We can derive two {\em generalized Jacobi identities} for double adjoint
actions,
\begin{equation}
\begin{array}{rcl}
x \ad (y \ad z) & = & (x y) \ad z\\
& =& ((x_{(1)} \ad y) x_{(2)}) \ad z\\
& =& (x_{(1)} \ad y) \ad (x_{(2)} \ad z),
\end{array}
\end{equation}
and,
\begin{equation}
\begin{array}{rcl}
(x \ad y) \ad z & = & (x_{(1)} y S(x_{(2)})) \ad z\\
& = & x_{(1)} \ad (y \ad (S(x_{(2)}) \ad z )).
\end{array}
\end{equation}
Both expressions become the ordinary Jacobi identity in the classical
limit and they are not independent: Using the fact that $\ad$ is an
action they imply each other.

In the following we would like to derive the quantum version of
(\ref{commutator}) with ``quantum commutator'' and
``quantum structure constants''.
The idea is to utilize the (passive) transformations that we have studied in
some detail in sections \ref{Bico} and \ref{Braid} to find an expression
for the corresponding active transformations or actions.
The effects of passive transformations are the inverse of active
transformations, so
here is the inverse or right adjoint action for a group:
\begin{equation}
h^{-1} \ad g = g \stackrel{\mbox{\scriptsize ad}}{\triangleleft} h
= S(h_{(1)}) g h_{(2)}.
\end{equation}
This gives rise to a (right) adjoint coaction in Fun(\mbox{\Deutsch G}):
$$A \mapsto S(A') A A',\z\mbox{i.e.}$$
\begin{equation}
\fun\ni\x {A^{i}}_{j} \mapsto {A^{k}}_{l} \otimes S({A^{i}}_{k}) {A^{l}}_{j}
\x\in \fun\otimes\fun;
\end{equation}
here we have written ``\fun'' instead of ``Fun(\mbox{\Deutsch G})''
because the coalgebra of \fun\ is in fact the same undeformed
coalgebra as the one of Fun(\mbox{\Deutsch G}).
In section \ref{Braid} we saw that the $Y$-matrix has particularly nice
transformation properties:
\begin{eqnarray*}
A & \mapsto & S(A') A:\z Y \x\mapsto\x 1 \otimes Y,\\
A &\mapsto &A A':\z Y \x\mapsto\x S(A') Y A'.
\end{eqnarray*}
It follows that:
\begin{equation}
A \x\mapsto\x S(A') A A':\z {Y^{i}}_{j} \x\mapsto\x {Y^{k}}_{l}
\otimes S({A^{i}}_{k}) {A^{l}}_{j}.
\label{adco}
\end{equation}
This is the ``unspecified'' adjoint right {\em coaction} for $Y$;
we recover the ``specific''  left adjoint {\em action},
$$x \ad {Y^{i}}_{j} = x_{(1)} {Y^{i}}_{j} S(x_{(2)}),$$
of an arbitrary $x \in \uqg$ by evaluating the second factor of the
adjoint coaction (\ref{adco}) on $x$:
\begin{equation}
x \ad {Y^{i}}_{j} = {Y^{k}}_{l} < x\:,\:S({A^{i}}_{k}) {A^{l}}_{j}>,
\z\mbox{for }\forall x \in \uqg.
\label{xadY}
\end{equation}
At the expense of  intuitive insight we can
alternatively derive a more general formula directly from
equations (\ref{adjoint}), (\ref{altxy}), and (\ref{dayb}),
\begin{equation}
\begin{array}{rcl}
x \ad Y_{b} & = & x_{(1)} Y_{b} S(x_{(2)})\\
          & = & (Y_{b})^{(1)}<x_{(1)},(Y_{b})^{(2)'}> x_{(2)} S(x_{(3)})\\
          & = & (Y_{b})^{(1)}<x_{(1)},(Y_{b})^{(2)'}> \epsilon (x_{(2)}) \\
          & = & (Y_{b})^{(1)}<x,(Y_{b})^{(2)'}> \\
          & = & Y_{b_{(2)}}<x , S(b_{(1)}) b_{(3)}>;
\end{array}
\label{adyb}
\end{equation}
note the appearance of the (right) adjoined coaction \cite{W2} in \fun,
\begin{equation}
\Delta ^{\mbox{\scriptsize Ad}}(b) = b_{(2)} \otimes S(b_{(1)}) b_{(3)},
\end{equation}
in this formula.

We have found exactly what we were looking for in a {\em quantum Lie algebra};
the adjoint action (\ref{xadY}) or (\ref{adyb}) ---
which is the generalization of the classical commutator --- of elements
of \uqg\ on elements in a certain subset of \uqg\ evaluates to a
{\em linear} combination of elements of that subset.
So we do not really have to use the whole
universal enveloping algebra when dealing with quantum groups but can
rather consider a  subset spanned by elements of the
general form $Y_{b} \equiv <\Y , b \otimes i\!d>$, $\Y \in
\mbox{span}\{B_{2}\}$;
we will call this subset the ``quantum Lie algebra''
$\mbox{\Deutsch g}_{q}$ of the quantum group.
Now we need to find a basis of generators with the right classical limit.

\subsection{$R$-Matrix Approach} \label{S:RMA}

Let us first evaluate (\ref{xadY}) in the case where $x$ is a matrix
element of $Y$. We introduce the short hand,
\begin{equation}
{\bigA^{(kl)}}_{(ij)} \equiv S({A^{i}}_{k}) {A^{l}}_{j},
\label{adrep}
\end{equation}
for the adjoint representation and find,
\begin{equation}
Y_{A} \ad Y_{B} = Y_{C} < Y_{A} , {\bigA^{C}}_{B}>,
\end{equation}
where, again, capital letters stand for pairs of indices.
The evaluation of the inner product $<Y_{A} , {\bigA^{C}}_{B}>
=: C_{A}{}^{C}{}_{B}$ is not
hard even though we do not have an explicit expression for the
coproduct of $Y$; we  simply use the commutation relation (\ref{YA}) of
$Y$ with $A$ and the left and right vacua defined in section \ref{ComRel}:
\begin{equation}
\begin{array}{rcl}
<Y_{1},SA^{T}_{2} A_{3}> & = & <Y_{1} SA^{T}_{2} A_{3}>\\
                 & = & <SA^{T}_{2} (R_{21}^{-1})^{T_{2}}
                        Y_{1} A_{3} (R_{12}^{T_{2}})^{-1}>\\
                 & = & <SA^{T}_{2} (R_{21}^{-1})^{T_{2}} A_{3}
                        R_{31} Y_{1} R_{13} (R_{12}^{T_{2}})^{-1}>\\
                 & = & (R_{21}^{-1})^{T_{2}} R_{31}  R_{13}
                       (R_{12}^{T_{2}})^{-1},\\
\Rightarrow C_{(ij)}{}^{(kl)}{}_{(mn)} & = &  \left(
(R_{21}^{-1})^{T_{2}} R_{31}  R_{13} (R_{12}^{T_{2}})^{-1}
\right)^{ikl}{}_{jmn}.
\end{array}
\label{C}
\end{equation}

The matrix $Y$ becomes the identity matrix in the classical limit, so
$X \equiv (I-Y)/\lambda $ is a better choice; it has the additional
advantage that it has zero counit  and its coproduct (\ref{DeltaX})
resembles the coproduct of
classical differential operators and therefore allows us to write the
adjoint action (\ref{adcom}) as a {\em generalized commutator}:
\begin{equation}
\begin{array}{rcl}
Y_{A} \ad X_{B} & = & {Y_{A}}_{(1)} X_{B} S({Y_{A}}_{(2)}) \\
            & = & {O_{A}}^{D} X_{B} S(Y_{D})\\
            & = & {O_{A}}^{D} X_{B} S({O_{D}}^{E})
                  (\underbrace{I_{E} - \lambda X_{E}}_{Y_{E}} + \lambda
X_{E})\\
            & = & Y_{A} X_{B} + ({O_{A}}^{E} \ad X_{B}) \lambda X_{E}\\
            & = & Y_{A} X_{B} + \lambda <{O_{A}}^{E},{\bigA^{D}}_{B}> X_{D}
X_{E},\\
&\makebox[4mm][l]{with: ${O_{D}}^{E} I_{E} = Y_{D},\x S({O_{D}}^{E}) Y_{E} =
I_{D};$}&\\
\Rightarrow X_{A} \ad X_{B} & = & X_{A} X_{B} - <{O_{A}}^{E},{\bigA^{D}}_{B}>
X_{D} X_{E}.
\end{array}
\end{equation}
Following the notation of reference \cite{B} we introduce the $N^{4} \times
N^{4}$ matrix,
\begin{eqnarray}
\hat{\bigR}^{DE}{}_{AB} & :=&  <{O_{A}}^{E},{\bigA^{D}}_{B}>,\\
\hat{\bigR}^{(mn)(kl)}{}_{(ij)(pq)} & = &
\left(({R_{31}}^{-1})^{T_{3}} R_{41} R_{24}
({R_{23}}^{T_{3}})^{-1}\right)^{ilmn}%
{}_{kjpq},
\end{eqnarray}
but realize when considering the above calculation that $\bigR$ is
not the ``R-matrix in the adjoint representation'' --- that
would be $<\R , {\bigA^{E}}_{A} \otimes {\bigA^{D}}_{B}>$
--- but rather the R-matrix for the braided commutators of
$\mbox{\Deutsch g}_{q}$, giving the commutation relations of the generators
a form resembling an (inhomogeneous) quantum plane.

Now we can write down the generalized Cartan equations of a quantum Lie
algebra $\mbox{\Deutsch g}_{q}$:
\begin{equation}
X_{A} \ad X_{B} = X_{A} X_{B} - \hat{\bigR}^{DE}{}_{AB} X_{D} X_{E} = X_{C}
f_{A}{}^{C}{}_{B},
\label{AadB}
\end{equation}
where, from equation (\ref{C}),
\begin{equation}
f_{A}{}^{C}{}_{B} = (I_{A} I^{C} I_{B} - C_{A}{}^{C}{}_{B})/\lambda .
\label{adrep2}
\end{equation}

\subsection{General Case}

Equation (\ref{AadB}) is strictly only valid for systems of $N^{2}$ generators
with an $N^{2} \times N^{2}$ matrix $\hat{\bigR}$ because $X \in M_{N}(\mbox{%
\Deutsch g}_{q})$ in our construction. Some of these $N^{2}$ generators and
likewise some of the matrix elements of $\hat{\bigR}$ could of course
be zero, but let us anyway consider the more general case of
equation (\ref{adyb}). We will assume a set of $n$ generators $X_{b_{i}}$
corresponding to  a set of $n$ linearly independent functions
$\{ b_{i} \in \fun \, | \,  i = 1,\ldots,n \}$ and an element of the pure braid
group $\mbox{\schreib X} \in \mbox{span}(B_{2})$ via:
\begin{equation}
X_{b_{i}} = < \mbox{\schreib X} , b_{i} \otimes i\!d >.
\end{equation}
We will usually require that all generators have vanishing counit.
A sufficient condition on the $b_{i}$'s ensuring linear closure
of the generators $X_{b_{i}}$ under adjoint action (\ref{adyb}) is,
\begin{equation}
\Delta ^{\mbox{\scriptsize Ad}}(b_{i}) = b_{j} \otimes \mbox{\kreuz
M}^{j}{}_{i}
+ k_{l} \otimes k^{l}_{i},
\label{basfun}
\end{equation}
where $\mbox{\kreuz M}^{j}{}_{i} \in M_{n}(\fun)$ and $k_{l},\,k^{l}_{i} \in
\fun$
such that $< \mbox{\schreib X} , k_{l} \otimes i\!d > = 0$.
The generators will then transform like,
\begin{equation}
\DA(X_{b_{i}}) = X_{b_{j}} \otimes \mbox{\kreuz M}^{j}{}_{i};
\end{equation}
from $(\DA \otimes i\!d)\DA(X_{b_{i}})$
$= (i\!d \otimes \Delta )\DA(X_{b_{i}})$
and $(i\!d \otimes \epsilon )
\DA(X_{b_{i}}) = X_{b_{i}}$ immediately
follows\footnote{This assumes that the $X_{b_{i}}$'s
are linearly independent.} $\Delta (\mbox{\kreuz M})
= \mbox{\kreuz M} \dot{\otimes} \mbox{\kreuz M}$,
$\epsilon (\mbox{\kreuz M}) = I$ and
consequently $S(\mbox{\kreuz M}) = \mbox{\kreuz M}^{-1}$.
\mbox{\kreuz M} is the {\em adjoint matrix representation}.
We find,
\begin{equation}
X_{b_{k}} \ad  X_{b_{i}} = X_{b_{j}} < X_{b_{k}} , \mbox{\kreuz M}^{j}{}_{i}>,
\label{genxadx}
\end{equation}
as a generalization of (\ref{AadB}) with structure constants
$f_{k}{}^{j}{}_{i} = < X_{b_{k}} , \mbox{\kreuz M}^{j}{}_{i}>$.
Whether $X_{b_{k}} \ad  X_{b_{i}}$ can be reexpressed as a deformed
commutator depends on the coproducts of the $X_{b_{i}}$'s and
hence  on the particular choice of
{\schreib X} and $\{ b_{i} \}$.

Equations (\ref{adco}) and (\ref{adrep}) -- (\ref{adrep2}) apply directly
to $Gl_{q}(N)$ and $Sl_{q}(N)$ and other quantum groups in matrix form with
(numerical) $R$-matrices. Such quantum groups have been studied in great
detail in the literature; see e.g. \cite{RTF,B,CC} and references therein.
In the next subsection we would like to discuss the
2-dimensional quantum euclidean algebra as an example that illustrates
some subtleties in the general picture.

\subsection{Bicovariant Generators for $e_{q}(2)$} \label{S:BGfe}

In \cite{W3} Woronowicz introduced the functions on the deformed
$E_{q}(2)$. This and
the corresponding algebra $U_{q}(e(2))$ were explicitly constructed
in chapter~1 using a contraction procedure; here is a short summary:
$m$, $\mb$ and $\theta = \overline{\theta }$ are generating elements of the
Hopf
algebra Fun$(E_{q}(2))$, which satisfy:
\begin{equation}
\begin{array}{l}
m \mb = q^{2} \mb m,\z e^{i \theta } m = q^{2} m e^{i \theta },\z
e^{i \theta } \mb = q^{2} \mb e^{i \theta },\\
\Delta (m) = m \otimes 1 + e^{i \theta } \otimes m,\z
\Delta (\mb) = \mb \otimes 1 + e^{-i \theta } \otimes \mb,\\
\Delta (e^{i \theta }) = e^{i \theta } \otimes e^{i \theta },\z
S(m) = -e^{-i \theta } m,\z S(\mb) = -e^{i \theta } \mb,\\
S(\theta ) = -\theta ,\z
\epsilon (m) = \epsilon (\mb) = \epsilon (\theta ) = 0.
\end{array}
\label{Eqtwo}
\end{equation}
Fun$(E_{q}(2))$ coacts on the complex coordinate function $z$ of the euclidean
plane as $\DA(z) = z \otimes e^{i \theta } + 1 \otimes m$; i.e. $\theta $
corresponds to rotations, $m$ to translations.
The dual Hopf algebra $U_{q}(e(2))$ is generated by $J = \overline{J}$
and $P_{\pm } = \overline{P_{\mp }}$ satisfying:
\begin{equation}
\begin{array}{l}
[J,P_{\pm }] = \pm P_{\pm },\z [P_{+},P_{-}] = 0,\\
\Delta (P_{\pm }) = P_{\pm } \otimes q^{J} + q^{-J} \otimes P_{\pm },\z
\Delta (J) = J \otimes 1 + 1 \otimes J,\\
S(P_{\pm }) = -q^{\pm 1} P_{\pm },\z S(J) = -J,\z \epsilon (P_{\pm }) =
\epsilon (J) = 0.
\end{array}
\label{eqtwo}
\end{equation}
The duality between Fun$(E_{q}(2))$ and $U_{q}(e(2))$ is given by:
\begin{eqnarray}
\lefteqn{<{P_{+}}^{k} {P_{-}}^{l} q^{m J}\:,\: e^{i \theta a} m^{b} \mb^{c}> =
}\z\z
\nonumber \\
& & (-1)^{l} q^{-1/2(k-l)(k+l-1) + l(k-1)} q^{(k+l-m) a}
[k]_{q} ! [l]_{q^{-1}} ! \delta _{lb} \delta _{kc},
\label{duaeqtwo}
\end{eqnarray}
where $k,l,b,c \in \mbox{\kreuz N}_{0}$,\x $m,a \in \mbox{\kreuz Z}$, and,
$$ [x]_{q} ! = \prod _{y=1}^{x} \frac {q^{2y} - 1}{q^{2} - 1},
\z [0]_{q}! = [1]_{q}! = 1.$$
Note that $P_{+} P_{-}$ is central in $U_{q}(e(2))$; i.e. it is a casimir
operator.
$U_{q}(e(2))$ does not have a (known) universal \R, so we have to construct
an element {\schreib X} of span$(B_{2})$ from the casimir $P_{+} P_{-}$:
\begin{equation}
\begin{array}{rcl}
\mbox{\schreib X} & := & \frac {1}{q-q^{-1}}\{ \Delta (P_{+} P_{-})
- (P_{+} P_{-} \otimes 1) \} \\
& = & \frac {1}{q-q^{-1}}\{ P_{+} P_{-} \otimes (q^{2J} -1) +
P_{+} q^{-J} \otimes q^{J} P_{-} \\
& &\mbox{} + P_{-} q^{-J} \otimes q^{J} P_{+} +q^{-2J}
\otimes P_{+} P_{-}\} .
\end{array}
\end{equation}
\mbox{\schreib X} commutes with $\Delta (x)$ for all $x \in U_{q}(e(2))$
because $P_{+} P_{-}$ is a casimir. We introduced the second term
$(P_{+} P_{-} \otimes 1)$ in
\mbox{\schreib X} to ensure $(i\!d \otimes \epsilon )\mbox{\schreib X} = 0$
so that we are guaranteed to get bicovariant generators with zero counit.
Now we need a set of functions which transform like (\ref{basfun}).
A particular simple choice is $a_{0} := e^{i \theta } - 1$,
$a_{+} := m$, and $a_{-} := e^{i \theta } \mb$. These functions transform
under the adjoint coaction as:
\begin{equation}
\Delta ^{\mbox{\scriptsize Ad}} (a_{0}, a_{+}, a_{-}) =
(a_{0}, a_{+}, a_{-}) \dot{\otimes} \left(
\begin{array}{ccc}
1 & e^{-i \theta } m & -e^{i \theta } \mb\\
0 & e^{-i \theta }   & 0 \\
0 & 0               & e^{i \theta }
\end{array} \right).
\end{equation}
Unfortunately we notice that $a_{0}$ and thereby $X_{a_{0}}$ are invariant,
forcing $X_{a_{0}}$ to be a casimir independent of the particular choice
of \mbox{\schreib X}. Indeed we find $X_{a_{0}} = q P_{+} P_{-}$,
$X_{a_{+}} = -\sqrt {q}/(q-q^{-1}) q^{J} P_{+}$, and
$X_{a_{-}} = q/(q-q^{-1}) q^{J} P_{-}$, making this an incomplete choice of
bicovariant generators for $e_{q}(2)$.
An ansatz with four functions $b_{0} := (e^{i \theta } - 1)^{2}$,
$b_{1} := -m e^{i \theta } \mb$, $b_{+} := -(e^{i \theta } - 1) m$, and
$b_{-} := q^{-2}(e^{i \theta } - 1) e^{i \theta } \mb$ gives:
\begin{equation}
\Delta ^{\mbox{\scriptsize Ad}} (b_{0}, b_{1}, b_{+}, b_{-}) =
(b_{0}, b_{1}, b_{+}, b_{-}) \dot{\otimes} \left(
\begin{array}{cccc}
1 & \mb m & -e^{-i \theta } m & -q^{-2}e^{i \theta } \mb\\
0 & 1 & 0 & 0\\
0 & - \mb & e^{-i \theta } & 0\\
0 & - m & 0 & e^{i \theta }
\end{array} \right).
\label{ADEQTWO}
\end{equation}
The corresponding bicovariant generators are:
\begin{equation}
\begin{array}{l}
X_{b_{0}} = q (q^{2} - 1) P_{+} P_{-},\z X_{b_{1}} = (q-q^{-1})^{-1}(q^{2J}
-1),\\
X_{b_{+}} = q^{J} P_{+},\z X_{b_{-}} = q q^{J} P_{-}.
\end{array}
\end{equation}
In the classical limit $(q \rightarrow 1)$
these generators become
``zero'', $J$, $P_{+}$, and $P_{-}$ respectively\footnote{The same generators
and their transformation properties
can alternatively be obtained by contracting the bicovariant
calculus on $SU_{q}(2)$.}. The coproducts of the bicovariant generators
have the form expected for differential operators
{\scriptsize \begin{equation}
\Delta \left(\begin{array}{c}
        X_{0}\\X_{1}\\X_{+}\\X_{-}\end{array}\right)
= \left(\begin{array}{c}
        X_{0}\\X_{1}\\X_{+}\\X_{-}\end{array}\right) \otimes 1
+ \left(\begin{array}{cccc}
        \lambda  SX_{1} + 1 & \lambda  X_{0}     & \lambda (\lambda  SX_{1} +
1) & \lambda  X_{+} (\lambda  SX_{1} + 1) \\
        0         & \lambda  X_{1} + 1 & 0            & 0                \\
        0         & \lambda  X_{+}     & 1            & 0                \\
        0         & \lambda  X_{-}     & 0            & 1
\end{array}\right)
\otimes \left(\begin{array}{c}
        X_{0}\\X_{1}\\X_{+}\\X_{-}\end{array}\right).
\end{equation}}
The commutation relations of the generators
follow directly from (\ref{eqtwo}), their adjoint actions are calculated from
(\ref{genxadx}), (\ref{duaeqtwo}), and (\ref{ADEQTWO}) and finally
the commutation relations of the generators with the
functions can be obtained from (\ref{commrel}), (\ref{Eqtwo}) and
(\ref{eqtwo}).

\section{General Vector Fields}
\label{genervf}
In this section we will give a ``quantum geometric'' construction
of the action of general, $i.e.$ neither necessarily left or right invariant,
vector fields, thereby justifying the form of the action that we used
in the construction of the cross-product algebra of differential operators.

\subsection{Classical Left Invariant Vector Fields}

First, recall the left-invariant classical case:
The Lie algebra is spanned by left-invariant vector fields on the group
manifold of a Lie group $G$.
These are uniquely determined by the tangent space at $1$ (the
identity of $G$). Curves on $G$ can be naturally transported by left
(or right) translation {\em i.e.} $h \mapsto g h$ ($h \mapsto h g$).
This defines  a left transport $L_{g^{-1}}$ of the tangent vectors:\x
$L_{g^{-1}}(\chi _{1}) = \tilde{\chi }_{g}$. $\chi _{1}$ is the vector field
$\chi $ at the identity
of the group and $\tilde{\chi }_{g}$ is the new vector field $\tilde{\chi }$
evaluated at the point of the group manifold corresponding to the group
element $g$; if $\chi $ is left invariant then $\chi  = \tilde{\chi }$ and
in particular
\begin{equation}
L_{g^{-1}}(\chi _{1}) = \tilde{\chi }_{g} = \chi _{g}.
\end{equation}
An inner product for a vector field $\chi $ with a function $f$ can be defined
by acting with the vector field on the function and evaluating the
resulting function at the identity of the group:
\begin{equation}
<\chi  , f> := \left. \chi _{1} \tr f \right|_{1} \in k.
\end{equation}
If we know these values for all functions, we can reconstruct
the action of $\chi $ on a function $f$, $\chi _{g} \tr f|_{g}$,
at any (connected) point of the
group manifold. The construction goes as follows (see figure):\\ \\
\unitlength=1.00mm
\special{em:linewidth 0.4pt}
\linethickness{0.4pt}
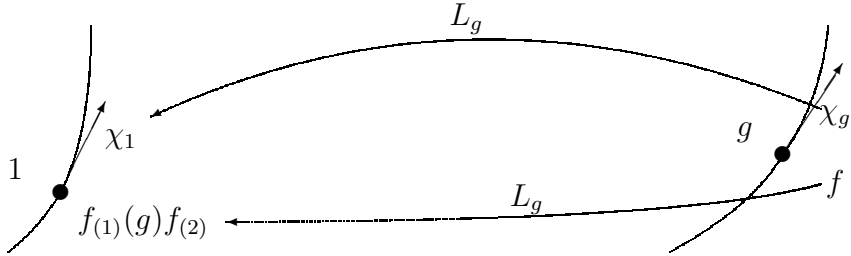
\begin{picture}(114.00,39.00)
\put(10.00,9.00){\vector(1,2){6.00}}
\put(10.00,9.00){\circle*{2.00}}
\put(106.00,14.00){\circle*{2.00}}
\put(4.00,12.00){\makebox(0,0)[cc]{1}}
\put(101.00,17.00){\makebox(0,0)[cc]{$g$}}
\put(113.00,19.00){\makebox(0,0)[cc]{$\chi_g$}}
\put(18.00,16.00){\makebox(0,0)[cc]{$\chi_1$}}
\put(113.00,10.00){\makebox(0,0)[cc]{$f$}}
\put(21.00,5.00){\makebox(0,0)[cc]{$f_{(1)}(g)f_{(2)}$}}
\put(106.00,14.00){\vector(2,3){8.00}}
\bezier{92}(10.00,9.00)(14.00,17.00)(14.00,31.00)
\bezier{44}(10.00,9.00)(7.00,4.00)(3.00,1.00)
\bezier{76}(106.00,14.00)(111.00,21.00)(112.00,31.00)
\bezier{84}(106.00,14.00)(102.00,7.00)(91.00,1.00)
\put(24.00,20.00){\vector(-2,-1){2.00}}
\bezier{388}(22.00,19.00)(65.00,39.00)(111.00,20.00)
\put(34.00,5.00){\vector(-1,0){2.00}}
\bezier{320}(32.00,5.00)(93.00,5.00)(111.00,10.00)
\put(64.00,32.00){\makebox(0,0)[cc]{$L_g$}}
\put(72.00,8.00){\makebox(0,0)[cc]{$L_g$}}
\end{picture}
\\ \\
We start at the point $g$, transport $f$ and $\chi $ back to the identity
by left translation and then evaluate them on each other. The result,
being a number, is invariant under translations and hence gives
the desired quantity. The left translation $Lg(f)$ of a function,
implicitly defined through $L_{g}(f)(h) = f(g h)$, finds an explicit
expression in Hopf algebra language
\begin{equation}
L_{g}(f) = f_{(1)}(g) f_{(2)},
\end{equation}
that we now use to express
\begin{equation}
\begin{array}{rcl}
\left. \chi _{g} \tr f\right|_{g} & = & \left. L_{g}(\chi )_{1} \tr f_{(1)}(g)
f_{(2)}\right|_{1}\\
        & = & \left. \chi _{1} \tr f_{(1)}(g) f_{(2)}\right|_{1}\\
        & = & f_{(1)}(g) <\chi  , f_{(2)}>,
\end{array}
\end{equation}
for a left-invariant vector field $\chi $.
If the drop $g$, we obtain the expression for the action of a vector
field on a function valid on the whole group manifold
\begin{equation}
\chi  \tr f = f_{(1)}<\chi  , f_{(2)}>,
\end{equation}
already familiar from the first chapter. The left and right
vacua by the way find the following `geometric' interpretation:
\begin{quote}
left vacuum $<$: ``Evaluate at the identity (of the group).''\\
right vacuum $>$: ``Evaluate on the unit function.''
\end{quote}

\subsection{Some Quantum Geometry}

Group elements ($g$) do not exist for quantum groups, everything has to
be formulated in terms of a Hopf algebra of functions. The group
operation is replaced by the coproduct of functions. A quantum group
has only few classical points. These correspond to elements of \U\ with
group-like coproducts, e.g. the quantum determinant of $Y$ in $Gl_{q}(2)$:
$\Delta $det${}_{q}Y = $det${}_{q}Y \otimes$det${}_{q}Y$. If we take care only
to
speak about functions in \A\ and its dual Hopf algebra \U, we can, however,
still
develop a  geometric picture for vector fields on quantum groups.
``Points'' will be {\em labeled} by elements of $\tilde{\U}$, which is
the same as \U\ but has the opposite multiplication; elements of
$\tilde{\U}$ are {\em right}-invariant. Lie derivatives along elements
of $\tilde{\U}$ take the place of left translations, while Lie
derivatives along elements of \U\ correspond to right translations.
Here is the quantum picture of the classical construction given
in the previous section:\\ \\
\unitlength=1.00mm
\special{em:linewidth 0.4pt}
\linethickness{0.4pt}
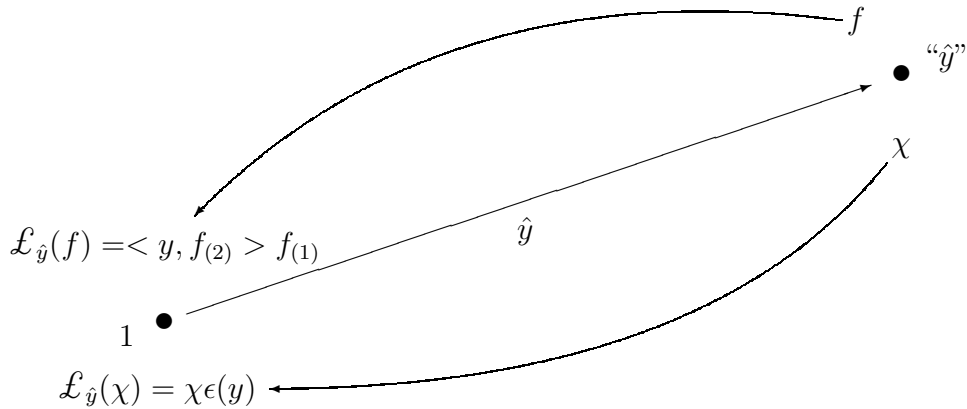
\begin{picture}(125.00,62.00)
\put(24.00,16.00){\vector(3,1){91.00}}
\put(119.00,48.00){\circle*{2.00}}
\put(21.00,15.00){\circle*{2.00}}
\put(125.00,50.00){\makebox(0,0)[cc]{``$\hat{y}$''}}
\put(16.00,13.00){\makebox(0,0)[cc]{1}}
\put(119.00,38.00){\makebox(0,0)[cc]{$\chi$}}
\put(113.00,55.00){\makebox(0,0)[cc]{$f$}}
\put(20.00,6.00){\makebox(0,0)[cc]{$\Li_{\hat{y}}(\chi) = \chi \epsilon(y)$}}
\put(21.00,25.00){\makebox(0,0)[cc]{$\Li_{\hat{y}}(f) = <y,f_{(2)}>f_{(1)}$}}
\put(69.00,27.00){\makebox(0,0)[cc]{$\hat{y}$}}
\put(36.00,6.00){\vector(-1,0){1.00}}
\bezier{384}(35.00,6.00)(93.00,6.00)(117.00,36.00)
\put(26.00,30.00){\vector(-1,-1){1.00}}
\bezier{400}(25.00,29.00)(57.00,62.00)(111.00,55.00)
\end{picture}
\\
Note that $\Li_{\hat{y}}(x) = x \epsilon (\hat{y})$ because $x$ is
left-invariant.
(More precise definitions of these Lie derivatives in connection with
right-projectors will be given in
section~\ref{S:RaLP}). Before we can read any equations off the picture we have
to invent a rule for multiple appearances of the same Hopf algebra
element in the same term:
\begin{quote}
Multiple occurrences of the same Hopf algebra element in a single term
are not allowed. One should use the parts of the coproduct of this
element instead --- starting with the last part of the coproduct and
collecting terms from the right to the left as one moves along the path
that the function is transported.
\end{quote}
Now can compute $x \tr f$ in complete analogy to the classical case
\begin{equation}
\begin{array}{rcl}
\left. x \tr f\right|_{``\hat{y}"} &=& \left. \Li_{\hat{y}_{(1)}}(x) \tr
\Li_{\hat{y}_{(2)}}(f)\right|_{1}\z(\equiv \left. \Li_{\hat{y}}(x \tr
f)\right|_{1})\\
        &=& \left. \epsilon (y_{(1)}) x \tr \Li_{\hat{y}_{(2)}}(f)\right|_{1}\\
        &=& \left. x \tr \Li_{\hat{y}}(f)\right|_{1}\\
        &=& \left. x \tr <y,f_{(1)}> f_{(2)}\right|_{1}\\
        &=& <y,f_{(1)}><x,f_{(2)}>
\end{array}
\end{equation}
or, for arbitrary $\hat{y}$:
\begin{equation}
x \tr f = f_{(1)}<x,f_{(2)}>,
\end{equation}
giving a geometric justification for the left action of \U\
on \A\ that we had introduced in chapter~1.

Now we would like to study the adjoint action in \U, which
can be interpreted as a quantum Lie bracket as we shall see.
Recall the classical construction: Functions and hence curves on
a group manifold can be transported along a vector field. With
the curves we implicitly also transport their tangent vectors.
This transport is called the Lie derivative of a (tangent) vector
along a vector field. Classically we find it to be equal to
the commutator (Lie bracket) of the two vector fields.
Here is how the computation goes in practice: Let $y$ be the
vector field along which the functions are transported and
let $x$ be the ``tangent'' vector field. Consider a function $f$
on the new curve and transport it along the following two equivalent
paths:
\begin{enumerate}
\item Go back along $y$ to the old curve, follow the old curve along $x$
and finally return along $y$ to the new curve.
\item Follow the new curve along $\Li_{y}(x)$.
\end{enumerate}
We have to invent a new rule for backward transport:\footnote{Note that
we follow the path of the transported function; forward hence means ``opposite
to the direction that the vector is pointing'', backward means ``along the
direction that the vector is pointing''.}
\begin{quote}
Moving a function back along a vector field $y$ is the same
as moving forward along the antipode $S(y)$ of that vector field.
(When moving a 1-form, one should use the inverse antipode.)
\end{quote}
The following picture illustrates the geometric construction of the
quantum Lie derivative of a left-invariant vector field along another
left-invariant vector field:\\
\unitlength=1.00mm
\special{em:linewidth 0.4pt}
\linethickness{0.4pt}
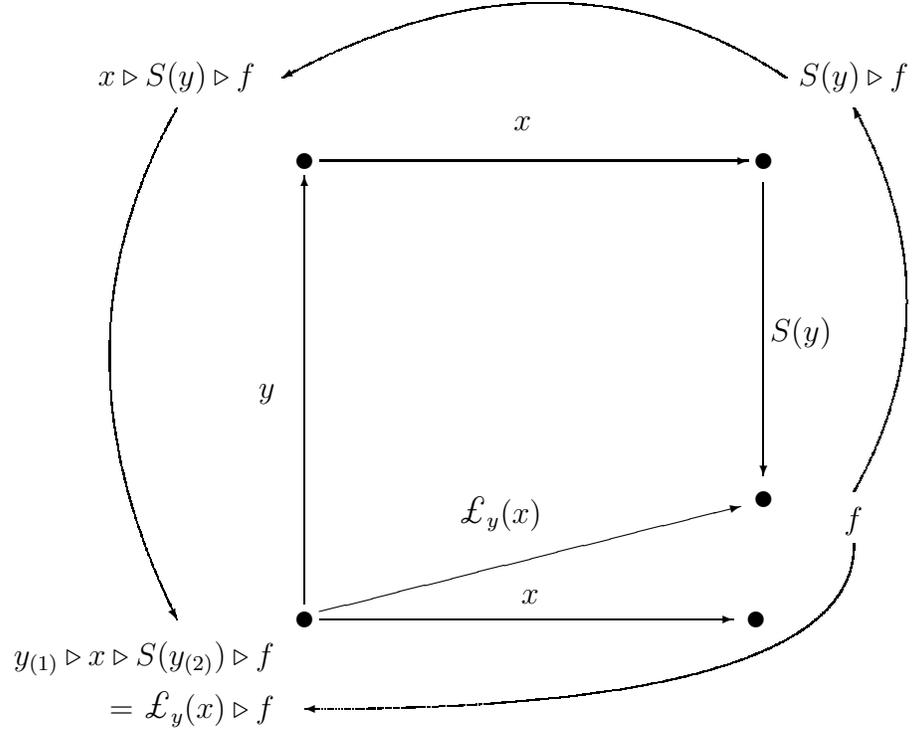
\begin{picture}(145.00,107.00)
\put(60.00,16.00){\vector(4,1){56.00}}
\put(58.00,15.00){\circle*{2.00}}
\put(119.00,31.00){\circle*{2.00}}
\put(119.00,73.00){\vector(0,-1){39.00}}
\put(119.00,76.00){\circle*{2.00}}
\put(60.00,76.00){\vector(1,0){57.00}}
\put(58.00,76.00){\circle*{2.00}}
\put(58.00,17.00){\vector(0,1){57.00}}
\put(60.00,15.00){\vector(1,0){55.00}}
\put(118.00,15.00){\circle*{2.00}}
\put(131.00,28.00){\makebox(0,0)[cc]{$f$}}
\put(131.00,87.00){\makebox(0,0)[cc]{$S(y) \tr f$}}
\put(41.00,87.00){\makebox(0,0)[cc]{$x \tr S(y) \tr f$}}
\put(54.00,12.00){\makebox(0,0)[rt]{$y_{(1)} \tr x \tr S(y_{(2)}) \tr f $}}
\put(124.00,53.00){\makebox(0,0)[cc]{$S(y)$}}
\put(87.00,81.00){\makebox(0,0)[cc]{$x$}}
\put(53.00,45.00){\makebox(0,0)[cc]{$y$}}
\put(84.00,29.00){\makebox(0,0)[cc]{$\Li_y(x)$}}
\put(88.00,18.00){\makebox(0,0)[cc]{$x$}}
\put(132.00,81.00){\vector(-1,2){1.00}}
\bezier{232}(131.00,83.00)(145.00,57.00)(131.00,32.00)
\put(57.00,88.00){\vector(-2,-1){2.00}}
\bezier{312}(55.00,87.00)(92.00,107.00)(122.00,87.00)
\put(40.00,17.00){\vector(1,-2){1.00}}
\bezier{308}(41.00,15.00)(23.00,51.00)(41.00,83.00)
\put(54.00,3.00){\makebox(0,0)[rc]{$= \Li_y(x) \tr f$}}
\put(59.00,3.00){\vector(-1,0){1.00}}
\bezier{380}(58.00,3.00)(131.00,3.00)(131.00,25.00)
\end{picture}
\vspace*{8 mm}\\
We read off this picture that
\begin{equation}
\begin{array}{rcl}
\Li_{y}(x) \tr f & =\, & y_{(1)} \tr x \tr S(y_{(2)}) \tr f\\
              & =:  & (y_{(1)} x S(y_{(2)}) ) \tr f,
\end{array}
\end{equation}
{\em i.e.} $\Li_{y}(x) = y_{(1)} x S(y_{(2)})  = y \ad x$.

\subsection{Action of General Vector Fields}

Our derivation of the action of a vector field on a function in the
previous section relied on the use of left translations in conjunction
with left-invariant vector fields. In this section we would like
to free ourselves from this limitation and show how to derive
the action of a general vector field --- neither necessarily left or right
invariant --- on a function using alternatively left or right
translations.

Left and right coactions $\AD$, $\DA$ contain the information about
transformation properties of vector fields. Here is how a
vector field transforms (classically) if we {\em left}-transport it
from a point $g$ on the group manifold back to the identity
\begin{equation}
\chi |_{g} \mapsto \chi ^{(1)'}(g)\cdot \chi ^{(2)} |_{1},\z\AD(\chi ) \equiv
\chi ^{(1)'}\otimes \chi ^{(2)};
\end{equation}
here is the behavior under a {\em right} translation:
\begin{equation}
\chi |_{g} \mapsto \chi ^{(1)}\cdot \chi ^{(2)'}(g) |_{1},\z\DA(\chi ) \equiv
\chi ^{(1)} \otimes \chi ^{(2)'}.
\end{equation}
If we now redo the construction of the previous section for
general vector fields $\chi $, both for left and right translations, we
get the following two equivalent results for actions on functions:
\begin{equation}
\fbox{$\chi (f) = \underbrace{<\chi ^{(1)},f_{(1)}>
\chi ^{(2)'}f_{(2)}}_{\mbox{from right translation}} = \underbrace{\chi ^{(1)'}
f_{(1)}
<\chi ^{(2)},f_{(2)}>}_{\mbox{from left translation}}$}.
\end{equation}
Technically there is an ordering ambiguity for $f$ and the primed
parts of $\chi $, but this can be easily resolved by requiring
$a(f) = a f$ for $a \in \A$ in both cases; both expressions are
written as left actions.
{}From this equation we can derive the following relations between
left and right coactions for $\chi  \in \A^{*}$:
\begin{equation}
\begin{array}{rcl}
\DA \chi  & = & e_{i} \otimes \chi ^{(1)'}f_{(1)}^{i} <\chi ^{(2)},f_{(2)}^{i}>
S f_{(3)}^{i}\\
        & = & e_{i(1)} \chi ^{(2)} S e_{i(2)} \otimes \chi ^{(1)'} f^{i}\\
        & = & (e_{i} \ad \,\otimes \,\,i\!d)\, \tau (\AD \chi ) (1 \otimes
f^{i}),
\end{array}
\end{equation}
\begin{equation}
\begin{array}{rcl}
\AD \chi  & = & \chi ^{(2)} f_{(3)}^{i} <\chi ^{(1)},f_{(2)}^{i}> S^{-1}
f_{(1)}^{i} \otimes e_{i}\\
        & = & \chi ^{(2)'} f^{i} \otimes S^{-1}(e_{i(2)}) \chi ^{(1)}
e_{i(1)}\\
        & = & (i\!d \otimes S^{-1} e_{i} \ad)\, \tau (\DA \chi ) (f^{i} \otimes
1).
\end{array}
\end{equation}
In this thesis we choose the {\em convention}
that elements in $\U \cong  \A^{*}$ be left-invariant.

\subsection{Right and Left Projectors} \label{S:RaLP}

In this section we will show how to obtain right-invariant
vector fields from left-invariant ones by allowing functional
coefficients. These right-invariant vector fields will live
in $\A\cross\U$ --- recall that elements of \U\ were chosen to
be left-invariant. Let $x$ be the left-invariant vector field and
$\hat{x}$ the corresponding right-invariant vector field.
These vector fields should coincide at the identity, i.e. for any
function $f$
\begin{equation}
\epsilon (\hat{x}) = \epsilon (x),\z <\hat{x},f> = <x,f>. \label{infx}
\end{equation}
For this to make sense we have to extend the definition of the inner
product a little bit to allow elements of $\A\cross\U$ in the first
space. Recalling the geometrical definition
\begin{equation}
<\phi  , f> := \left. \phi  \tr f \right|_{1},\z \phi  \in \A\cross\U,\, f \in
\A
\end{equation}
this is not hard: ($a,f \in \A,\,x\in \U$)
\begin{eqnarray}
<a x,f> &:=& \epsilon (a) <x,f>,\\
<x a,f> &:=& <x, a f>,
\end{eqnarray}
in perfect agreement with the formulation in terms of vacua.
Let $\DA(x) = x^{(1)} \otimes x^{(2)'} \in \U\otimes\A$; it is not hard
to see that
\begin{equation}
\hat{x} := S^{-1}(x^{(2)'}) x^{(1)} \label{hat}
\end{equation}
has the required properties and is right invariant
\begin{equation}
\begin{array}{rcl}
\DA(\hat{x}) & = & (S^{-1} x^{(2)'})_{(1)}
x^{(1)(1)} \otimes (S^{-1} x^{(2)'})_{(2)}
x^{(1)(2)'}\\
        & = & S^{-1}(x^{(2)'})_{(3)} x^{(1)} \otimes S^{-1}(x^{(2)'})_{(2)}
x^{(2)'}{}_{(1)}\\
        & = & S^{-1}(x^{(2)'})_{(2)} x^{(1)} \otimes 1 \epsilon
(x^{(2)'}{}_{(1)})\\
        & = & S^{-1}(x^{(2)'}) x^{(1)} \otimes 1\\
        & = & \hat{x} \otimes 1,
\end{array}
\end{equation}
but (of course) no longer left-invariant:
\begin{equation}
\begin{array}{rcl}
\AD(\hat{x}) & = & (S^{-1} x^{(2)'})_{(1)} \otimes (S^{-1} x^{(2)'})_{(1)}
x^{(1)}\\
            & = & S^{-1} x^{(2)'} \otimes \widehat{x^{(1)}}.
\end{array}
\end{equation}
We define $\hat{\U}$ to be the space $\{\hat{x} | x \in \U \}$.
It turns out that
the $\widehat{\x}$-operation is a projection operator from
$\A\cross\U$ to $\hat{\U}$; we will call it the right projector.
Three explicit expressions for such
right-invariant vector fields can be quickly derived:
\begin{equation}
\begin{array}{rcl}
\hat{x} & = & f^{i} (S^{-1}(e_{i}) \ad x)\\
        & = & f_{(3)}^{k} S^{-1} f_{(1)}^{k} <x, f_{(2)}^{k}> e_{k}
\end{array}
\end{equation}
and, for $\Upsilon _{b} = <\Upsilon ,b \otimes i\!d>$ with $\Upsilon $ being a
pure braid element,
\begin{equation}
\widehat{\Upsilon _{b}} = S^{-1}(b_{(3)}) b_{(1)} \Upsilon _{b_{(2)}}.
\end{equation}
Left- and right-invariant vector fields commute:
\begin{equation}
\begin{array}{rcl}
y \hat{x} & = & \hat{x}^{(1)} <y_{(1)},\hat{x}^{(2)'}> y_{(2)}\\
        & = & \hat{x} <y_{(1)}, 1> y_{(2)}\\
        & = & \hat{x} y.
\end{array}
\end{equation}
The right projector is an antimultiplicative operation:
\begin{equation}
\begin{array}{rcl}
\widehat{x y} & = & S^{-1}\left( (xy)^{(2)'}\right) (xy)^{(1)}\\
        & = & S^{-1} y^{(2)'} S^{-1} x^{(2)'} x^{(1)} y^{(1)}\\
        & = & S^{-1} y^{(2)'} \underbrace{\hat{x} y^{(1)}}_{\mbox{commute}}\\
        & = & S^{-1} y^{(2)'} y^{(1)} \hat{x}\\
        & = & \hat{y} \hat{x}.
\end{array}
\end{equation}
The right invariant vector fields form a Hopf algebra with the same
coproduct as \U\ because of (\ref{infx}), but opposite antipode and
multiplication:
\begin{equation}
\epsilon (\hat{x}) = x,\z \Delta \hat{x} = \widehat{x_{(1)}} \otimes
\widehat{x_{(2)}},\z
S(\hat{x}) = \widehat{S^{-1}x}.
\end{equation}
The Lie derivative of --- or the adjoint action on --- an element $\phi $ of
$\A\cross\U$ along a right invariant vector field comes out formally
equivalent to the left invariant version, when expressed in terms of
the new $\Delta $ and $S$:
\begin{equation}
\begin{array}{rcl}
\Li_{\hat{x}}(\phi ) & = & \hat{x} \ad \phi \\
        & = & \hat{x}_{(1)} \phi  S \hat{x}_{(2)}\\
        & = & \widehat{x_{(1)}} \phi  \widehat{S^{-1}x_{(2)}}.
\end{array}
\end{equation}
It immediately follows that \begin{equation}
\Li_{\hat{x}}(y) = 0,\z \mbox{for } y \in \U,
\end{equation}
in agreement with the geometrical picture.
Let us now compute the action of a right-invariant vector field on
a function $a$, using only the algebraic relations of the cross product
algebra and
the right vacuum:
\begin{equation}
\begin{array}{rcl}
\hat{x} a > & = & S^{-1}f^{i} a_{(1)} <e_{i} \ad x, a_{(2)}>\\
        & = & S^{-1}f^{i} a_{(1)} <e_{i} \otimes x, a_{(2)} S a_{(4)} \otimes
a_{(3)}>\\
        & = & a_{(4)} S^{-1}a_{(2)} a_{(1)} <x,a_{(3)}>\\
        & = & a_{(2)} <x , a_{(1)}>\\
        & = & <\hat{x} , a_{(1)} > a_{(2)},
\end{array}
\end{equation}
as expected from the geometrical considerations of the previous section.
The Hopf algebra $\hat{\U}$ mimics \U\ very closely. There is even
a canonical element $\widehat{\C}$ in $\A \otimes \hat{\U}$ that determines
{\em left}
coactions by conjugation:
\begin{eqnarray}
\AD(\hat{x}) & = & \widehat{\C} (1 \otimes \hat{x})
\widehat{\C^{-1}},\z\hat{x}\in \hat{\U},\\
\AD( a) & = & \widehat{\C} (1 \otimes a )
\widehat{\C^{-1}} \z a \in \A\\
        & = & a_{(1)} \otimes a_{(2)}.
\end{eqnarray}
By symmetry there is of course also a left projector $\check{ }$
\begin{equation}
\check{\phi } = S(\phi ^{(1)'}) \phi ^{(2)},
\end{equation}
that is most useful in the equality
\begin{equation}
x = x^{(2)'} \widehat{x^{(1)}}.
\end{equation}

\subsection{Applications}

Here is an example of  a typical manipulation using projectors
onto right-invariant vector fields:
$$\begin{array}{rcl}
x a & = & x a\\
\Leftrightarrow x^{(2)'}\widehat{x^{(1)}} a & = & x_{(1)}(a) x_{(2)}\\
\Leftrightarrow x^{(2)'}\widehat{x^{(1)}{}_{(1)}}(a)\widehat{x^{(1)}{}_{(2)}}
& = & x_{(1)}(a) x_{(2)}{}^{(2)'}\widehat{x_{(2)}{}^{(1)}}
\end{array} $$
Now use the $\A\cross\widehat{\U} \cong \A\otimes\widehat{\U}$
isomorphism, remove the ``$\widehat{\x}$'' over the second space and
switch spaces:
\begin{equation}
\begin{array}{rcl}
\Leftrightarrow x^{(1)}{}_{(2)}\otimes x^{(2)'}\widehat{x^{(1)}{}_{(1)}}(a) & =
&
x_{(2)}{}^{(1)}\otimes x_{(1)}(a) x_{(2)}{}^{(2)'}\\
\Leftrightarrow x^{(1)}{}_{(2)}\otimes x^{(2)'}<x^{(1)}{}_{(1)},a_{(1)}>
a_{(2)} & = &
x_{(2)}{}^{(1)}\otimes x_{(1)}(a) x_{(2)}{}^{(2)'}
\end{array}
\end{equation}
The expression that we have just derived is incidentally equivalent to
a proof that $\DA$ is a \A\cross\U-algebra homomorphism, only this
time we did not need to make any reference to linear infinite bases $\{e_{i}\}$
and $\{f^{i}\}$ of \U\ and \A, that do not necessarily exist. Let us now
complete the proof: Using the fact that $a\in\A$ was arbitrary, we take
it to be the second part of the coproduct of some other element
$b\in\A$ and multiply our expression by $b_{(1)}$ in the first space
\begin{equation}
\begin{array}{rcl}
\Leftrightarrow x^{(1)}{}_{(1)}(b_{(1)}) x^{(1)}{}_{(2)}
\otimes x^{(2)'} b_{(2)} & = &
b_{(1)} x_{(2)}{}^{(1)}\otimes x_{(1)}(b_{(2)}) x_{(2)}{}^{(2)'}\\
\Leftrightarrow \DA(x) \DA(b) & = & \DA(x_{(1)}(b) ) \DA(x_{(2)}) \,=\,
\DA(x a).\z\Box
\end{array}
\end{equation}
This example shows that the projections introduced in this section
are powerful tools in formal computations. The manipulations in the
given example were not quite as elegant as the corresponding
ones using the canonical element,
but the projectors are much more versatile tools and they do not require
the existence of linear countable infinite bases that were implicitly
assumed for the canonical element.

For further applications please see the covariance proofs in part II of
this thesis.

\chapter{A Quantum Mechanical Model}

\label{C:AQMM}
In this chapter we would like to illustrate at the example of a
simple toy model one possible way how quantum groups might find use in
physics. Quantum mechanics is a remarkably good theory as far
as experimental verification is concerned, so we will not attempt
to modify its most basic features as for instance the canonical
commutation relations. We instead want to focus on a generalization
of unitary transformations. These transformations form groups in
quantum mechanics; we will investigate --- at the example of time
evolution --- what happens if we generalize these transformations
to be elements of Hopf algebras. The introduction of deformed
{\em Poincare} symmetry in physics is expected to lead to similar
new phenomena.
We will in particular embed the
operator algebra of a simple quantum mechanical model
in a Hopf algebra with possibly non-trivial coproduct
and propose generalized time evolution equations.
We find that probability is conserved in this formulation
but pure states can evolve
into mixed ones (and vice versa); microscopic entropy is only conserved
for a special {\em stable} state.
The theory could
be interpreted as quantum mechanics for open systems.

\section*{Introduction}

There have been a number of proposals for a deformation of ordinary
quantum mechanical systems using quantum groups. In particular
systems with quantum group symmetries $e.g.$ \cite{PaSa,Ce}
and with deformed
canonical commutation relations $e.g.$ \cite{Ch,Z3} have been investigated in
some detail. Here we would like to focus on deformed time evolution
equations, i.e. deformations of the Heisenberg equations of motion
(Heisenberg picture)
and of the Liouville equation for the density operator (Schr\"odinger
picture).
It turns out to be fruitful to consider both pictures (H.p./ S.p.)
simultaneously.
Let us list some basic requirements on time evolution equations:
\begin{itemize}
\item The equations have to be linear.
\item Time evolution should be multiplicative.
\item Hermiticity must be preserved.
\item Probability must be preserved, i.e. the trace of the density
matrix must be constant.
\item Probabilities must be positive at all times.
\end{itemize}
All these requirements are fulfilled by unitary time evolution:
\begin{eqnarray}
X(t) & = & [U(t)]^{-1} X(0) U(t),\z \mbox{(H.p.)},\\
\rho (t) & = & U(t) \rho (0) [U(t)]^{-1},\z \mbox{(S.p.)},
\end{eqnarray}
with $[U(t)]^{+} = [U(t)]^{-1}$.
In this paper we would like to argue that the above equations
are not the only possible ones satisfying all the listed
requirements; in order to find more general equations we,
however, need to extend the operator algebra to a Hopf algebra.\\
Generalizations of unitary time evolution have been studied before
in the 70's in the context of completely positive maps and dynamical
semi-groups. Lindblad \cite{Li} found the general form for generators
of such semi-groups, however, without being able to give a {\em cause}
for the modified time evolution equation because he does not make any
reference to an underlying structure --- like Hopf algebras or
non-commutative geometry in our case.

\section{Schr\"odinger Picture}

Let us briefly review density matrices in ``classical'' quantum mechanics:
All observables are described by operators $X$ constant in time,
states are given as time dependent density matrices $\rho (t)$.
Expectation values are calculated as usual via
\begin{equation}
<X>_{\rho (t)} = tr(X \rho (t)),
\end{equation}
where the trace is cyclic ($tr(x y) = tr(y x)$).
The eigenvalues of the density matrix are the probabilities of
the pure components of the mixed state. In a diagonal basis
\begin{equation}
\rho  = \sum _{i} p_{i} |i><i|,\z 0 \leq p_{i} \leq 1.
\end{equation}
The sum of the eigenvalues
of the density matrix must hence be one
\begin{equation}
tr(\rho ) = \sum _{i} p_{i} = 1,\z\mbox{\it(normalization)}
\end{equation}
independent of time. Note that
\begin{equation}
tr(\rho ^{2}) = \sum _{i} p_{i}^{2} \leq 1;
\end{equation}
the equality is only satisfied for a pure ($\rho  = |\psi ><\psi |$) state. All
mixed states have $tr(\rho ^{2}) < 1$.
Unitary time evolution not only preserves the trace of $\rho $, {\em i.e.}
it conserves probability,
\begin{equation}
tr(U \rho  U^{-1}) = tr(\rho ) = 1
\end{equation}
but also conserves entropy:
\begin{equation}
tr\left( (U \rho  U^{-1}) (U \rho  U^{-1})\right) = tr(\rho ^{2}).
\end{equation}
It preserves hermiticity of $\rho $ because of $U^{\dagger } = U^{-1}$ and
is multiplicative: $U(t_{1} + t_{2}) = U(t_{1})\cdot U(t_{2}).$
Our task is now to find a generalized time
evolution for the density matrix with all those properties except
for the conservation of entropy. To satisfy {\em linearity} and
{\em multiplicativity} we choose time evolution to be realized through
the action (see chapter 1) of some new time evolution operator
$\ut$
\begin{equation}
\rho (t) = \ut \op \rho ,\z \ut = \ut(t).
\end{equation}
To leave freedom for deformations we ask $\ut$ to be an element
of a Hopf algebra \U\ (rather than a group) and propose the following left
action:
\begin{equation}
\fbox{$\rho (t) = \ut_{(2)} \rho S(\ut_{(1)})$} .
\end{equation}
Due to $S(\ut_{(1)}) \ut_{(2)} = \epsilon (\ut)$ and the cyclicity of the trace
this time evolution equation {\em conserves probability}
\begin{equation}
tr(\ut_{(2)} \rho  S(\ut_{(1)}) ) = tr(S(\ut_{(1)}) \ut_{(2)} \rho )
= tr(\rho )\cdot \epsilon (\ut),
\end{equation}
if we impose
\begin{equation}
\fbox{$\epsilon (\ut) = 1$}.
\end{equation}
In order to {\em conserve hermiticity} we have to impose
\begin{equation}
\fbox{$\ut^{\dagger } = S(\ut)$},
\end{equation}
because then
\begin{equation}
\begin{array}{rcl}
\left( \rho (t) \right)^{\dagger } & = & \left( \ut_{(2)} \rho
              S(\ut_{(1)})\right)^{\dagger }\\
        & = & S^{-1}(\ut^{\dagger }_{(1)}) \rho ^{\dagger } \ut^{\dagger
}_{(2)}\\
        & = & S^{-1}(\ut^{\dagger }) \op \rho ^{\dagger }\\
        & = & \ut \op \rho ^{\dagger }.
\end{array}
\end{equation}
Entropy is however no longer necessarily conserved:
\begin{equation}
tr\left( \rho (t)\cdot \rho (t) \right) \neq tr( \rho _{0} \cdot \rho
_{0}),\z\mbox{\it in
general}.
\end{equation}
{\bf Example:} ``Classical'' Quantum Mechanics is a special case with
\begin{equation}
\begin{array}{c}
\Delta (U) = U \otimes U,\z S(U) = U^{-1},\z \epsilon (U) = 1,\\
\rho (t)\, =\, U_{(2)} \rho  S(U_{(1)})\, =\, U \rho  U^{-1},\\
U^{\dagger }\, =\, S(U)\, =\, U^{-1}.
\end{array}
\end{equation}

\section{Heisenberg Picture}

Now we stick all the time evolution into the observables, leaving
the density matrix time invariant.
The time evolution equation for the operators
easily follows from the one for the density matrix using the
cyclic nature of the trace and the fact that the time evolution
of the expectation values should be independent of the particular
picture. We find:
\begin{equation}
\fbox{$X(t) =  X(0) \da \ut(t) \equiv  S(\ut_{(1)}) X(0) \ut_{(2)}$}.
\end{equation}
Two consistency requirements give the same conditions on $\ut$
\begin{eqnarray}
1(t) = 1 & \Rightarrow & \epsilon (\ut) = 1,\\
\left( X(t)\right)^{\dagger } = X^{\dagger } \da \ut & \Rightarrow &
\ut^{\dagger } = S(\ut),
\end{eqnarray}
as were already obtained in the previous section.

\section{Infinitesimal Transformation}

One great thing about working with Hopf algebras is that finite
and infinitesimal transformations are unified in the sense that they
have the exact same form. The infinitesimal version of our time
evolution equation must have the form
\begin{equation}
\frac {d \rho }{d t} = \frac {\Ht}{i \hbar } \op \rho  = \frac {1}{i \hbar }
\Ht_{(2)} \rho  S(\Ht_{(1)}),\label{inftrafo}
\end{equation}
where $i$ is purely conventional and we have inserted $\hbar $ to give
$\Ht$ units of energy.
The conditions on $\Ht$ are slightly different from the ones on
$\ut$:\footnote{The first condition may possibly be interpreted as requiring
a zero energy ground state.}
\begin{eqnarray}
\epsilon (\Ht) & = & 0\\
\Ht^{\dagger } & = & - S(\Ht).
\end{eqnarray}
How do we obtain the time evolution operator $\Ht$ from the (hermitian)
Hamiltonian $H$? Here is a {\bf Conjecture:}
\begin{equation}
\Ht = \frac {1}{2}\left(H - S^{-1}(H)\right),\z H^{\dagger } = H.
\end{equation}
(The $2$ might be a ``quantum-2''.) This choice for $\Ht$ will
automatically satisfy both conditions. Finite time translations
can be recovered by Taylor expansion
\begin{equation}
\rho (t)  =  \left.\sum _{n = 0}^{\infty } \frac {t^{n}}{n!}
\frac {d^{n} \rho }{d t^{n}}\right|_{t=0}\,
= \sum _{n = 0}^{\infty } \frac {t^{n}}{n!} \left(
        \left(\frac {\Ht}{i \hbar }\right)\op\,\right)^{n}\rho\,
        =  e^{\frac {\Ht}{i \hbar } t} \op \rho ,
\end{equation}
where we have used the multiplicative properties of successive actions.
Note that this is an ordinary exponential function, not a q-deformed
one.

In the following section we will study a system with a finite number
of eigenstates. In this case equation (\ref{inftrafo}) can be converted
into a matrix equation by taking the inner product with a matrix $A \in
M_{n}(\A)$ as follows:
\begin{equation}
\begin{array}{rcl}
\displaystyle \frac {d <\rho ,A^{i}{}_{l}>}{d t}
& = &  \frac {1}{i \hbar } <\Ht_{(2)} \rho  S \Ht_{(1)}, A^{i}{}_{l}>\\
& = & \frac {1}{i \hbar }
        <\Ht , S(A^{k}{}_{l}) A^{i}{}_{j}><\rho  , A^{j}{}_{k}>
\end{array}
\end{equation}
or, in a short hand,
\begin{equation}
\frac {d \rho ^{(il)}}{d t} = \frac {1}{i \hbar } \Ht^{(il)}{}_{(jk)} \rho
^{(jk)}.
\end{equation}
This matrix equation can easily be exponentiated to give an
explicit solution
\begin{equation}
\rho ^{(il)}(t) = \exp\left( \frac {\Ht}{i \hbar } \right)^{(il)}{}_{(jk)} \rho
_{0}^{(jk)}
\end{equation}
for $\rho $. In practice one would now express $\rho $ in terms of eigenvectors
of $\Ht^{(il)}{}_{(jk)}$ so that the matrix exponential diagonalizes with
the exponentials of $\frac {1}{i \hbar }$ times $\Ht$'s eigenvalues along its
diagonal.

\section{A Simple 2-Level System}

Consider a single particle in a double well potential
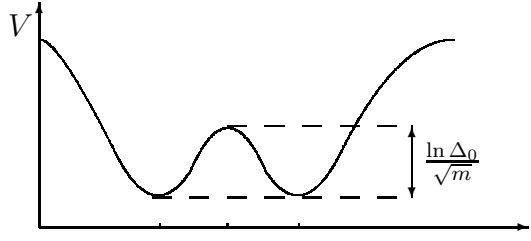
\begin{figure}
\unitlength=0.50mm
\special{em:linewidth 0.4pt}
\linethickness{0.4pt}
\begin{picture}(223.00,62.00)
\put(93.00,2.00){\vector(0,1){60.00}}
\put(93.00,2.00){\vector(1,0){130.00}}
\bezier{200}(133.00,17.00)(143.00,40.00)(153.00,17.00)
\bezier{160}(133.00,17.00)(123.00,2.00)(113.00,22.00)
\bezier{160}(153.00,17.00)(163.00,2.00)(173.00,22.00)
\bezier{156}(113.00,22.00)(98.00,52.00)(93.00,52.00)
\bezier{196}(173.00,22.00)(188.00,52.00)(203.00,52.00)
\put(143.00,29.00){\line(1,0){5.00}}
\put(153.00,29.00){\line(1,0){5.00}}
\put(163.00,29.00){\line(1,0){5.00}}
\put(173.00,29.00){\line(1,0){5.00}}
\put(183.00,29.00){\line(1,0){5.00}}
\put(123.00,10.00){\line(1,0){5.00}}
\put(133.00,10.00){\line(1,0){5.00}}
\put(143.00,10.00){\line(1,0){5.00}}
\put(153.00,10.00){\line(1,0){5.00}}
\put(163.00,10.00){\line(1,0){5.00}}
\put(173.00,10.00){\line(1,0){5.00}}
\put(183.00,10.00){\line(1,0){5.00}}
\put(192.00,19.00){\vector(0,-1){9.00}}
\put(192.00,19.00){\vector(0,1){10.00}}
\put(125.00,2.00){\line(0,1){1.00}}
\put(162.00,2.00){\line(0,1){1.00}}
\put(143.00,2.00){\line(0,1){1.00}}
\put(195.00,19.00){\makebox(0,0)[lc]{$\frac{\ln \Delta_0}{\sqrt{m}}$}}
\put(88.00,56.00){\makebox(0,0)[cc]{$V$}}
\end{picture}

\caption{Double Well} \label{doubwell}
\end{figure}
(Fig.~\ref{doubwell}) with a barrier
of height $\sim \frac {\ln \Delta _{0}}{\sqrt {m}}$. If we are only interested
in which dip
the particle is localized then we are dealing with a 2-level system.
A phenomenological hamiltonian that describes tunneling through
the barrier is easily written down in terms of the $x$-Pauli matrix:
\begin{equation}
H = \Delta _{0} \sigma _{x} = \Delta _{0}(\sigma _{+} + \sigma
_{-}).\z\mbox{\em (tunneling only)}
\end{equation}
Instead of viewing the Pauli matrices as the fundamental representation
of $su(2)$ we would like to consider $su_{q}(2)$ with $q \in (0,1]$
as given in (\ref{HXX}).
All irreducible representations of $su_{q}(2)$, e.g.
\begin{equation}
\begin{array}{ll}
\mbox{2-dim:}& \sigma _{z} = \left(\begin{array}{cc}1 & 0\\0 & -1
\end{array}\right),\z \sigma _{+}=\left(\begin{array}{cc}0 & 1\\0 & 0
\end{array}\right),\z \sigma _{-}=\left(\begin{array}{cc}0 & 0\\1 & 0
\end{array}\right),\\[9mm]
\mbox{3-dim:}& J_{z} =  \left(\begin{array}{ccc}1&0&0\\0&0&0\\0&0&-1
\end{array}\right),\x
J_{+}=\left(\begin{array}{ccc}0&\sqrt {2}&0\\0&0&\sqrt {2}\\0&0&0
\end{array}\right),\x
J_{-}=\left(\begin{array}{ccc}0&0&0\\\sqrt {2}&0&0\\0&\sqrt {2}&0
\end{array}\right),\\
\vdots &
\end{array}
\end{equation}
are undeformed. This makes it easy to derive a matrix representation
of the time evolution operator:
\begin{equation}
\Ht = \frac {1}{2}(H - S^{-1}(H) ) = \frac {1}{2}(\sigma _{+} + \sigma _{-}
+q^{-1}\sigma _{+} +q\sigma _{-})\Delta _{0} \propto \left(
\begin{array}{cc}0&1\\q&0\end{array}\right).
\end{equation}
We will ignore the proportionality constant because it can always be
incorporated in $\Delta _{0}$. The time evolution equation in matrix form
is
\begin{equation}
\frac {d \rho }{d t} = \frac {1}{i \hbar }
\left(\left(\begin{array}{cc}0&1\\q&0\end{array}\right) \rho
\left(\begin{array}{cc}q^{\frac {1}{2}}&0\\0&q^{-\frac
{1}{2}}\end{array}\right) -
\left(\begin{array}{cc}q^{\frac {1}{2}}&0\\0&q^{-\frac
{1}{2}}\end{array}\right) \rho
\left(\begin{array}{cc}0&q\\1&0\end{array}\right) \right),
\end{equation}
which reduces to the correct classical limit
\begin{equation}
\frac {d \rho }{d t} = \frac {1}{i \hbar } \left[\sigma _{x} , \rho
\right],\z\mbox{\em (classical)}
\end{equation}
as $q \rightarrow 1$.
Plugging the hamiltonian $H$ into the matrix time evolution equation
in the Heisenberg picture,
\begin{equation}
\frac {d X}{d t} = \frac {1}{i \hbar }
\left(\left(\begin{array}{cc}0&q\\1&0\end{array}\right) X
\left(\begin{array}{cc}q^{\frac {1}{2}}&0\\0&q^{-\frac
{1}{2}}\end{array}\right) -
\left(\begin{array}{cc}q^{\frac {1}{2}}&0\\0&q^{-\frac
{1}{2}}\end{array}\right) X
\left(\begin{array}{cc}0&1\\q&0\end{array}\right) \right),
\end{equation}
gives incidentally
\begin{equation}
\frac {d H}{d t} = 0,
\end{equation}
{\em i.e.} energy {\em is} conserved in our toy model.

\subsection{Time Evolution and Mixing}

It is instructive to look at an actual computation
of the evolution of a system that is
in an eigenstate $|+>$ of $\sigma _{z}$ at $t = 0$; the corresponding
density matrix
\begin{equation}
\rho _{0} = \left(\begin{array}{cc}1&0\\0&0\end{array}\right)\z\mbox{\it
(initial pure state)}
\end{equation}
is that of a pure state ($tr(\rho ) = tr(\rho ^{2}) = 1)$. Interesting are the
eigenvalues $p_{1}, p_{2}$ of $\rho (t)$ as a function of time.
\begin{figure}
        \begin{minipage}{75mm} 
\unitlength=0.50mm
\special{em:linewidth 0.4pt}
\linethickness{0.4pt}
\begin{picture}(140.00,115.00)
\put(5.00,5.00){\vector(1,0){130.00}}
\put(10.00,0.00){\vector(0,1){110.00}}
\put(50.00,5.00){\line(0,1){2.00}}
\put(90.00,5.00){\line(0,1){2.00}}
\put(130.00,5.00){\line(0,1){2.00}}
\put(10.00,105.00){\line(1,0){2.00}}
\put(10.00,55.00){\line(1,0){2.00}}
\put(50.00,4.00){\makebox(0,0)[ct]{2}}
\put(90.00,4.00){\makebox(0,0)[ct]{4}}
\put(130.00,4.00){\makebox(0,0)[ct]{6}}
\put(9.00,55.00){\makebox(0,0)[rc]{$0.5$}}
\put(9.00,105.00){\makebox(0,0)[rc]{1}}
\put(140.00,5.00){\makebox(0,0)[cc]{$t$}}
\put(10.00,115.00){\makebox(0,0)[cc]{$p_{1,2}$}}
\put(10.00,105.00){\line(1,0){120.00}}
\put(124.00,99.00){\vector(-1,1){4.00}}
\put(127.00,96.00){\makebox(0,0)[cc]{$p_1$}}
\put(124.00,11.00){\vector(-1,-1){4.00}}
\put(127.00,14.00){\makebox(0,0)[cc]{$p_2$}}
\end{picture}

        \caption{q = 1}     \label{qeqone} \end{minipage}
        \begin{minipage}{75mm} 
\unitlength=0.50mm
\special{em:linewidth 0.4pt}
\linethickness{0.4pt}
\begin{picture}(140.00,115.00)
\bezier{136}(10.00,5.00)(20.00,5.00)(34.00,24.00)
\bezier{112}(34.00,24.00)(47.00,45.00)(50.00,45.00)
\bezier{112}(50.00,45.00)(53.00,45.00)(66.00,24.00)
\bezier{136}(66.00,24.00)(80.00,5.00)(90.00,5.00)
\bezier{136}(90.00,5.00)(100.00,5.00)(114.00,24.00)
\bezier{112}(114.00,24.00)(127.00,45.00)(130.00,45.00)
\bezier{136}(10.00,105.00)(20.00,105.00)(34.00,86.00)
\bezier{112}(34.00,86.00)(47.00,65.00)(50.00,65.00)
\bezier{112}(50.00,65.00)(53.00,65.00)(66.00,86.00)
\bezier{136}(66.00,86.00)(80.00,105.00)(90.00,105.00)
\bezier{136}(90.00,105.00)(100.00,105.00)(114.00,86.00)
\bezier{112}(114.00,86.00)(127.00,65.00)(130.00,65.00)
\put(5.00,5.00){\vector(1,0){130.00}}
\put(10.00,0.00){\vector(0,1){110.00}}
\put(50.00,5.00){\line(0,1){2.00}}
\put(90.00,5.00){\line(0,1){2.00}}
\put(130.00,5.00){\line(0,1){2.00}}
\put(10.00,105.00){\line(1,0){2.00}}
\put(10.00,55.00){\line(1,0){2.00}}
\put(50.00,4.00){\makebox(0,0)[ct]{2}}
\put(90.00,4.00){\makebox(0,0)[ct]{4}}
\put(130.00,4.00){\makebox(0,0)[ct]{6}}
\put(9.00,55.00){\makebox(0,0)[rc]{$0.5$}}
\put(9.00,105.00){\makebox(0,0)[rc]{1}}
\put(140.00,5.00){\makebox(0,0)[cc]{$t$}}
\put(10.00,115.00){\makebox(0,0)[cc]{$p_{1,2}$}}
\end{picture}

        \caption{$q^{2} = 0.7$} \label{qeqhalf} \end{minipage}
\end{figure}
They are the
probabilities of the respective pure states in the mixture.
For $q = 1$ (Fig.~\ref{qeqone}) nothing much happens,
but for e.g. $q = \sqrt {0.7} \approx 0.5$ (Fig.~\ref{qeqhalf})
the system oscillates between a pure and a partially mixed state.
A behavior like that does not appear in ordinary quantum case and
opens up interesting possibilities for, say in the present case,
a phenomenological quantum mechanical description of just one
part of a coupled system. Here we do not want to plunge too deep
into possible interpretations but would just like to point
out some new phenomena that appear when laws of physics are deformed.
Just out of curiousity let us find the $q$ for which the system
becomes totally mixed. Plotting $p_{1,2}$ against $q$ (Fig.~\ref{taptwo})
at fixed time
\begin{figure}
        \begin{minipage}{75mm} 
\unitlength=0.50mm
\special{em:linewidth 0.4pt}
\linethickness{0.4pt}
\begin{picture}(120.00,115.00)
\put(10.00,0.00){\vector(0,1){110.00}}
\put(10.00,105.00){\line(1,0){2.00}}
\put(10.00,55.00){\line(1,0){2.00}}
\put(9.00,55.00){\makebox(0,0)[rc]{$0.5$}}
\put(9.00,105.00){\makebox(0,0)[rc]{1}}
\put(10.00,115.00){\makebox(0,0)[cc]{$p_{1,2}$}}
\put(5.00,5.00){\vector(1,0){110.00}}
\put(60.00,5.00){\line(0,1){2.00}}
\put(110.00,5.00){\line(0,1){2.00}}
\put(120.00,5.00){\makebox(0,0)[cc]{$q$}}
\put(60.00,4.00){\makebox(0,0)[ct]{$0.5$}}
\put(110.00,4.00){\makebox(0,0)[ct]{1}}
\bezier{332}(10.00,5.00)(30.00,5.00)(68.00,55.00)
\bezier{332}(10.00,105.00)(30.00,105.00)(68.00,55.00)
\bezier{264}(68.00,55.00)(98.00,15.00)(110.00,5.00)
\bezier{264}(68.00,55.00)(98.00,95.00)(110.00,105.00)
\end{picture}

        \caption{$t \approx 1.9$}     \label{taptwo} \end{minipage}
        \begin{minipage}{75mm} 
\unitlength=0.50mm
\special{em:linewidth 0.4pt}
\linethickness{0.4pt}
\begin{picture}(140.00,115.00)
\put(5.00,5.00){\vector(1,0){130.00}}
\put(10.00,0.00){\vector(0,1){110.00}}
\put(50.00,5.00){\line(0,1){2.00}}
\put(90.00,5.00){\line(0,1){2.00}}
\put(130.00,5.00){\line(0,1){2.00}}
\put(10.00,105.00){\line(1,0){2.00}}
\put(10.00,55.00){\line(1,0){2.00}}
\put(50.00,4.00){\makebox(0,0)[ct]{2}}
\put(90.00,4.00){\makebox(0,0)[ct]{4}}
\put(130.00,4.00){\makebox(0,0)[ct]{6}}
\put(9.00,55.00){\makebox(0,0)[rc]{$0.5$}}
\put(9.00,105.00){\makebox(0,0)[rc]{1}}
\put(140.00,5.00){\makebox(0,0)[cc]{$t$}}
\put(10.00,115.00){\makebox(0,0)[cc]{$p_{1,2}$}}
\bezier{292}(10.00,5.00)(30.00,5.00)(49.00,55.00)
\bezier{292}(49.00,55.00)(68.00,105.00)(88.00,105.00)
\bezier{292}(10.00,105.00)(30.00,105.00)(49.00,55.00)
\bezier{292}(49.00,55.00)(68.00,5.00)(88.00,5.00)
\bezier{336}(88.00,5.00)(109.00,6.00)(130.00,65.00)
\bezier{340}(88.00,105.00)(109.00,105.00)(130.00,45.00)
\end{picture}

        \caption{$q^{2} = 1/3$} \label{qcrit} \end{minipage}
\end{figure}
$t \approx 1.9$\footnote{This value was found by iteration.}
we find $q_{critical} = \sqrt {1/3}$; see also Fig.~\ref{qcrit}.
The significance
of this number is unknown.

\subsection{Stable State}

An interesting question is whether there exists a stable (mixed) state
that is invariant under the deformed time evolution. This is indeed
the case and has to do with the square of the antipode: The square of
the antipode is an inner automorphism in $U_{q}(su(2))$ implemented
by elements $u$ and $v = S(u)$ via conjugation \cite{Df}
\begin{equation}
S^{2}(x) = u x u^{-1} = v^{-1} x v,\z \forall x \in U_{q}(su(2)).
\end{equation}
Let us try $v$ as a density operator:
\begin{equation}
\begin{array}{rcl}
v(t) & := & \ut_{(2)} v S(\ut_{(1)})\\
        & = & v S^{2}(\ut_{(2)}) S(\ut_{(1)})\\
        & = & v \epsilon (\ut)\\
        & = & v.
\end{array}
\end{equation}
Thus $v$ has the desired properties. Its 2-dimensional matrix
representation
\begin{equation}
v = \frac {1}{1+q^{2}} \left(\begin{array}{cc}q^{2}&0\\0&1\end{array}\right)
\end{equation}
looks like a thermal state for a hamiltonian with dominant part
proportional to $\sigma _{z}$, {\em i.e.}
\addtocounter{footnote}{1}
\footnotetext{Or: Time average of $\Delta _{0}
\approx 0$.} \addtocounter{footnote}{-1}
\begin{equation}
H = \Delta _{1} \sigma _{z} + \Delta _{0} \sigma _{x},\z \Delta _{1} \gg \Delta
_{0}\footnotemark,
\end{equation}
and suggests
\begin{equation}
q = \exp(-\Delta _{1}/k T),\z q\in (0,1].
\end{equation}
Higher matrix representations of $v$ give additional support
for this hypothesis:
\begin{equation}
\mbox{3-dim:} \z v \propto \left(\begin{array}{ccc}q^{4}&0&0\\0&q^{2}&0\\0&0&1
\end{array}\right),\z J_{z} = \left(\begin{array}{ccc}1&0&0\\0&0&0\\0&0&-1
\end{array}\right);\z\mbox{\it e.t.c}.
\end{equation}

\part{Differential Geometry on Quantum Spaces}

%
%
%
%
%
\chapter{Quantum Spaces}

\section{Quantum Planes}

A classical plane can be fully described by the commutative algebra of
(coordinate) functions over it. This algebra is typically covariant
under the action of some symmetry group, and derivatives on it satisfy
an undeformed product rule.
A quantum plane in contrast to this is
covariant under a quantum group whose non-commutative algebra of
functions \A\  also forces the
algebra of functions on the q-plane \fum\ to seize to commute.
The transformations of \fum\
and of the dual algebra of quantum derivatives $\tqm$ is most
easily described in terms of \A-coactions on coordinate functions
and partial derivatives
\begin{eqnarray}
\DA x^{i} & = & x^{j} \otimes S t^{i}{}_{j},\\
\DA \partial _{i} & = & \partial _{j} \otimes S^{2} t^{j}{}_{i},
\end{eqnarray}
which we sometimes write in short matrix form as
\begin{eqnarray}
x & \to  & t^{-1}\cdot  x,\\
\partial  & \to  & \partial \cdot S^{2}t.
\end{eqnarray}
{\em Remark:}\x The ``$S$'' was inserted here to make these transformations
{\em right} coactions, the $S^{2}$ is needed for covariance (see below).\\
{\em Remark:} One can use $t^j{}_i$ in place of $S t^i{}_j$. Then $x \to x
\cdot t$ and
$\partial \to St \cdot \partial$. The choice is purely conventional.

\subsection{Product Rule for Quantum Planes}

Having made the ring of functions non-commutative,
we must now also modify the product rule in order to
retain covariant equations.
We make the following ansatz (see \cite{WZ})
\begin{equation}
\partial _{i} x^{k} = \partial _{i}(x^{k}) + L_{i}{}^{j}(x^{k}) \partial _{j},
\label{prodrule}
\end{equation}
where $\partial _{i}(x^{k}) = \delta^k_i$ and
$L_{i}{}^{j}$ is a linear operator that describes the braiding of $\partial
_{i}$
as it moves through $x^{k}$.
In place of the coordinate function
$x^{k}$ one could write any other
function in \fum\ and in particular (formal) power series in the
coordinate functions. When we consider products of coordinate
functions we immediately see that $L$ satisfies
\begin{equation}
L_{i}{}^{j}(x y) = L_{i}{}^{l}(x) L_{l}{}^{j}(y),\z L_{i}{}^{j}(1) = \partial
_{i}^{j},
\end{equation}
which can be reinterpreted in Hopf algebra language as
$\Delta  L = L \dot{\otimes} L$ and $\epsilon (L) = I$; $S L = L^{-1}$ follows
naturally.
We are hence let to believe  that  $L$ should belong to  some Hopf algebra,
the Braiding Hopf Algebra.
In the case of linear quantum groups $L$ is for instance
an element of the quasitriangular
Hopf algebra $\U$ of the quantum symmetry group.
Considering multiple derivatives gives additional conditions that
can be summarized by requiring that
\begin{equation}
\UD \partial _{i} = L_{i}{}^{j} \otimes \partial _{j}
\end{equation}
be a Hopf algebra coaction, {\em i.e.}
\begin{equation}
\UD(\partial  \partial ') = \UD(\partial ) \UD(\partial '),\x
(i\!d \otimes \UD) \UD
= (\Delta  \otimes i\!d) \UD,\x (\epsilon  \otimes i\!d) \UD = i\!d.
\end{equation}
For arbitrary functions $f$ and derivatives $\partial $ we
find a generalized product rule
\begin{equation}
\fbox{$\partial  f = \partial (f) + \partial _{1'}(f) \partial _{2}$},
\end{equation}
where ${}_{\cal U}\Delta  \partial
\equiv  \partial _{1'} \otimes \partial _{2}$. Covariance of the product rule
(\ref{prodrule}) under coactions is expected to give
strong conditions on $L_{i}{}^{j}$.

{\em Remark:} The formula for the product rule (\ref{prodrule}) was
inspired by the form of the multiplication of two elements $\xi ,\phi $ in
the cross product algebra \A\cross\U
\begin{equation}
\xi  \phi  = \phi ^{(1)} <\xi _{1'},\phi ^{(2)'}> \xi _{2},
\end{equation}
where $\DA(\phi ) = \phi ^{(1)} \otimes \phi ^{(2)'}$  and $\UD(\xi ) = \xi
_{1'} \otimes \xi _{2}$
(see chapter~\ref{C:BC}).

\subsection[Covariance]{Covariance of: $\partial _{i} f = \partial _{i}(f) +
L_{i}{}^{j}(f) \partial _{j}$}
\label{covdfl}

We need to use an {\bf inductive} approach: We start by requiring that
\begin{equation}
\DA(\partial _{i}(x^{j})) = \DA \partial _{i}(\DA x^{j}).\z\mbox{\it (anchor)}
\end{equation}
This is in fact satisfied, because we
already have $\DA x^{j} = x^{l} \otimes S t^{j}{}_{l}$ and
iff $\DA \partial _{j} = \partial _{l} \otimes
S^{2} t^{l}{}_{j}$ then:
$\DA \partial _{i}(\DA x^{j})
= \partial _{k}(x^{l}) \otimes S^{2} t^{k}{}_{i} S t^{j}{}_{l}
= \delta _{k}^{l} \otimes S^{2} t^{k}{}_{i}
S t^{j}{}_{l} = \delta _{i}^{j} \otimes 1$, in agreement
with $\DA(\partial _{i}(x^{j})) = \DA(\delta _{i}^{j}) =
\delta _{i}^{j} \otimes 1$. That was the anchor;
now the induction to higher powers
in the coordinate functions:
Assume that the action of $\partial_i$ on $f$ is covariant:
\begin{equation}
\DA(\partial _{i}(f)) = \DA \partial _{i}(\DA f),\label{dapif}
\end{equation}
where $f$ is a function of the coordinate functions $x^{i}$.
Try to proof covariance of the $\partial_i$--$f$ commutation relation, \ie\
\begin{equation}
(\protect\ref{dapif})\:\stackrel{?}{\Rightarrow}\:\DA(\partial _{i} f)
= \DA(\partial _{i})\cdot \DA(f).\z\mbox{\it
(induction)}
\end{equation}
After some computation we find
\begin{equation}
\DA(L_i{}^j(f))(1 \otimes S^2t^k{}_j) \stackrel{!}{=}
L_l{}^k(f^{(1)}) \otimes S^2t^l{}_i f^{(2)'},\label{first}
\end{equation}
where $\DA(f) \equiv f^{(1)} \otimes f^{(2)'}$. This simplifies further if we
know how
$L_i{}^j$ acts on $f$. If the braiding Hopf algebra acts like the covariance
quantum group, then
$L_i{}^j(f) = f^{(1)}<L_i{}^j,f^{(2)'}>$, $L_i{}^j \in \A^*$ and
(\ref{first}) becomes
\begin{equation}
\left( L_{i}{}^{j}(f^{(2)'}) S^{2}t^{k}{}_{j}
- S^{2}t^{l}{}_{i} \widehat{L_{l}{}^{k}}(f^{(2)'}) \right)
\otimes f^{(1)} = 0,
\end{equation}
where $\widehat{\x}:\smash \to \smash$ is the projector onto right-invariant
vector fields: $\widehat{x} = S^{-1}(x^{(2)'}) x^{(1)}$
with $\DA(x) \equiv x^{(1)} \otimes x^{(2)}$, such that
$\widehat{L_l{}^k}(f^{(2)'}) =$ $<L_l{}^k,f^{(2)'}> f^{(3)'}$.
This is satisfied if
\begin{equation}
L_{i}{}^{j}(a) S^{2}t^{k}{}_{j}
= S^{2}t^{l}{}_{i} \widehat{L_{l}{}^{k}}(a),\z\forall a \in \A.
\end{equation}
(The reverse is true only if \A\ is generated by  $[S t^{i}{}_{j}]$ --- or
$[t^i{}_j]$, if the choose the
convention $\DA(x) = x \cdot t$.)
In the case where the braiding Hopf algebra is quasitriangular, there
are (exactly) two natural choices
\begin{equation}
L_{j}{}^{i} \propto \left\{\begin{array}{l} S^{-1} L^{-}{}^{i}{}_{j}
\equiv  <\R,S^{2}t^{i}{}_{j} \otimes i\!d>\\
S^{-1} L^{+}{}^{i}{}_{j} \equiv  <\R,i\!d \otimes S
t^{i}{}_{j}>\end{array}\right.
\end{equation}
that satisfy the above equation and all other requirements
(coproduct, {\em e.t.c.}).

For the Wess-Zumino quantum plane \cite{WZ} the action of $L$ on the
coordinate functions is linear and of first degree in those functions,
so we can use the coaction $\DA$ to express it:
\begin{equation}
L_{i}{}^{j}(x^{k}) = <L_{i}{}^{j},S t^{k}{}_{l}> x^{l} \propto
\left\{\begin{array}{l} r^{kj}{}_{li} x^{l}\\
(r^{-1})^{jk}{}_{il} x^{l}\end{array}\right.\label{lplane}
\end{equation}
in perfect agreement with \cite{WZ}.
(The overall multiplicative constant ($\frac{1}{q}$) is not fixed by covariance
considerations but is given by the characteristic equation of $\hat{r}$
and the requirement that $\widehat{C}^{kj}{}_{li} \equiv <L_i{}^k,St^j{}_l>$
should have an eigenvalue $-1$.)

\section{Quantum Groups}

A quantum group is a quantum plane covariant under itself. However, it has more
structure and the coactions $\DA$ and $\UD$ are now completely
determined by the multiplication in \U\ and \A: Let $\phi  \in \A\cross\U$
and $\DA \phi  \equiv  \phi ^{(1)} \otimes \phi ^{(2)'}$; then
\begin{equation}
        \Lix{{}}(\phi ) = \chi  \ad \phi  \equiv  \chi _{(1)} \phi  S \chi
_{(2) }
        = \phi ^{(1)} <\chi ,\phi ^{(2)'}>,\z\forall \chi  \in \U
\end{equation}
determines $\DA$. The coaction $\UD$ is simply the coproduct
$\Delta :\U \to  \U \otimes \U$, so that the product rule becomes
\begin{equation}
x a = a_{(1)} <x_{(1)},a_{(2)}> x_{(2)},
\end{equation}
where $x \in \U,\: a \in \A$. This defines the multiplicative structure
in the so called cross product algebra \cite{SWZ3} $\smash$.
Interestingly, equation (\ref{lplane}) does not apply in the
case of a quantum
group: In that case $t$ is replaced by the
adjoint representation $T$ and $L$ becomes $O$, a part in the coproduct
of the basic generators.
Not all elements of $T$ are linearly
independent. There is a trivial partial sum $T^{(ii)}{}_{(kl)} = 1 \delta
_{(kl)}$;
the same sum for $O$, $O^{(ii)}{}_{(kl)} =: Y_{(kl)}$, is in general
non-trivial
thus leading to a contradiction. An explanation for
this is that quantum groups have more structure than quantum planes.
They already contain an intrinsic braiding and do not leave any freedom
for external input such as $\R$ in equation (\ref{lplane}); the product rule is
in fact automatically
covariant by the construction of the cross product algebra. There are, however,
some indications that $O$ and $T$ might be related to a universal
$\tilde{\R}$ that lives in the sub-Hopf algebra of \A\, generated by the
elements of $T$.

{}From the discussion of the quantum planes we would like to keep the idea
of a finite number of so-called bicovariant generators $\chi _{i}$ that
close under adjoint action
        $\chi _{i} \ad \chi _{j} = \chi _{k} f_{i}{}^{k}{}_{j}$
and span an invariant subspace of \U, $i.e.$
        $\DA \chi _{j} = \chi _{k} \otimes T^{k}{}_{j}$.
We  call quantum groups with such generators Quantum Lie
Algebras.
In following section we will give  more
precise definitions of quantum Lie algebras.

\chapter{Cartan Calculus}

\subsection{Cartan Identity}

The central idea behind Connes Universal Calculus \cite{Co1} in the context of
non-com\-mu\-ta\-tive geometry was to retain from the classical differential
geometry
the nilpotency of $\dl$
\begin{equation}
\dl^{2} = 0
\end{equation}
and the undeformed Leibniz rule
for $\dl$\footnote{We use parentheses
to delimit operations like \dl, $\I_{x}$ and $\Li_{x}$, {\em e.g.}
$\dl a = \dl(a)
+ a \dl$.  However, if the limit of the operation is clear from the context,
we will
suppress the parentheses, {\em e.g.} $\dl(\I_{x} \dl a) \equiv
\dl(\I_{x}(\dl(a)))$.}
\begin{equation}
\dl \alpha  = \dl(\alpha ) + (-1)^{p} \alpha  \dl    \label{LEIBNIZ}
\end{equation}
for any $p$-form $\alpha $. The exterior derivative $\dl$ is a scalar making
this equation hard to deform, except for a possible multiplicative
constant in the second term.
Here we want to base the construction of a differential calculus on
quantum groups on two additional classical formulas: to extend the definition
of a Lie derivative from functions and vector fields to forms
we postulate
\begin{equation}
\Li \circ \dl = \dl \circ \Li; \label{LIE}
\end{equation}
this is essential for a geometrical interpretation along the lines
of chapter~\ref{C:VoQG}. The second formula that we can --- somewhat
surprisingly ---
keep undeformed in the quantum case is originally due to Henri Cartan
\begin{equation}
\Lix{i} = \Ix{i} \dl + \dl \Ix{i},\z\mbox{\it (Cartan
Identity)}\label{CARTAN}
\end{equation}
where $\chi _{i}$ are the generators of some quantum Lie algebra.
The only possibility to deform this equation and not violate its
covariance is to introduce multiplicative deformation parameters
$\kappa ,\lambda $ for the two terms on the right hand side of (\ref{CARTAN})
such
that now $\Lix{i} = \kappa  \Ix{i} \dl + \lambda  \dl \Ix{i}$. For a function
$a \in \A$ that gives
$$\Lix{i}(a) = \kappa  \Ix{i}(\dl a)$$
($\Ix{i}$ vanishes
on functions), for $\dl a$ we find
$$\Lix{i}(\dl a) = \lambda  \dl(\Ix{i}(\dl a))$$
and finally together
$$\Lix{i}(\dl a) = \frac {\lambda }{\kappa } \dl(\Lix{i}(a)),$$
in contrast to (\ref{LIE}) unless $\frac {\lambda }{\kappa } = 1$, in which
case we can easily
absorb either $\kappa $ or $\lambda $ into $\Ix{{}}$. Being now (hopefully)
convinced
of our two basic equations (\ref{LIE}) and (\ref{CARTAN}) we want
to turn to the generators $\chi _{i}$ next.\\
Several discussions with P. Aschieri helped clarifying the relation
between the material presented in the next section and Woronowicz's
theory.

\section{Quantum Lie Algebras}

A quantum Lie algebra is a Hopf algebra \U\ with a finite-dimensional
biinvariant sub vector space $\tq$ spanned by generators $\{\chi _{i}\}$
with coproduct
\begin{equation}
\Delta  \chi _{i} = \chi _{i} \otimes 1 + O_{i}{}^{j} \otimes \chi _{j}.
\end{equation}
More precisely we will call this a quantum Lie algebra of {\bf type II}.
Let $\{\omega ^{j} \in \tq^{*}\}$ be a dual basis of 1-forms corresponding to a
set of
functions $b^{j} \in \A$ via $\omega ^{j} \equiv  S b^{j}_{(1)} \dl
b^{j}_{(2)}$; {\it i.e.}
\begin{eqnarray}
\AD(\chi _{i}) &=& 1\otimes \chi _{i}, \nonumber \\
\DA(\chi _{i}) &=& \chi _{j} \otimes T^{j}{}_{i},\x T^{j}{}_{i}\in\fun,\\
\I_{\chi _{i}}(\omega ^{j}) &  = & -<\chi _{i},S b^{j}> = \delta
^{j}_{i},\label{ebdual}\\
\AD(\omega ^{i}) &=& 1\otimes \omega ^{i},\\
\DA(\omega ^{i}) &=& \omega ^{j} \otimes S^{-1} T^{i}{}_{j}.
\end{eqnarray}
If the functions $b^{i}$ also close under adjoint coaction $\Delta ^{Ad}(b^{i})
= b^{j}
\otimes S^{-1} T^{i}{}_{j}$, we will call the corresponding quantum Lie algebra
one of {\bf type I}.
Getting a little ahead of ourself's let us mention that we can derive
an expression for the
exterior derivative of a function from the Cartan identity
(\ref{CARTAN}) in terms of these bases
\begin{equation}
\dl(a) = \omega ^{i}(\chi _{i} \tr a) = \omega ^{i} \Lix{i}(a) \label{exder}
\end{equation}
and that this leads to the following $f-\omega $ commutation relations
\cite{W2}
\begin{equation}
f \omega ^{i} = \omega ^{j} (O_{j}{}^{i} \tr f).
\end{equation}

\subsection{Generators, Metrics and the Pure Braid Group}

How does one practically go about finding the basis of generators $\{\chi
_{i}\}$
and the set of functions $\{b^{i}\}$ that define the basis of 1-forms
$\{\omega ^{i}\}$?
Here we would like to present a method that utilizes pure braid group
elements as introduced in the first part of this thesis.

Let us recall that a pure braid element $\Ups$ is an element of
$\U\hat{\otimes}\U$ that commutes with all coproducts of
elements of \U, $i.e.$
\begin{equation}
\Ups \Delta (y) = \Delta (y) \Ups,\z \forall y \in \U.
\end{equation}
$\Ups$ maps elements of \A\ to elements of \U\ with special
transformation properties under the right coaction:
\begin{equation}
\begin{array}{c}
\Ups:\A \to  \U\,:\x b \mapsto \Ups_{b} \equiv <\Ups,b \otimes
i\!d>;\\
\DA(\Ups_{b}) = \Ups_{b_{(2)}} \otimes S(b_{(1)}) b_{(3)} = <\Ups \otimes i\!d,
\tau ^{23}(\Delta ^{Ad}(b) \otimes i\!d)>.
\end{array}
\end{equation}
An element $\Ups$ of the pure braid group
defines furthermore a bilinear quadratic form on \A
\begin{equation}
(\x,\x): \A \otimes \A \to  k\,:\z a \otimes b \mapsto (a,b) = -<\Ups,a
\otimes S(b)> \in k,
\end{equation}
with respect to which we can construct orthonormal $(b_{i},b^{j}) = \delta
_{i}^{j}$
bases $\{b_{i}\}$ and $\{b^{j}\}$ of functions that in turn will define
generators $\chi _{i} := \Ups_{b_{i}}$  and 1-forms $\omega ^{j} :=
S(b^{j}_{(1)}) \dl b^{j}_{(2)}$.
Typically one can choose span$\{b_{i}\}$ = span$\{b^{j}\}$; then one
starts by constructing one set, say $\{b_{i}\}$, of functions
that close under adjoint coaction
\begin{equation}
\Delta ^{Ad} b_{i} = b_{j} \otimes T^{j}{}_{i}.
\end{equation}
If the numerical matrix
\begin{equation}
\fbox{$\eta _{ij} := -<\Ups,b_{i} \otimes Sb_{j}>$}\z\mbox{\it (metric)}
\end{equation}
is invertible, $i.e.$ det$(\eta ) \neq 0$, then we can use its inverse
$\eta ^{ij} := (\eta ^{-1})_{ij}$ to raise indices
\begin{equation}
b^{i} = b_{j} \eta ^{ji}.
\end{equation}
This metric is invariant --- or $T$ is orthogonal --- in the sense
\begin{equation}
\begin{array}{rcl}
\eta _{ji}     & = & -<S\chi _{j},b_{i}>\\
        & = & -<S\chi _{j},b_{k}> ST^{k}{}_{l} T^{l}{}_{i}\\
        & = & -<\chi _{k},Sb_{l}> T^{k}{}_{j} T^{l}{}_{i}\\
        & = & \eta _{kl} T^{k}{}_{j} T^{l}{}_{i},
\end{array}
\end{equation}
where we have used the Hopf algebraic identity
\begin{equation}
<\DA(\chi ), Sa \otimes i\!d> = S(<S\chi  \otimes i\!d,\Delta ^{Ad}(a)>),
\end{equation}
which we will proof in an appendix to this section.
Once we have obtained a metric $\eta $, we can truncate the pure braid element
$\Ups$ and work instead with:
\begin{equation}
\Ups \to  \Ups_{trunc} =
-S(\chi _{i}) \otimes \chi ^{i} = -S(\chi _{i}) \otimes \chi _{j} \eta
^{ji},\z\mbox{\it (truncated pure
braid element)}
\end{equation}
which also commutes with all coproducts. In part I of these thesis we
have shown how to construct casimir operators from elements of the pure
braid group. For the truncated  pure braid element that gives the
quadratic casimir:
\begin{equation}
[\:\cdot  \circ \tau  \circ (S^{-1} \otimes i\!d)](\Ups_{trunc}) = \eta ^{ji}
\chi _{j}
\chi _{i}.\z\mbox{\it (casimir)}
\end{equation}
Now we would like to show that we have actually
obtained a quantum Lie algebra of type I:\footnote{Note, that
$\Ups$ has to be carefully chosen to insure the correct number of
generators. Furthermore, we
still have to check the coproduct of the generators. If they are not of
the form $\Delta  \chi _{i} = \chi _{i} \otimes 1 + O_{i}{}^{j} \otimes \chi
_{j}$ then we can still
consider a calculus with deformed Leibniz rule (see next section).}
\begin{equation}
-<\chi _{i},S b^{j}> = - <\Ups,b_{i} \otimes Sb^{j}> = -<\Ups,b_{i} \otimes
Sb_{k}> \eta ^{kj} = \eta _{ik}
\eta ^{kj} = \delta _{i}^{j},
\end{equation}
\begin{equation}
\DA(\chi _{i}) = \Ups_{b_{i(2)}} \otimes S(b_{i(1)}) b_{i(3)} = \Ups_{b_{j}}
\otimes T^{j}{}_{i} =
\chi _{j} \otimes T^{j}{}_{i}
\end{equation}
and
\begin{equation}
\Delta ^{Ad}(b^{i})  =  \Delta ^{Ad}(b_{j}) \eta ^{ji}
         =  b_{k} \otimes T^{k}{}_{j} \eta ^{ji}
         =  b_{k} \otimes \eta ^{kl} \eta _{ln} T^{n}{}_{j} \eta ^{ji}
         =  b^{k} \otimes S^{-1} T^{i}{}_{k}.
\end{equation}

\subsubsection{Examples}

\paragraph{The $r$-matrix approach:}
Often one can take $b_{i} \in $span$\{t^{n}{}_{m}\}$, where $t^{n}{}_{m}$ is a
quantum
matrix in the defining representation of the quantum group under
consideration. If we are dealing with a quasitriangular Hopf algebra,
a natural choice for the pure braid element is
\begin{equation}
\Ups_{r} = \frac {1}{\lambda }\left(1 \otimes 1 - \R^{21} \R^{12}\right),
\end{equation}
where the term $\R^{21} \R^{12}$ has been introduced and extensively studied
by Reshetikhin \& Semenov-Tian-Shansky \cite{RSTS} and later by Jurco
\cite{J}, Majid \cite{pM}
and Schupp, Watts \& Zumino \cite{SWZ3}. These choices of $b_{i}$s and $\Ups$
lead to the $r$-matrix approach to differential geometry on quantum
groups. The metric is
\begin{equation}
\eta  = -<X_{1},St_{2}> = \frac {1}{\lambda
}\left(\left[\left(r_{21}{}^{-1}\right)^{t_{2}}
\left(r_{12}{}^{t_{2}}\right)^{-1}\right]^{t_{2}} - 1 \right),\label{eta}
\end{equation}
where $X_{1} = <\Ups_{r},t_{1} \otimes i\!d>$ and $r_{12} = <\R,t_{1} \otimes
t_{2}>$. In
the case of GL${}_{q}(2)$ we find\footnote{In its reduced form, this matrix
agrees \cite{pW} with a metric obtained along more standard lines from
quantum traces (except perhaps in the casimir sector $X^1{}_1 + q^{-2}
X^2{}_2$).
The formulation in terms of the pure braid element has
the great advantage that it does not require the existence of an element
like $u$ that implements the square of the antipode.}
\begin{equation}
\eta _{\mbox{\tiny GL${}_{q}(2)$}} = - \left(
        \begin{array}{cccc}
                q^{-3} & 0 & 0 & 0 \\
                0 & 0 & q^{-1} & 0 \\
                0 & q^{-3} & 0 & 0 \\
                0 & 0 & 0 & q^{-1}
        \end{array}\right).
\end{equation}
Now we will evaluate the metric in the case of GL${}_q(n)$.
The $\hat{r}$-matrix of GL${}_q(n)$ satisfies a characteristic equation
\begin{equation}
\hat{r}^2 - \lambda \hat{r} - 1 = 0
\end{equation}
which we can use in the form
\begin{equation}
r^{-1}_{21} = r_{12} - \lambda P_{12},
\end{equation}
where $P^{ij}{}_{kl} = \delta^i_l \delta^j_k$ is the permutation matrix,
to replace $(r^{-1}_{21})^{t_2}$ in equation (\ref{eta}).
That gives
\begin{equation}
\begin{array}{rcl}
\eta_{12} & = & - \left(P_{12}{}^{t_2} \left( (r_{12}{}^{t_2})^{-1}
\right) \right)^{t_2}\\
        & = & - \mbox{tr}_3\left(P_{23} (r_{23}{}^{t_3})^{-1} \right) P_{12}\\
        & = & - D_2 P_{12}.
\end{array}\label{etagln}
\end{equation}
In the last step we have used
\begin{equation}
D \equiv <u,t> = \mbox{tr}_2\left(P (r^{t_2})^{-1} \right),
\end{equation}
where $u \equiv \cdot (S \otimes i\!d) \R^{21}$ is the element of \U\
that implements the square of the antipode.
With the explicit formula ($\eta_{12} = - D_2 P_{12}$) for the metric
we immediately find an expression \cite{SWZ3} for the exterior
derivative $\dl$ on functions in terms of $X$ and the Maurer-Cartan form
$\Omega = t^{-1} \dl t$:
\begin{equation}
\dl = - \mbox{tr}( D^{-1} \Omega X). \z \mbox{\it (on functions)}
\end{equation}

The pure braid approach to the construction of quantum Lie algebras is
however particularly important in cases (like the 2-dim quantum euclidean
group)
where there is no quasitriangular Hopf algebra and where the $b_i$s
are not given by the elements of $t^i{}_j$.

\paragraph{The 2-dim quantum euclidean group}
is an example of a quantum Lie algebra that has no universal $\R$
and where the set of functions $\{b_{i}\}$ does not arise from the
matrix elements of some quantum matrix. In section~\ref{S:BGfe} we constructed
such a set of functions
\begin{equation}
b_{0} =(e^{i\theta }-1)^{2},\x b_{1}=-m e^{i\theta }\overline{m},\x
b_{+}=-(e^{i\theta }-1)m,\x
b_{-}=q^{-2}(e^{i\theta }-1)e^{i\theta }\overline{m},
\end{equation}
and a pure braid element
\begin{equation}
\Ups_{e}= \frac {1}{\lambda }\{ P_{+} P_{-} \otimes (q^{2J} -1) +
P_{+} q^{-J} \otimes q^{J} P_{-} + P_{-} q^{-J} \otimes q^{J} P_{+} +q^{-2J}
\otimes P_{+} P_{-}\}
\end{equation}
by hand. Now we can put the new machinery to work and calculate
the (invertible) metric
\begin{equation}
\eta _{\mbox{\tiny E${}_{q}(2)$}} = \left(
        \begin{array}{cccc}
                0 & 1 & 0 & 0 \\
                1 & 0 & 0 & 0 \\
                0 & 0 & 0 & -1 \\
                0 & 0 & -q^{-2} & 0
        \end{array}\right),
\end{equation}
which immediately gives an expression for $\dl$ on functions:
\begin{equation}
\dl = \omega _{0} \chi _{1} + \omega _{1} \chi _{0} - q^{2}\omega _{+} \chi
_{-} - \omega _{-} \chi _{+}.
\end{equation}

\subsection{Various Types of Quantum Lie Algebras}

The functions $c^{j} := -S b^{j}$ play the role of coordinate functions
their span$\{c^{j}\} =: R^{\perp }$ is the vector space dual to the quantum
tangent space $\tq$,
such that
\begin{equation}
\begin{array}{rcl}
1 \oplus \tq \oplus \tq^{\perp } & = & \U\\
1 \oplus R^{\perp } \oplus R & = & \A
\end{array}
\end{equation}
as vector spaces, with\footnote{We write here vector spaces in place of
their elements in an  obvious notation.}
\begin{equation}
<\tq , R> = 0,\z <\tq^{\perp },R^{\perp }> = 0.
\end{equation}
Let $\widetilde{R^{\perp }} =$span$\{b^{i}\}$ and $\widetilde{R}$ be the spaces
obtained from $R^{\perp }$ and $R$ by application of $S^{-1}$ on all of their
elements.
In the following we will state various desirable properties that
different kinds of quantum Lie algebras might have; we will comment
on their significance and we will derive the corresponding expressions
in the dual space. The proofs are given in an appendix to this section.
\begin{equation}
i)\z\DA \tq \subset \tq \otimes \A \z\Leftrightarrow\z
\Delta ^{Ad} \widetilde{R} \subset \widetilde{R} \otimes \A
\end{equation}
The left hand side states the right invariance of $\tq$, which is
important for the covariance of the cartan identity (\ref{CARTAN})
and the invariance of the realization (\ref{exder}) of $\dl$.
The right hand side is essential to Woronowicz's formulation
of the differential calculus because it allows to consistently
set $\omega _{\widetilde{R}} = 0$.
\begin{equation}
ii)\z \Delta  \tq \subset \U \otimes (\tq \oplus 1) \z\Leftrightarrow\z
\A R = R
\end{equation}
The left hand side is necessary to ensure the existence of $f-\omega $
commutation relations that are consistent with an undeformed
Leibniz rule for $\dl$. It also implies a quadratic quantum commutator
for the $\chi _{i}$:
\begin{equation}
\chi _{k} \ad \chi _{l}
\equiv \Lix{k}(\chi _{l}) = \chi _{b} \chi _{c} (\delta ^{c}_{k}\delta ^{b}_{l}
-
\hat{R}^{cb}{}_{kl})
= \chi _{a} <\chi _{k},T^{a}{}_{l}> = \chi _{a} f_{k}{}^{a}{}_{l},
\end{equation}
where
\begin{equation}
\hat{R}^{cb}{}_{kl} = <O_{k}{}^{b},T^{c}{}_{l}>
\end{equation}
is the so-called ``big R-matrix''.
If $ii)$ is not satisfied we have the choice of giving
up the $f-\omega $ commutation relations, so that the algebra of forms $\Lambda
$
is only a left \A-module, or we can try a generalized Leibniz rule for $\dl$.
The right hand side of the equation is equivalent to $\widetilde{R} \A =
\widetilde{R}$
and states that $\widetilde{R}$ is a right
\A-ideal; it is the second fundamental ingredient
of Woronowicz's theory. If the Leibniz rule is satisfied then $ii)$ follows
from
$\omega_r = 0 \Rightarrow r \in \widetilde{R} \oplus 1$:
Let $a \in \A$, then
\begin{equation}
\omega_{r a} = S(a_{(1)}) S(r_{(1)}) \dl(r_{(2)} a_{(2)}) = S(a_{(1)}) \omega_r
a_{(2)}
+ \epsilon(r) \omega_a = 0,
\end{equation}
$\epsilon(r a) = \epsilon(r)\epsilon(a) = 0$ and hence $r a \in \widetilde{R}$.
$\A R = R$ is in agreement with the intuitive picture that
the ideal $R$ is spanned by polynomials in the $c^{i}$ of order 2 or higher,
$i.e.$ span$\{e_{i}\} \approx  \{1,c^{i},c^{i} c^{j}, \ldots\}$.
\begin{equation}
iii)\z \Delta ^{Ad} \widetilde{R^{\perp }} \subset \widetilde{R^{\perp }}
\otimes \A \z\Leftrightarrow\z
\DA \tq^{\perp } \subset \tq^{\perp } \otimes \A
\end{equation}
The right hand side keeps us out of trouble with covariance when
we set $\I_{\tq^{\perp }} = 0$.
The left hand side is a {\em sufficient} condition for
$\DA(\tq^{*}) \subset \tq^{*} \otimes \A$. Quantum Lie algebras that satisfy
$iii)$ have particular nice properties in connection
with pure braid elements and a (Killing) metric.
That merits a special name for them:
\begin{quote}
Quantum Lie Algebra of {\bf type I\ }: $i)$,$ii)$,$iii)$\\
Quantum Lie Algebra of {\bf type II}: $i)$,$ii)$
\end{quote}
We will mainly be dealing with  type I, in fact, all examples of
quantum group calculi known to me are of this type.
Quantum Lie algebras of type II are mathematically equivalent
to Woronowicz's \cite{W2} theory.
\begin{equation}
iv)\z \Delta  R^{\perp } \subset \A \otimes (R^{\perp } \oplus 1)
\z\Leftrightarrow\z
\U \tq^{\perp } = \tq^{\perp }
\end{equation}
The LHS enables us to define {\bf partial derivatives} instead of
left-invariant ones: It implies
$\Delta  c^{i} = M^{i}{}_{j} \otimes c^{j} + c^{i} \otimes 1$ with
$\Delta  M = M \dot{\otimes} M$, $S M = M^{-1}$, $\epsilon (M) = I$ and then
$\chi _{k} c^{i} = M^{i}{}_{k} + M^{i}{}_{j} <O_{k}{}^{l},c^{j}> \chi _{l} +
c^{i} \chi _{k}$,
such that $\partial _{n} := S^{-1}M^{k}{}_{n} \chi _{k}$ gives a commutation
relation
\begin{equation}
\partial _{n} c^{i} = \delta _{n}^{i} +
\left( S^{-1}M^{k}{}_{n} M^{i}{}_{j}
<O_{k}{}^{l},c^{j}> M^{m}{}_{l}
+ S^{-1}M^{k}{}_{n} c^{i} M^{m}{}_{k}\right) \partial _{m}
\end{equation}
worthy of a partial derivative.
 (In the case of GL${}_q(n)$ we can use (\ref{etagln}) to show
that $c^{(mn)} = (D^{-1})^n{}_k St^k{}_m$, $M^{(mn)}{}_{(ij)} = St^i{}_m
\delta^n_j$, and
$\partial_{(ij)} = t^i{}_k X^k{}_j$.) The exterior derivative (on functions)
becomes
\begin{equation}
\dl = \omega^i \chi_i = \dl(c^j) S^{-1}(M^i{}_j) M^n{}_i \partial_n = \dl(c^n)
\partial_n.
\end{equation}
\begin{equation}
v)\z \Delta  R^{\perp } \subset (R^{\perp } \oplus 1) \otimes \A
\z\Leftrightarrow\z \tq^{\perp } \U = \tq^{\perp }
\end{equation}
This and $ii)$ imply quadratic $\chi -c$ commutation relations that
close in terms of the elements of $\tq$ and $R^{\perp }$.
The right hand sides of $iv)$ and $v)$ state that $\tq$ is a
left (right) \U-ideal, which supports the picture
of a Poincare-Birkhoff-deWitt type basis for \U\ in terms
of the $\chi _{i}$, $i.e.$ $\{1,\chi _{i},\chi _{i} \chi _{j},\ldots\}$.
Here and in the discussion following $ii)$ we have to be careful
though with higher order conditions on the generators.

\subsection{Universal Calculus}

Given (infinite) linear bases $\{ e_{i}\}$ of \U\ and
$\{f^{i}\}$ of \A\ we can always construct new counit-free elements
$\vec{e_{i}} := e_{i} - 1 \epsilon (e_{i})$  and $\vec{f^{i}} := f^{i} - 1
\epsilon (f^{i})$
that span (infinite) spaces $\tq^{u}$ and $R^{\perp u}$ respectively satisfying
properties $i)$ through $v)$; in fact $1 \oplus \tq^{u} = \U$ and
$1 \oplus R^{\perp u} = \A$ as vector spaces.
The $f-\omega $ commutation relations, however, become trivial in that they
are equivalent to the Leibniz rule for $\dg$\footnote{To distinguish
this calculus from quantum Lie algebras we use the symbol $\dg$
instead of $\dl$ for the exterior derivative}; we are hence dealing
with a Connes type calculus \cite{pZ}, a ``Universal Calculus on
Hopf Algebras''.  It is interesting to see what happens to
the formula for the partial derivatives in this limit:

\subsubsection{A Subbialgebra
and the Vacuum Projection
Operator}

To simplify notation we will assume that the infinite bases of \U\
and \A\ have been arranged in such a way that $e_{0} = 1_{{\cal U}}$,
$f^{0} = 1^{{\cal A}}$ and $e_{i}$, $f^{i}$ with
$\epsilon (e_{i}) = \epsilon (f^{i}) = 0$  for $i = 1,\ldots,\infty $ span
$\tq$ and $R^{\perp }$
respectively. Greek indices $\alpha ,\beta ,\ldots$ will run from
$0$ to $\infty $ whereas Roman indices $i,j,k,\ldots$ will only
take on values from $1$ to $\infty $ unless otherwise stated.
A short calculation gives
\begin{equation}
\Delta  f^{i} = M^{i}{}_{k} \otimes f^{k} + f^{i} \otimes 1,\z M^{i}{}_{k} =
f^{i}_{(1)}<e_{k},f^{i}_{(2)}>
\end{equation}
and
\begin{equation}
\Delta  M = M \dot{\otimes} M,\z S(M) = M^{-1},\z \epsilon (M) = I.
\end{equation}
Using the definition from the previous section we will now write down
partial derivatives
\begin{equation}
\partial _{n} = S^{-1}(M^{l}{}_{n}) e_{l},\z (l\geq  1 !)
\end{equation}
which take on a peculiar form when using the explicit expression for $M$
\begin{equation}
\begin{array}{rcl}
\partial _{n}      & = & S^{-1}(f^{l}_{(1)})<e_{n},f^{l}_{(2)}> e_{l}\\
        & = & S^{-1}(f^{\alpha }_{(1)})<e_{n},f^{\alpha }_{(2)}> e_{\alpha }\\
        & = & S^{-1}(f^{\alpha }) <e_{n},f^{\beta }> e_{\alpha } e_{\beta } \\
        & = & S^{-1}(f^{\alpha }) e_{\alpha } e_{n}\\
        & = & E e_{n},
\end{array}
\end{equation}
where we have introduced the ``vacuum projector'' $E$ in the last step.
It was first discovered (quite accidently) in collaboration with
C. Chryssomalakos \cite{CSW} and has interesting properties like
\begin{eqnarray}
E a & = & E \epsilon (a), \z a \in \A,\\
x E & = & E \epsilon (x), \z x \in \U,\\
E^{2} & = & E.
\end{eqnarray}
Prof. B. Zumino \cite{pZ} pointed out that the classical expression of
$E$ is related to a Taylor expansion.
Note also that
\begin{equation}
E = \partial _{0} - 1.
\end{equation}
As expected we can express $\dg$ on functions in terms of partial
derivatives
\begin{equation}
\dg(f) = \dg(f^{i}) \partial _{i}(f).
\end{equation}
The partial derivatives are of course no longer left invariant,
but it turns out that we can actually define a coproduct for them
making the space $E \U = \{Ey;y \in \U\} \subset \smash$ a
unital bialgebra.
Inspired by
\begin{equation}
E y f = <y_{(1)},f> E y_{(2)} = (E y_{(1)})(f) E y_{(2)}
\end{equation}
we define
\begin{equation}
\Delta _{E}(E y) = E y_{(1)} \otimes E y_{(2)},\z \epsilon _{E}(E y) = \epsilon
(y),\z 1_{E} = E,
\end{equation}
in consistency with the axioms for a bialgebra.
$E \U$ is however not a Hopf algebra because it does not have an
antipode --- at least not with respect to the multiplication in $\smash$
--- so $E \U$ might be of use as an example of a quantum plane.

\subsubsection{Quantum Lie Algebras in a Universal Calculus}

If the span $\tq^{u}$of the generators
$\{e_{a} | a = 1,\ldots,\infty \}$ of the universal calculus
contains a finite dimensional subspace, $\tq$ spanned by
$\{\chi _{i} | i = 1,\ldots,N\}$, that satisfies axioms
$i)$ and $ii)$ then one may ask how to obtain the finite calculus
from the infinite one. Let $\dg$ be the exterior derivative of the
universal calculus and $\dl$ the exterior derivative of the finite
calculus. One might be tempted to try an ansatz like
\begin{equation}
\dg = \dl + \dl^{\perp },
\end{equation}
where $\dg = \omega ^{a} e_{a}$ and $\dl = \omega ^{i} \chi _{i}$  on
functions. This equation
is covariant if axiom $iii)$ is also satisfied, but we run into
problems with the $f-\omega $ commutation relations. From the Leibniz rule
for $\dg$ we obtain
\begin{equation}
f \omega ^{i} = \omega ^{j} O_{j}{}^{i}(f) + \omega ^{r} \Theta
_{r}{}^{i}(f),\z i= 1,\ldots,N;\x r = N+1,\ldots,\infty,
\end{equation}
$i.e.$ the $f-\omega $ commutation relations do not close within the
finite calculus. So unless one decides to do without a bicovariant calculus
we have to make the second term vanish. The naive choice is to try
and set $\Theta $ equal to zero. This could be nicely expressed in terms
of another axiom
$$ \Delta  \tq^{\perp } \subset \U \otimes (\tq^{\perp } \oplus 1)
\z\Leftrightarrow\z
\A R^{\perp } = R^{\perp },$$
but the right hand side neither has a classical limit nor does it lend
itself to a description of $\A$ in terms of a Poincare-Birkhoff-deWitt
basis. The only choice left is to set the forms $\omega ^{r}$ corresponding
to functions in $R$ (recall: $<\tq,R> = 0$)
equal to zero. Following Woronowicz's approach
we hence set
\begin{equation}
\omega _{R} = 0 \z\Rightarrow\z \dg  \to  \dl.
\end{equation}

\subsubsection{Deformed Leibniz Rule?}

Here we want to briefly mention what might happen if axiom $ii)$ is
not satisfied. We will still have $\omega _{R} = 0$ in consistency with
axiom $i)$ but the generators $\chi _{i}$ now have coproducts
\begin{equation}
\Delta  \chi _{i} = \chi _{i} \otimes 1 + O_{i}{}^{j} \otimes \chi _{j} +
\Theta _{i}{}^{r} \otimes e_{r},\z i,j= 1,\ldots,N;\x r = N+1,\ldots,\infty
\end{equation}
that do not close in $\U \otimes (\tq \oplus 1)$. After some
thought we can convince ourselfs that we should use $f-\omega $ commutation
relation
\begin{equation}
f \omega ^{i} = \omega ^{j} B_{j}{}^{i}(f),
\end{equation}
with a braiding matrix $B_{j}{}^{i} \in \U$ that satisfies
$\Delta(B) = B \dot{\otimes} B$, $S(B) = B^{-1}$, $\epsilon(B) = I$ and
a bicovariance condition for all $f \in \A$
\begin{equation}
T^l{}_j \widehat{B_l{}^k}(f) = B_j{}^i(f) T^k{}_i,
\end{equation}
where $T$ is the adjoint representation.
We will then need to change the Leibniz rule for $\dl$ to
\begin{equation}
\dl f = \dl(f) + (S B_{k}{}^{i} O_{i}{}^{j}) \tr f \omega ^{k} \chi_{j}
+ (S B_{k}{}^{i} \Theta _{i}{}^{r}) \tr f \omega ^{k} e_{r}.
\end{equation}
This is a fully bicovariant first order differential calculus with
a deformed Leibniz rule. It might be of use in reducing the number
of forms in quantum calculi to the classical number.

\subsection*{Appendix}

Here we will give fairly detailed proofs of propositions $i)$ and $ii)$
and symbolic proofs of the related propositions $iii)$ through $v)$.
\paragraph{Proof of $i)$. }
We start by proofing a lemma about the relation of coactions in
\U\ and \A:
\begin{equation}
\begin{array}{rcl}
S^{-1}(x^{(2)'})<x^{(1)},S a> & = & S^{-1}(x^{(2)'}) S(a_{(2)}) <x^{(1)},S
a_{(1)}> a_{(3)}\\
        & = & S^{-1}(x^{(2)'}) x^{(1)}(S a_{(1)}) a_{(2)} \\
        & = & \widehat{x}(S a_{(1)}) a_{(2)}\\
        & = & <x , S a_{(2)}> S a_{(1)} a_{(3)}.\z\Box\label{alid}
\end{array}
\end{equation}
Another useful identity:
\begin{equation}
<x^{(1)},f> x^{(2)'} = <x , f_{(2)}> f_{(1)} S(f_{(3)}),\z\forall x \in \U,\x
f \in \A.\label{usef}
\end{equation}
\subparagraph{$i)$ ``$\Rightarrow$'':} Assume
$\DA \tq \subset \tq \otimes \A $, then for $\forall x \in \tq,\x S(a)
\in R$
\begin{equation}
0    =  <x^{(1)},S a> S^{-1}x^{(2)'}
         =  <x, S a_{(2)}> S(a_{(1)}) a_{(3)},
\end{equation}
so that $S a_{(2)} \otimes S(a_{(1)}) a_{(3)} \subset (R \oplus 1) \otimes \A$,
but $\epsilon (S a_{(2)}) S(a_{(1)}) a_{(3)} = \epsilon (S a) = 0$ and hence
$S a_{(2)} \otimes S(a_{(1)}) a_{(3)} \subset R \otimes \A$, or
\begin{equation}
\Delta ^{Ad}(a) \equiv a_{(2)} \otimes S(a_{(1)}) a_{(3)} \subset \widetilde{R}
\otimes \A.\z\Box
\end{equation}
\subparagraph{$i)$ ``$\Leftarrow$'':} Assume
$\Delta ^{Ad} \widetilde{R} \subset \widetilde{R} \otimes \A$, then again for
$\forall x \in \tq,\x a \in \widetilde{R}$
\begin{equation}
0 = < x, S a_{(2)}> S(a_{(1)}) a_{(3)} = <x^{(1)},S a> S^{-1}x^{(2)'},
\end{equation}
so that $x^{(1)} \otimes S^{-1} x^{(2)'}\subset (\tq \oplus 1) \otimes \A$;
with
$0 = <x,1> = <x^{(1)},1> x^{(2)'}$  from (\ref{usef}) that gives
$x^{(1)} \otimes S^{-1}x^{(2)'} \subset \tq \otimes \A$ and also
\begin{equation}
\DA x = x^{(1)} \otimes x^{(2)'} \subset \tq \otimes \A.\z\Box
\end{equation}

\paragraph{Proof of $ii)$. }
\subparagraph{$ii)$ ``$\Rightarrow$'': } For all $x \in \tq$, $a \in \A$ and
$r \in R$ assume $\Delta  x \in \U \otimes (\tq \oplus 1)$, then
\begin{equation}
<x, a r> = <\Delta  x ,a \otimes r > = 0,
\end{equation}
which implies $a r \in (R \oplus 1)$ or, taking into account that
$\epsilon (a r) = \epsilon (a) \epsilon (r) = 0$,
\begin{equation}
a r \in R.\z\Box
\end{equation}
\subparagraph{$ii)$ ``$\Leftarrow$'': } Assume that for all $x \in \tq,\x r
\in R$ there exists a $r' \in R$ such that $r' = a r$; then we find
\begin{equation}
0 = <x,r'> = <x,a r> = <\Delta  x, a \otimes r>
\end{equation}
which can be restated as
\begin{equation}
\Delta  x \in \U \otimes (\tq \oplus 1).\z\Box
\end{equation}
\paragraph{Symbolic proof of $iii)$. }
\begin{equation}
0 = <\tq^{\perp } \otimes i\!d ,(S \otimes i\!d)\circ \Delta ^{Ad}
\widetilde{R^{\perp }}> =
<(i\!d \otimes S^{-1}) \circ \DA \tq^{\perp }, S \widetilde{R^{\perp }} \otimes
i\!d>
\end{equation}
\paragraph{Symbolic proof of $iv)$. }
\begin{equation}
0 = <R^{\perp },\tq^{\perp }> = <R^{\perp }, \U \tq^{\perp }> = <\Delta
R^{\perp },\U \otimes \tq^{\perp }> =
<\A \otimes (R^{\perp } \oplus 1), \U \otimes \tq^{\perp }>
\end{equation}
\paragraph{Symbolic proof of $v)$. }
\begin{equation}
0 = <R^{\perp },\tq^{\perp }> = <R^{\perp }, \tq^{\perp } \U> = <\Delta
R^{\perp },\tq^{\perp } \otimes \U > =
<(R^{\perp } \oplus 1) \otimes \A,\tq^{\perp } \otimes \U>
\end{equation}

\section[Calculus of Functions, Vector Fields and Forms]{Calculus
of Functions, Vector Fields\\ and Forms}

The purpose of this section is to generalize the Cartan calculus of
ordinary {\em commutative} differential geometry to the case of quantum
Lie algebras.
As in the classical case, the Lie derivative of a function is
given by the action of the corresponding vector field, {\em i.e.}
\begin{equation}
\begin{array}{l}
\Li_{x}(a) = x \tr a =  a_{(1)} < x, a_{(2)} >,\\
\Li_{x} a = a_{(1)} < x_{(1)}, a_{(2)} > \Li_{x_{(2)}}.
\end{array} \label{XA}
\end{equation}
The action on products is given through the coproduct of $x$
\begin{equation}
x \tr a b = (x_{(1)} \tr a) (x_{(2)} \tr b). \label{XAB}
\end{equation}
The Lie derivative along $x$ of an element $y \in  \U$ is given by the
adjoint action in \U:
\begin{equation}
\Li_{x}(y) = x \ad y =  x_{(1)} y S(x_{(2)}).
\label{xady}
\end{equation}
To find the action of $\Ix{i}$ we can now attempt to use the Cartan
identity (\ref{CARTAN})\footnote{The idea is
to use this identity as long as it is consistent
and modify it if needed.}
\begin{equation}
\begin{array}{rcl}
\chi _{i} \tr a & = & \Lix{i}(a)\\
        & = & \Ix{i}(\dl a) + \dl(\Ix{i} a).
\end{array}
\end{equation}
As the inner derivation $\Ix{i}$ contracts 1-forms and is zero on
0-forms like $a$, we find
\begin{equation}
\Ix{i}(\dl a) = \chi _{i} \tr a = a_{(1)} < \chi _{i}, a_{(2)} >.
\label{incomplete}
\end{equation}
An equation like this could not be true for any $x \in \U$ because from the
Leibniz
rule for \dl\ we
have $\dl(1) = \dl(1\cdot  1) = \dl(1) 1 + 1 \dl(1) = 2 \dl(1)$
and any $\I_{x}$ that gives a non-zero result
upon contracting $\dl(1)$ will hence lead to a contradiction. From
(\ref{incomplete}) we see that the troublemakers would be  $x \in \U$
with $\epsilon (x) \neq 0$, but as $\epsilon (\chi _{i}) = 0$ we have nothing
to worry about.
Without loss of generality we can now set
\begin{equation}
\dl(1) \equiv 0\x\mbox{and}\x \I_{1} \equiv  0.
\end{equation}
Next consider for any form $\al$
\begin{equation}
\begin{array}{rcl}
\Lix{i}(\dl \al ) & = & \dl(\Ix{i} \dl \al )
                      + \Ix{i}(\dl \dl \al )\\
                & = & \dl(\Lix{i} \al ) + 0,
\end{array}
\label{Ld}
\end{equation}
which shows that Lie derivatives commute with the exterior
derivative; $\Lix{i} \dl = \dl  \Lix{i}$. We will later need to extend
this equation to all elements of \U:
\begin{equation}
\Li_{x} \dl = \dl  \Li_{x}.
\end{equation}
{}From this and (\ref{XA}) we find
\begin{equation}
\Li_{x} \dl(a) = \dl(a_{(1)}) < x_{(1)}, a_{(2)} > \Li_{x_{(2)}}.
\end{equation}
To find the complete commutation relations of $\Ix{i}$ with functions
and forms rather than just its action on them,
we next compute the action of $\Lix{i}$ on a product of functions
$a$, $b$ $\in \A$
\begin{equation}
\begin{array}{rcl}
\Lix{i}(a b) & = & \Ix{i} \dl(a b)\\
           & = & \Ix{i}(\dl(a) b + a \dl(b))
\end{array}
\end{equation}
and compare with equation (\ref{XAB}).
Recalling that the $\chi _{i}$  have coproducts of the form
\begin{equation}
\Delta \chi _{i} = \chi _{i} \otimes 1 + O_{i}{}^{j} \otimes \chi _{j},\z
O_{i}{}^{j} \in
\U,\label{DO}
\end{equation}
we obtain
\begin{equation}
\begin{array}{rcl}
\I_{\chi _{i}} a &=& (O_{i}{}^{j} \tr a) \;\I_{\chi _{j}}\\
&=& \Li_{O_{i}{}^{j}}(a) \;\I_{\chi _{j}},
\end{array}
\end{equation}
if we assume that the commutation relation of $\I_{\chi _{i}}$ with $\dl(a)$ is
of the general form
\begin{equation}
\I_{\chi _{i}} \dl(a) = \underbrace{\I_{\chi _{i}}(\dl a)}_{\in \A} +
\mbox{``braiding term''}\cdot \I_{\chi _{?}}\,.
\end{equation}
A calculation of $\Li_{\chi _{i}}(\dl(a) \dl(b))$ along similar lines
gives in fact
\begin{equation}
\begin{array}{rcl}
\I_{\chi _{i}} \dl(a) &=& (\chi _{i} \tr a) - \dl(O_{i}{}^{j} \tr a)
\;\I_{\chi _{j}}\\
&=&\I_{\chi _{i}}(\dl a) - \Li_{O_{i}{}^{j}}(\dl a) \;\I_{\chi _{j}}
\end{array}
\end{equation}
and we propose for any $p$-form $\al$:
\begin{equation}
\I_{\chi _{i}} \al = \I_{\chi _{i}}(\al) +  (-1)^{p} \Li_{O_{i}{}^{j}}(\al)\;
\I_{\chi _{j}}.
\end{equation}

Missing in our list are commutation relations of Lie derivatives with
vector fields and inner derivations.
It was shown earlier in chapter~\ref{C:BC} that the product in \U\ can be
expressed
in terms of a right coaction $\DA: \U \rightarrow \U \otimes \A$,
denoted $\DA(y) = y^{(1)} \otimes y^{(2)'}$, such that
$x y = y^{(1)} <x_{(1)},y^{(2)'}> x_{(2)}$.
In the context of (\ref{xady}), this gives
\begin{eqnarray}
\Li_{x}(y) &=& x_{(1)} y S(x_{(2)}) = y^{(1)} <x,y^{(2)'}>,\\
\Li_{x} \Li_{y} &=& \Li_{\mbox{\small \pounds}_{x_{(1)}}(y)} \Li_{x_{(2)}} =
\Li_{y^{(1)}} <x_{(1)},y^{(2)'}> \Li_{x_{(2)}}.
\end{eqnarray}
For the special case $\chi _{i},\chi _{j} \in \tq$ that becomes
\begin{equation}
\begin{array}{rcl}
\Lix{i} \Lix{k} & = & \Lix{i}(\Lix{k}) + \Lio{i}{j}(\Lix{k})\Lix{j}\\
        & = & \Lix{l} f_{i}{}^{l}{}_{k}  + \Lix{a}  \Lix{b} \hat{R}^{ab}{}_{ik}
\end{array}
\end{equation}
and --- using the Cartan identity ---
\begin{equation}
\begin{array}{rcl}
\Lix{i} \Ix{k} & = & \Lix{i}(\Ix{k}) + \Lio{i}{j}(\Ix{k})\Ix{j}\\
               & = & \Ix{l} f_{i}{}^{l}{}_{k} + \Ix{a}  \Lix{b}
\hat{R}^{ab}{}_{ik},
\end{array}
\end{equation}
where
\begin{equation}
\hat{R}^{ab}{}_{ik} = <O_{i}{}^{b} , T^{a}{}_{k}>.
\end{equation}

\subsection{Maurer-Cartan Forms}

The most general left-invariant 1-form can be written \cite{W2}
\begin{equation}
\om_{b} := S(b_{(1)}) \dl(b_{(2)}) = - \dl(S b_{(1)} ) b_{(2)}
\end{equation}
\begin{equation}
(\mbox{\em left-invariance:}\x\AD(\om_{b}) =
S(b_{(2)}) b_{(3)} \otimes S(b_{(1)}) \dl(b_{(4)})
= 1 \otimes \om_{b} ),
\end{equation}
corresponding to a function $b \in \A$. If this function happens to
be $t^{i}{}_{k}$, where $t \in M_{m}(\A)$ is an $m \times m$ matrix
representation of \U\
with $\Delta (t^{i}{}_{k}) =$\mbox{$t^{i}{}_{j} \otimes t^{j}{}_{k}$} and
$S(t)=t^{-1}$, we obtain
the well-known Cartan-Maurer form $\om_{t} = t^{-1} \dl(t) =: \Omega $. Here is
a
nice formula for the exterior derivative of $\om_{b}$:
\begin{equation}
\begin{array}{rcl}
\dl(\om_{b}) & = & \dl(S b_{(1)} ) \dl(b_{(2)})\\
&=& \dl(S b_{(1)} ) b_{(2)} S(b_{(3)}) \dl(b_{(4)})\\
&=& - \om_{b_{(1)}} \om_{b_{(2)}}.
\end{array}
\end{equation}
The Lie derivative is
\begin{equation}
\begin{array}{rcl}
\Li_{\chi }(\om_{b})&=&\Li_{\chi _{(1)}}(S b_{(1)} ) \Li_{\chi _{(2)}}(\dl
b_{(2)}) \\
&=&<\chi _{(1)},S(b_{(1)})> S(b_{(2)}) \dl(b_{(3)}) <\chi _{(2)},b_{(4)}>\\
&=&\om_{b_{(2)}} <\chi ,S(b_{(1)})b_{(3)}> \\
&=&<\chi _{(1)},S(b_{(1)})> \om_{b_{(2)}} <\chi _{(2)},b_{(3)}>.
\end{array} \label{XOM}
\end{equation}
For $\chi  = \chi _{i}$ and $b = t^{k}{}_{n}$ this becomes a quantum
commutator:
\begin{equation}
\begin{array}{rcl}
\Lix{i}(t) & = & <\chi _{i},S t>\cdot \Omega  +
        <O_{i}{}^{j},S t>\cdot \Omega \cdot <S^{-1} \chi _{j} ,S t>\\
& = & <\chi _{i},S t>\cdot \Omega  -
        <O_{i}{}^{j},S t>\cdot \Omega \cdot <S^{-1}O_{j}{}^{k} ,S t>\cdot <\chi
_{k},S t>\\
& = & <\chi _{i},S t>\cdot \Omega  -
        \Lio{i}{k}(\Omega )\cdot <\chi _{k},S t>
\end{array}
\end{equation}
and, if we denote a $S t$-matrix representation for the moment by
``$\widetilde{\x}$'',
\begin{equation}
\Lix{{}}(t) = \tilde{\chi }\cdot t - \tilde{O}\cdot t\cdot \tilde{O}^{-1}\cdot
\tilde{\chi }
=: \left[ \tilde{\chi } , t \right]_{q}.
\end{equation}
The contraction of left-invariant forms with $\Ix{{}}$
--- {\em i.e.} by a {\em left-invariant} $x \in \U$ ---
gives a number in the field $k$ rather than a function
in \A\ as was the case for $\dl(a)$. (The result must be a number because
the only invariant function is 1.)
\begin{equation}
\begin{array}{rcl}
\Ix{{}}(\om_{b}) & = & \Ix{{}}(- \dl(S b_{(1)} ) b_{(2)})\\
&=& - \Ix{{}}(\dl S b_{(1)} ) b_{(2)}\\
& = & -<\chi ,S(b_{(1)})> S(b_{(2)}) b_{(3)}\\
&=& -<\chi ,S(b)>.
\end{array} \label{IOM}
\end{equation}
As an exercise and to check consistency we will compute the
same expression in a different way:
\begin{equation}
\begin{array}{rcl}
\Ix{i}(\om_{b}) & = & \Ix{i}(S b_{(1)} \dl(b_{(2)}))\\
& =&<O_{i}{}^{j},S(b_{(1)})> S(b_{(2)}) \Ix{j}(\dl b_{(2)})\\
& = & <O_{i}{}^{j},S(b_{(1)})> S(b_{(2)}) b_{(3)} <\chi _{j},b_{(4)}>\\
& =& <O_{i}{}^{j},S(b_{(1)})><\chi _{j},b_{(2)}>\\
&=&  -<\chi _{i},S(b)>.
\end{array}
\end{equation}

\subsubsection{The Exterior Derivative on Functions}

We would like to express the exterior derivative of a function $f$
in terms of the basis of 1-forms $\{\omega ^{i}\}$ with functional
coefficients.
There are two natural ans\"atze: $\dl(f) = \omega ^{j} a_{j}$  and $\dl(f) =
b_{j} \omega ^{j}$
with appropriate $a_{j}, b_{j} \in \A$. Applying the Cartan identity
(\ref{CARTAN}) to $f$ we find
$$ \Lix{i}(f) = a_{i} = \Lio{i}{j}(b_{j}),$$
giving two alternate expressions for $\dl(f)$ :
\begin{equation}
\dl(f) = \omega ^{j} \Lix{j}(f) = - \Li_{S \chi _{j}}(f) \omega ^{j}.
\end{equation}
The Woronowicz and Castellani groups use the second expression, while
we prefer the first one because it allows us to write $\dl$ as
an operator $\omega ^{j} \chi _{j}$ on \A. An operator expression just like
this,
but written in terms of partial derivatives, is at least classically valid
on all forms. (For quantum planes that also holds \cite{pZ}).
Combining the two expressions for $\dl$ one easily derives the
well-known $f-\omega $ commutation relations
\begin{equation}
f \omega ^{i} = \omega ^{j} \Lio{j}{i}(f).
\end{equation}
The classical limit is given by
$O_{j}{}^{i} \rightarrow 1 \delta ^{i}_{j}$, so that  forms
commute with functions.

\paragraph{On the Invariance of $\dl = \omega _{b^{j}} \chi _{j}.\z$}
Recall: $\DA(\om^{i}) = \om_{b^{i}_{(2)}} \otimes
S(b^{i}_{(1)}) b^{i}_{(3)} = -\om^{j} \otimes <S\chi _{j},b^{i}_{(2)}>
Sb^{i}_{(1)} b^{i}_{(3)}$. Assuming
$\DA \chi _{i} = \chi _{j} \otimes T^{j}{}_{i}$\x(axiom $i)\,$)  we would like
to show
\begin{equation}
\DA(\omega _{b^{i}} \chi _{i}) = \om_{b^{i}_{(2)}} \chi _{i}^{(1)} \otimes
S(b^{i}_{(1)}) b^{i}_{(3)} \chi _{i}^{(2)'} =
\omega ^{i} \chi _{i} \otimes 1,
\end{equation}
$i.e.$
\begin{equation}
\DA(\om^{i}) = \om^{j} \otimes S^{-1}(T^{i}{}_{j}),
\end{equation}
or equivalently
\begin{equation}
-<S \chi _{k},b^{i}_{(2)}> S(b^{i}_{(1)}) b^{i}_{(3)}  = -S^{-1}(<\chi
_{k}^{(1)} , S b^{i}> \chi _{k}^{(2)'}).
\end{equation}
This turns out to be a purely Hopf algebraic identity for {\em any}
$x \in  \U,\x a \in \A$ (see equation \ref{alid}):
\begin{equation}
S^{-1}(x^{(2)'})<x^{(1)},S a> =  <x , S a_{(2)}> S a_{(1)} a_{(3)}.
\end{equation}

\subsection{Tensor Product Realization of the Wedge}

{}From (\ref{XOM}) and (\ref{IOM}) we find commutation relations
for $\I_{\chi _{i}}$ with $\om^{j}$,
\begin{equation}
\begin{array}{rcl}
\I_{\chi _{i}} \om^{j}
&=& \delta _{i}^{j} - \Li_{O_{i}{}^{k}}(\om^{j}) \I_{\chi _{k}}\\
&=& \delta _{i}^{j} - \om^{m} <O_{i}{}^{k},S^{-1}(T^{j}{}_{m})> \I_{\chi _{k}},
\end{array} \label{IOMI}
\end{equation}
which can be used to define the wedge product $\wedge$ of forms as some
kind of antisymmetrized tensor product\footnote{So far we have
suppressed the $\wedge$-symbol; to avoid confusion we will reinsert it
in this paragraph.}:  as in the classical case we make an ansatz for
the product of two forms in terms of tensor products
\begin{equation}
\om^{i} \wedge \om^{j} = \om^{i} \otimes \om^{j} - \hat{\sigma }^{ij}{}_{mn}
\om^{m} \otimes \om^{n},
\end{equation}
with as yet unknown numerical constants $\hat{\sigma }^{ij}{}_{mn} \in
k$, and define $\I_{\chi _{i}}$ to act on this product by contracting
in the first tensor product space, {\em i.e.}
\begin{equation}
\I_{\chi _{i}}(\om^{j} \wedge \om^{k}) = \delta _{i}^{j} \om^{k} -
\hat{\sigma }^{jk}{}_{mn} \delta _{i}^{m} \om^{n}.
\end{equation}
But from (\ref{IOMI}) we already know how to compute this,
namely
\begin{equation}
\begin{array}{rcl}
\I_{\chi _{i}}(\om^{j} \wedge \om^{k})&=&\delta _{i}^{j} \om^{k} -
\Li_{O_{i}{}^{\ell }}(\om^{j}) \delta _{\ell }^{k}\\
&=& \delta _{i}^{j} \om^{k} - \om^{m} <O_{i}{}^{k},S^{-1}(T^{j}{}_{m})>,
\end{array}
\end{equation}
and by comparison we find
\begin{equation}
\hat{\sigma }^{ij}{}_{mn} = <O_{m}{}^{j},S^{-1}(T^{i}{}_{n})>,
\end{equation}
or
\begin{equation}
\begin{array}{rcl}
\om^{i} \wedge \om^{j}
&=& \om^{i} \otimes \om^{j} - <O_{m}{}^{j},S^{-1}(T^{i}{}_{n})>
\om^{m} \otimes \om^{n}\\
&=& (I-\hat{\sigma })^{ij}{}_{mn}\om^{m} \otimes \om^{n} \\
&=& \om^{i} \otimes \om^{j} - \om^{k} \otimes \Li_{O_{k}{}^{j}}(\om^{i}).
\end{array}
\end{equation}
These equations can be used to obtain the (anti)commutation relations
between the $\om^{i}$s; by using the characteristic equation for
$\hat{\sigma }$, projection matrices orthogonal to the antisymmetrizer $I-
\hat{\sigma }$ can be found, and these will annihilate $\om^{i} \wedge
\om^{j}$.  The resulting equations will determine how to commute the
1-forms. In some rare cases the $\omega -\omega $ commutation relations
are of higher than second order. We are then forced to consider
orthogonal projectors to the operator $W$, introduced below.
There is another reason why we want to emphasize the tensor product
realization of the wedge product rather than commutation relations
given in terms of projection operators:
In the case of quantum groups in the A, B, C and D series
$\hat{\sigma }$ typically has one eigenvalue equal to 1, so there is exactly
one projection operator $P_{0}$ \cite{pW} orthogonal to $(1 - \hat{\sigma })$,
but while $(1 - \hat{\sigma })$ has a sensible classical limit ---
it becomes $(1 - P)$ where $P$ is the permutation matrix --- $P_{0}$, on the
other hand
might change discontinuously as q reaches 1 if $(1-\hat{\sigma })$ had other
eigenvalues $\lambda _{i}$ that become equal to 1 in that limit because
the corresponding projection operators $P_{i}$ will now {\em all} be orthogonal
to $(1 - P) = \left.(1-\hat{\sigma })\right|_{q=1}$. The approach of the group
in M\"unchen trying to circumvent this problem in the case of SO${}_{q}(3)$
was to impose additional conditions on the wedge product ``by hand'',
requiring that all projection operators $P_{i}$  (see above) vanish on it.
In the present context we would have to simultaneously
impose similar conditions on products of inner derivations {\em and}
check consistency of the resulting equations on  a case by case basis.

\paragraph{Example: Maurer-Cartan-Equation}

\begin{equation}
\begin{array}{rcl}
\dl \omega ^{j}  & = & \dl \omega _{b^{j}} = -\omega _{b^{j}_{(1)}} \wedge
\omega _{b^{j}_{(2)}}\\
        & = & -\omega _{S^{-1}(Sb^{j}_{(1)} b^{j}_{(3)})} \otimes \omega
_{b^{j}_{(2)}}\\
        & = & -\omega ^{k} \otimes
                \omega ^{l} <-S\chi _{k},S^{-1}(Sb^{j}_{(1)}
b^{j}_{(3)})><-S\chi _{l},b^{j}_{(2)}>\\
        & = & -\omega ^{k} \otimes \omega ^{l} <\underbrace{(S^{-1}\chi
_{k})_{(1)} \chi _{l}
                S(S^{-1}\chi _{k})_{(2)}}_{S^{-1}\chi _{k} \ad \chi
_{l}},Sb^{j}> \\
        & = & -f'_{k}{}^{j}{}_{l} \omega ^{k} \otimes \omega ^{l}.
\end{array}
\end{equation}
In the previous equation we have introduced the adjoint action of
a left-invariant vector field on another vector field. A short
calculation gives
\begin{equation}
S^{-1}\chi _{k} \ad \chi _{l}
= \chi _{b} \chi _{c} (\delta ^{c}_{k}\delta ^{b}_{l} - \hat{\sigma
}^{cb}{}_{kl})
= \chi _{a} <S^{-1}\chi _{k},T^{a}{}_{l}>=  \chi _{a} f'{}_{k}{}^{a}{}_{l}
\end{equation}
as compared to
\begin{equation}
\chi _{k} \ad \chi _{l}
\equiv \Li_{\chi _{k}}(\chi _{l}) = \chi _{b} \chi _{c} (\delta ^{c}_{k}\delta
^{b}_{l} -
\hat{R}^{cb}{}_{kl})
= \chi _{a} <\chi _{k},T^{a}{}_{l}> = \chi _{a} f_{k}{}^{a}{}_{l},
\end{equation}
with
$\hat{R}^{cb}{}_{kl} = <O_{k}{}^{b},T^{c}{}_{l}>$.
The two sets of structure constants are related by
\begin{equation}
f_{k}{}^{a}{}_{l} = -f'_{i}{}^{a}{}_{l} R^{ij}{}_{kl}.
\end{equation}
Please see \cite{CaMo}
for a detailed discussion of such structure constants.

Using the same method as for $\omega $
we can also obtain a tensor product decomposition
of products of inner derivations
\begin{equation}
\I_{\chi _{m}} \wedge
\I_{\chi _{n}} =
\I_{\chi _{m}} \otimes \I_{\chi _{n}} - \hat{\sigma }^{ij}{}_{mn}
\I_{\chi _{i}} \otimes \I_{\chi _{j}},
\end{equation}
defined to act on 1-forms by contraction in the first tensor product
space.
This can again be used to find (anti)commutation relations for the
$\I$s via projection matrices as mentioned above.
\\{\em Remark:} The tensor product decomposition of the
wedge product is invariant
under linear changes of the $\{\chi _{i}\}$ basis, but it does
depend on our choice of quantum tangent bundle.
With the extreme
choice of $\U = $span$\{e_{i}\}$ (viewed as a vector space) for instance we
get a Connes type ``Universal Cartan Calculus''.

\paragraph{The ``Anti-Wedge'' Operator. }
There is actually an operator $W$  that recursively translates
wedge products into the tensor product representation:
\begin{equation}
\begin{array}{l}
W: \Lambda ^{p}_{q} \rightarrow
{\cal T}^{*}_{q} \otimes \Lambda ^{p-1}_{q},\x p \geq 1,\\
W(\alpha ) = \om^{n} \otimes \I_{\chi _{n}}(\alpha ),
\end{array}
\end{equation}
for any p-form $\alpha $. Two examples:
\begin{equation}
\begin{array}{rcl}
\om^{j} \wedge \om^{k}
        & = & \om^{n} \otimes \I_{\chi _{n}}(\om^{j} \wedge \om^{k})\\
        & = & \om^{n} \otimes
                (\delta ^{j}_{n} \om^{k} - \Li_{O_{n}{}^{m}}(\om^{j})
                \delta ^{k}_{m})\\
        & = & \om^{j} \otimes \om^{k} - \om^{n} \otimes
                \Li_{O_{n}{}^{k}}(\om^{j})\\
        & = & \om^{j} \otimes \om^{k} - \om^{n} \otimes \om^{m}
                \hat{\sigma }^{jk}{}_{nm}
\end{array}
\end{equation}
and, after a little longer computation that uses $W$ twice,
\begin{equation}
\begin{array}{rcl}
\om^{a} \wedge \om^{b} \wedge \om^{c}
        & = & \om^{a} \otimes (\om^{b} \wedge \om^{c})
                - \om^{i} \otimes (\om^{j} \wedge \om^{c}) \hat{\sigma
}^{ab}{}_{ij}\\
        &  & + \om^{i} \otimes (\om^{j} \wedge \om^{k}) \hat{\sigma
}^{al}{}_{ij}
                \hat{\sigma }^{bc}{}_{lk}\\
        & = & \om^{a} \otimes \om^{b} \otimes \om^{c}
                - \om^{a} \otimes \om^{j} \otimes \om^{k} \hat{\sigma
}^{bc}{}_{jk}\\
        &   & -\om^{i} \otimes \om^{j} \otimes \om^{c} \hat{\sigma
}^{ab}{}_{ij}
                + \om^{i} \otimes \om^{j} \otimes \om^{k}
                \hat{\sigma }^{lc}{}_{jk} \hat{\sigma }^{ab}{}_{il}\\
        &   & + \om^{i} \otimes \om^{j} \otimes \om^{k}
                \hat{\sigma }^{al}{}_{ij} \hat{\sigma }^{bc}{}_{lk}
                - \om^{i} \otimes \om^{j} \otimes \om^{k}
                \hat{\sigma }^{an}{}_{il} \hat{\sigma }^{bc}{}_{nm} \hat{\sigma
}^{lm}{}_{jk}.
\end{array}
\end{equation}
In some cases this expression can be further simplified with the help of
the characteristic equation of $\hat{\sigma }$.

\subsection{Summary of Relations in the Cartan Calculus} \label{S:SoRCC}

\paragraph{Commutation Relations}

For any $p$-form $\alpha $:
\begin{eqnarray}
\dl \alpha   & = & \dl(\alpha ) + (-1)^{p} \alpha  \dl\\
\I_{\chi _{i}} \alpha  & = & \I_{\chi _{i}}(\alpha ) + (-1)^{p}
\Li_{O_{i}{}^{j}}(\alpha ) \I_{\chi _{j}}\\
\Li_{\chi _{i}} \alpha  & = & \Li_{\chi _{i}}(\alpha ) +
\Li_{O_{i}{}^{j}}(\alpha ) \Li_{\chi _{j}} \label{il1}
\end{eqnarray}

\paragraph{Actions}

For any function $f \in \A$, 1-form $\omega _{f} \equiv Sf_{(1)} \dl f_{(2)}$
and
vector field $\phi  \in \A\cross\U$:
\begin{eqnarray}
\Ix{i}(f) & = & 0\\
\Ix{i}(\dl f) & = & \dl f_{(1)}<\chi _{i},f_{(2)}>\\
\Ix{i}(\omega _{f}) & = & -<\chi _{i},S f>\\
\Lix{{}}(f) & = & \chi (f) = f_{(1)}<\chi ,f_{(2)}>\\
\Lix{{}}(\omega _{f})& = & \omega _{f_{(2)}} <\chi ,S(f_{(1)}) f_{(3)}>\\
\Lix{{}}(\phi )& = & \chi _{(1)} \phi  S(\chi _{(2)}) \label{il2}
\end{eqnarray}

\paragraph{Graded Quantum Lie Algebra of the Cartan Generators}

\begin{eqnarray}
\dl \dl & = & 0\\
\dl \Lix{{}} & = & \Lix{{}} \dl\\
\Li_{\chi _{i}} & = & \dl \I_{\chi _{i}} + \I_{\chi _{i}} \dl\\
\left[\Lix{i},\Lix{k} \right]_{q} & = & \Lix{l} f_{i}{}^{l}{}_{k}\\
\left[\Lix{i},\Ix{k} \right]_{q} & = & \Ix{l} f_{i}{}^{l}{}_{k}
\end{eqnarray}
The quantum commutator $[\, , \,]_{q}$ is here defined as follows
\begin{equation}
\left[\Lix{i}, \Box \right]_{q} :=
\Lix{i} \Box - \Lio{i}{j}(\Box) \Lix{j}.
\end{equation}
This quantum Lie algebra becomes infinite dimensional as soon as we
introduce derivatives along general vector fields (see below).

\subsection{Braided Cartan Calculus}

There are several graphical representations of the relations
that we derived in the previous sections. One that
emphasizes the nature of differential operators is
illustrated here at the example of equation (\ref{il1}):\\
\begin{center}
\unitlength=1.00mm
\special{em:linewidth 0.4pt}
\linethickness{0.4pt}
\begin{picture}(88.00,25.00)
\put(3.00,25.00){\makebox(0,0)[lc]{$\Lix{i} \alpha \beta$}}
\put(28.00,25.00){\makebox(0,0)[cc]{=}}
\put(28.00,15.00){\makebox(0,0)[cc]{=}}
\put(28.00,5.00){\makebox(0,0)[cc]{=}}
\put(38.00,25.00){\makebox(0,0)[cc]{$\Lix{i}$}}
\put(45.00,25.00){\makebox(0,0)[cc]{$\alpha$}}
\put(52.00,25.00){\makebox(0,0)[cc]{$\beta$}}
\put(45.00,21.00){\vector(0,1){2.00}}
\put(59.00,25.00){\makebox(0,0)[cc]{+}}
\put(66.00,25.00){\makebox(0,0)[cc]{$\Lix{i}$}}
\put(73.00,25.00){\makebox(0,0)[cc]{$\alpha$}}
\put(80.00,25.00){\makebox(0,0)[cc]{$\beta$}}
\put(80.00,21.00){\vector(0,1){2.00}}
\put(45.00,15.00){\makebox(0,0)[cc]{$\Lix{i}(\alpha)\,\beta$}}
\put(59.00,15.00){\makebox(0,0)[cc]{+}}
\put(70.00,15.00){\makebox(0,0)[cc]{$\Lio{i}{j}(\alpha)$}}
\put(81.00,15.00){\makebox(0,0)[cc]{$\Lix{j}$}}
\put(88.00,15.00){\makebox(0,0)[cc]{$\beta$}}
\put(88.00,11.00){\vector(0,1){2.00}}
\put(45.00,5.00){\makebox(0,0)[cc]{$\Lix{i}(\alpha)\,\beta$}}
\put(59.00,5.00){\makebox(0,0)[cc]{+}}
\put(70.00,5.00){\makebox(0,0)[cc]{$\Lio{i}{j}(\alpha)$}}
\put(84.00,5.00){\makebox(0,0)[cc]{$\Lix{j}(\beta)$}}
\put(38.00,23.00){\line(0,-1){2.00}}
\put(38.00,21.00){\line(1,0){7.00}}
\put(66.00,23.00){\line(0,-1){2.00}}
\put(66.00,21.00){\line(1,0){14.00}}
\put(81.00,13.00){\line(0,-1){2.00}}
\put(81.00,11.00){\line(1,0){7.00}}
\end{picture}

\end{center}
There is another graphical representation that is special in
as it shows
that we are in fact dealing with a graded and braided
Lie algebra in the sense of \cite{Md4}. Recall that in the braided
setting the coproducts and antipodes of the
generators $\{\chi _{i}\}$ take on the
classical linear form
\begin{equation}
\Delta  \chi _{i} = \chi _{i} \otimes 1 + 1 \otimes \chi _{i},\z S \chi _{i} =
- \chi _{i}\z\mbox{\it (braided)},
\end{equation}
while the multiplication  of tensor products acquires braiding
$$
(a \otimes b) \cdot  (c \otimes d) = a \Psi (b \otimes c) d \x\in W \otimes V,
$$
described by a ``braided-transposition''\cite{Md4}  operator
$\Psi _{V,W}:V \otimes W \to  W \otimes V$. This notation suggests that
the braiding is of a symmetric nature with respect to the two spaces $V$
and $W$. In the present case it turns out to be more fruitful to
assign all braiding to the generators $\chi _{i}$ --- or linear combinations
of them --- as they move through various objects. The general
braiding rule can be stated symbolically as
\begin{equation}
\Psi :\x \chi _{i} \otimes \Box \mapsto \Lio{i}{j}(\Box) \otimes \chi _{j},
\end{equation}
where $\chi _{i}$  could be part of an object like $\Li$ or $\I$. If
$\chi _{i}$ is part of $\I$, $i.e.$  of degree -1, there will be an
additional $(-1)^{p}$ grading, depending on the degree $p$ of $\Box$.
Here is a summary of all braid relations involving Cartan generators:
For $\Box \in \{\Lix{k},\Ix{k},\dl,\mbox{vector
fields},\mbox{forms},\mbox{functions}\}$
\begin{equation}
\Psi :\, \Lix{i} \otimes \Box  \mapsto  \Lio{i}{j}(\Box) \otimes \Lix{j},
\end{equation}
for $\Box \in \{\dl,\mbox{vector
fields},\mbox{forms},\mbox{functions}\}$
\begin{equation}
\Psi :\, \Ix{i} \otimes \Box  \mapsto  (-1)^{p} \Lio{i}{j}(\Box) \otimes
\Ix{j},
\end{equation}
and finally
\begin{equation}
\Psi :\, \dl \otimes \dl \mapsto - \dl \otimes \dl.
\end{equation}
Let us now look at the graphical representation of
the adjoint action (\ref{il2})
$(\chi _{i},\phi ) \mapsto \Lix{i}(\phi ) = \chi _{i(1)} \phi  S(\chi
_{i(2)})$:\\
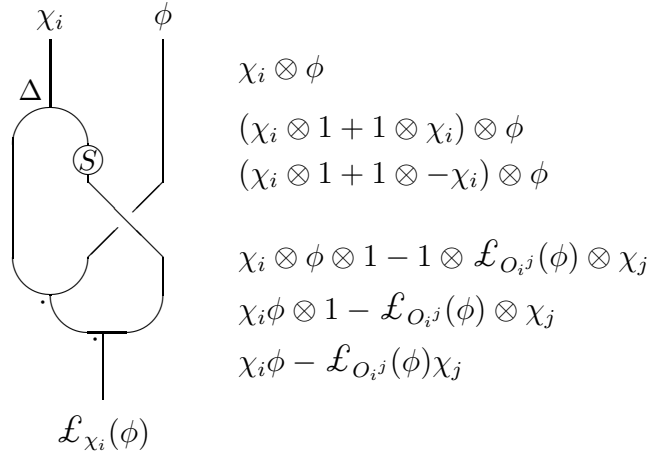
\begin{center}
\unitlength=1.00mm
\special{em:linewidth 0.4pt}
\linethickness{0.4pt}
\begin{picture}(31.00,56.00)
\put(6.00,25.00){\oval(10.00,10.00)[b]}
\put(11.00,38.00){\circle{4.00}}
\put(6.00,40.00){\oval(10.00,10.00)[t]}
\put(11.00,38.00){\makebox(0,0)[cc]{{\small $S$}}}
\put(5.00,46.00){\makebox(0,0)[rb]{{\small $\Delta$}}}
\put(5.00,19.00){\makebox(0,0)[cc]{$\cdot$}}
\put(12.00,14.00){\makebox(0,0)[cc]{$\cdot$}}
\put(6.00,56.00){\makebox(0,0)[cb]{$\chi_i$}}
\put(21.00,56.00){\makebox(0,0)[cb]{$\phi$}}
\put(13.00,4.00){\makebox(0,0)[ct]{$\Lix{i}(\phi)$}}
\put(6.00,45.00){\line(0,1){9.00}}
\put(21.00,35.00){\line(0,1){19.00}}
\put(11.00,36.00){\line(0,-1){1.00}}
\put(13.00,15.00){\line(0,-1){9.00}}
\put(13.50,20.00){\oval(15.00,10.00)[b]}
\put(21.00,25.00){\line(0,-1){5.00}}
\put(1.00,25.00){\line(0,1){16.00}}
\put(31.00,50.00){\makebox(0,0)[lc]{$\chi_i \otimes \phi$}}
\put(31.00,42.00){\makebox(0,0)[lc]{$(\chi_i \otimes 1 + 1 \otimes \chi_i)
\otimes \phi$}}
\put(31.00,36.00){\makebox(0,0)[lc]{$(\chi_i \otimes 1 + 1 \otimes -\chi_i)
\otimes \phi$}}
\put(31.00,25.00){\makebox(0,0)[lc]{$\chi_i \otimes \phi \otimes 1 - 1 \otimes
\Lio{i}{j}(\phi) \otimes \chi_j$}}
\put(31.00,11.00){\makebox(0,0)[lc]{$\chi_i \phi - \Lio{i}{j}(\phi) \chi_j$}}
\put(31.00,18.00){\makebox(0,0)[lc]{$\chi_i \phi \otimes 1 - \Lio{i}{j}(\phi)
\otimes \chi_j$}}
\put(11.00,35.00){\line(1,-1){10.00}}
\put(11.00,25.00){\line(1,1){4.00}}
\put(21.00,35.00){\line(-1,-1){4.00}}
\end{picture}

\end{center}
(In the right column we have translated the various graphical
manipulations into their algebraic counterparts.)
Taking this diagram as the definition of a braided (and graded) commutator
we can now express all Cartan relations in graphical form:\\
\paragraph{Lie derivatives.} Note that $\Lio{i}{j}(\dl) = \delta _{i}^{j} \dl$
because $\dl$ is invariant.\\
\unitlength=1.00mm
\special{em:linewidth 0.4pt}
\linethickness{0.4pt}
\begin{picture}(96.00,66.00)
\put(5.00,28.00){\oval(10.00,10.00)[b]}
\put(10.00,41.00){\circle{4.00}}
\put(5.00,43.00){\oval(10.00,10.00)[t]}
\put(10.00,41.00){\makebox(0,0)[cc]{{\small $S$}}}
\put(4.00,49.00){\makebox(0,0)[rb]{{\small $\Delta$}}}
\put(4.00,22.00){\makebox(0,0)[cc]{$\cdot$}}
\put(11.00,17.00){\makebox(0,0)[cc]{$\cdot$}}
\put(5.00,59.00){\makebox(0,0)[cb]{$\Lix{i}$}}
\put(20.00,59.00){\makebox(0,0)[cb]{$\Box$}}
\put(12.00,7.00){\makebox(0,0)[ct]{$\left[\Lix{i},\Box\right]_q$}}
\put(5.00,48.00){\line(0,1){9.00}}
\put(20.00,38.00){\line(0,1){19.00}}
\put(10.00,39.00){\line(0,-1){1.00}}
\put(12.00,18.00){\line(0,-1){9.00}}
\put(12.50,23.00){\oval(15.00,10.00)[b]}
\put(20.00,28.00){\line(0,-1){5.00}}
\put(0.00,28.00){\line(0,1){16.00}}
\put(30.00,53.00){\makebox(0,0)[lc]{$\Lix{i} \otimes \Ix{k}$}}
\put(30.00,45.00){\makebox(0,0)[lc]{$(\Lix{i} \otimes 1 + 1 \otimes \Lix{i})
\otimes \Ix{k}$}}
\put(30.00,39.00){\makebox(0,0)[lc]{$(\Lix{i} \otimes 1 + 1 \otimes
\Li_{-\chi_i}) \otimes \Ix{k}$}}
\put(30.00,28.00){\makebox(0,0)[lc]{$\Lix{i} \otimes \Ix{k} \otimes 1 - 1
\otimes \I_{\pounds_{O_i{}^j}(\chi_k)} \otimes \Lix{j}$}}
\put(30.00,14.00){\makebox(0,0)[lc]{$\Lix{i} \Ix{k} - ``\Lio{i}{j}(\Ix{k})"
\Lix{j}$}}
\put(30.00,21.00){\makebox(0,0)[lc]{$\Lix{i} \Ix{k} \otimes 1 -
``\Lio{i}{j}(\Ix{k})" \otimes \Lix{j}$}}
\put(10.00,38.00){\line(1,-1){10.00}}
\put(10.00,28.00){\line(1,1){4.00}}
\put(20.00,38.00){\line(-1,-1){4.00}}
\put(96.00,53.00){\makebox(0,0)[lc]{$\Lix{i} \otimes \dl$}}
\put(96.00,45.00){\makebox(0,0)[lc]{$(\Lix{i} \otimes 1 + 1 \otimes \Lix{i})
\otimes \dl$}}
\put(96.00,39.00){\makebox(0,0)[lc]{$(\Lix{i} \otimes 1 + 1 \otimes
\Li_{-\chi_i}) \otimes \dl$}}
\put(96.00,28.00){\makebox(0,0)[lc]{$\Lix{i} \otimes \dl \otimes 1 - 1 \otimes
\delta_i^j \dl \otimes \Lix{j}$}}
\put(96.00,14.00){\makebox(0,0)[lc]{$\Lix{i} \dl - \dl \Lix{i}$}}
\put(96.00,21.00){\makebox(0,0)[lc]{$\Lix{i} \dl \otimes 1 - \dl \otimes
\Lix{i}$}}
\put(30.00,7.00){\makebox(0,0)[lt]{$= \Ix{l}f_i{}^l{}_k$}}
\put(96.00,7.00){\makebox(0,0)[lt]{$= 0$}}
\end{picture}
\\
The relation $\left[\Lix{i},\Lix{k}\right]_{q} = \Lix{l} f_{i}{}^{l}{}_{k}$ has
a
very similar picture, so we did not show it here.
\paragraph{Inner derivations.} $\alpha $ is a $p$-form here.\\
\unitlength=1.00mm
\special{em:linewidth 0.4pt}
\linethickness{0.4pt}
\begin{picture}(101.00,66.00)
\put(5.00,28.00){\oval(10.00,10.00)[b]}
\put(10.00,41.00){\circle{4.00}}
\put(5.00,43.00){\oval(10.00,10.00)[t]}
\put(10.00,41.00){\makebox(0,0)[cc]{{\small $S$}}}
\put(4.00,49.00){\makebox(0,0)[rb]{{\small $\Delta$}}}
\put(4.00,22.00){\makebox(0,0)[cc]{$\cdot$}}
\put(11.00,17.00){\makebox(0,0)[cc]{$\cdot$}}
\put(5.00,59.00){\makebox(0,0)[cb]{$\Ix{i}$}}
\put(20.00,59.00){\makebox(0,0)[cb]{$\Box$}}
\put(12.00,7.00){\makebox(0,0)[ct]{$\left[\Ix{i},\Box\right]_q$}}
\put(5.00,48.00){\line(0,1){9.00}}
\put(20.00,38.00){\line(0,1){19.00}}
\put(10.00,39.00){\line(0,-1){1.00}}
\put(12.00,18.00){\line(0,-1){9.00}}
\put(12.50,23.00){\oval(15.00,10.00)[b]}
\put(20.00,28.00){\line(0,-1){5.00}}
\put(0.00,28.00){\line(0,1){16.00}}
\put(30.00,53.00){\makebox(0,0)[lc]{$\Ix{i} \otimes \alpha$}}
\put(30.00,45.00){\makebox(0,0)[lc]{$(\Ix{i} \otimes 1 + 1 \otimes \Ix{i})
\otimes \alpha$}}
\put(30.00,39.00){\makebox(0,0)[lc]{$(\Ix{i} \otimes 1 + 1 \otimes
\I_{-\chi_i}) \otimes \alpha$}}
\put(30.00,28.00){\makebox(0,0)[lc]{$\Ix{i} \otimes \alpha \otimes 1 - 1
\otimes (-1)^p \Lio{i}{j}(\alpha) \otimes \Ix{j}$}}
\put(30.00,14.00){\makebox(0,0)[lc]{$\Ix{i} \alpha - (-1)^p \Lio{i}{j}(\alpha)
\Ix{j}$}}
\put(30.00,21.00){\makebox(0,0)[lc]{$\Ix{i} \alpha \otimes 1 - (-1)^p
\Lio{i}{j}(\alpha) \otimes \Ix{j}$}}
\put(10.00,38.00){\line(1,-1){10.00}}
\put(10.00,28.00){\line(1,1){4.00}}
\put(20.00,38.00){\line(-1,-1){4.00}}
\put(101.00,53.00){\makebox(0,0)[lc]{$\Ix{i} \otimes \dl$}}
\put(101.00,45.00){\makebox(0,0)[lc]{$(\Ix{i} \otimes 1 + 1 \otimes \Ix{i})
\otimes \dl$}}
\put(101.00,39.00){\makebox(0,0)[lc]{$(\Ix{i} \otimes 1 + 1 \otimes
\I_{-\chi_i}) \otimes \dl$}}
\put(101.00,28.00){\makebox(0,0)[lc]{$\Ix{i} \otimes \dl \otimes 1 + 1 \otimes
\delta_i^j \dl \otimes \Ix{j}$}}
\put(101.00,14.00){\makebox(0,0)[lc]{$\Ix{i} \dl + \dl \Ix{i}$}}
\put(101.00,21.00){\makebox(0,0)[lc]{$\Ix{i} \dl \otimes 1 + \dl \otimes
\Ix{i}$}}
\put(30.00,7.00){\makebox(0,0)[lt]{$= \Ix{i}(\alpha)$}}
\put(101.00,7.00){\makebox(0,0)[lt]{$= \Lix{i}$}}
\end{picture}
\\
\paragraph{Exterior derivative.} Here we use that $\dl$ is a derivation
in the sense $``\Delta(\dl)" = \dl \otimes 1 + 1 \otimes \dl$.\\
\unitlength=1.00mm
\special{em:linewidth 0.4pt}
\linethickness{0.4pt}
\begin{picture}(90.00,66.00)
\put(5.00,28.00){\oval(10.00,10.00)[b]}
\put(10.00,41.00){\circle{4.00}}
\put(5.00,43.00){\oval(10.00,10.00)[t]}
\put(10.00,41.00){\makebox(0,0)[cc]{{\small $S$}}}
\put(4.00,49.00){\makebox(0,0)[rb]{{\small $\Delta$}}}
\put(4.00,22.00){\makebox(0,0)[cc]{$\cdot$}}
\put(11.00,17.00){\makebox(0,0)[cc]{$\cdot$}}
\put(5.00,59.00){\makebox(0,0)[cb]{$\dl$}}
\put(20.00,59.00){\makebox(0,0)[cb]{$\Box$}}
\put(12.00,7.00){\makebox(0,0)[ct]{$\left[\dl,\Box\right]_q$}}
\put(5.00,48.00){\line(0,1){9.00}}
\put(20.00,38.00){\line(0,1){19.00}}
\put(10.00,39.00){\line(0,-1){1.00}}
\put(12.00,18.00){\line(0,-1){9.00}}
\put(12.50,23.00){\oval(15.00,10.00)[b]}
\put(20.00,28.00){\line(0,-1){5.00}}
\put(0.00,28.00){\line(0,1){16.00}}
\put(30.00,53.00){\makebox(0,0)[lc]{$\dl \otimes \alpha$}}
\put(30.00,45.00){\makebox(0,0)[lc]{$(\dl \otimes 1 + 1 \otimes \dl) \otimes
\alpha$}}
\put(30.00,39.00){\makebox(0,0)[lc]{$(\dl \otimes 1 + 1 \otimes -\dl) \otimes
\alpha$}}
\put(30.00,28.00){\makebox(0,0)[lc]{$\dl \otimes \alpha \otimes 1 - 1 \otimes
(-1)^p \alpha \otimes \dl$}}
\put(30.00,14.00){\makebox(0,0)[lc]{$\dl \alpha - (-1)^p \alpha \dl$}}
\put(30.00,21.00){\makebox(0,0)[lc]{$\dl \alpha \otimes 1 - (-1)^p \alpha
\otimes \dl$}}
\put(10.00,38.00){\line(1,-1){10.00}}
\put(10.00,28.00){\line(1,1){4.00}}
\put(20.00,38.00){\line(-1,-1){4.00}}
\put(90.00,53.00){\makebox(0,0)[lc]{$\dl \otimes \dl$}}
\put(90.00,45.00){\makebox(0,0)[lc]{$(\dl \otimes 1 + 1 \otimes \dl) \otimes
\dl$}}
\put(90.00,39.00){\makebox(0,0)[lc]{$(\dl \otimes 1 + 1 \otimes -\dl) \otimes
\dl$}}
\put(90.00,28.00){\makebox(0,0)[lc]{$\dl \otimes \dl \otimes 1 + 1 \otimes \dl
\otimes \dl$}}
\put(90.00,14.00){\makebox(0,0)[lc]{$2 \dl \dl$}}
\put(90.00,21.00){\makebox(0,0)[lc]{$\dl \dl \otimes 1 + \dl \otimes \dl$}}
\put(30.00,7.00){\makebox(0,0)[lt]{$= \dl(\alpha)$}}
\put(90.00,7.00){\makebox(0,0)[lt]{$= 0$}}
\end{picture}

\subsection{Lie Derivatives Along General Vector Fields}

So far we have focused on Lie derivatives and inner derivations along
{\em left-invariant} vector fields, {\em i.e.}
along elements of $\tq$. The classical theory
allows functional coefficients, {\em i.e.} the
vector fields need not be left-invariant.
Here we may introduce derivatives along
elements in the
$\A\cross\tq$ plane by the following set of equations
valid on forms:
(note: $\epsilon (\chi ) = 0$ for $\chi \in {\cal T}_{q}$)
\begin{eqnarray}
\I_{f \chi } & = & f \I_{\chi },\\
\Li_{f \chi } & = & \dl  \I_{f \chi } + \I_{f \chi } \dl,\\
\Li_{f \chi } & = & f \Li_{\chi } + \dl(f) \I_{\chi }
\label{liefx},\\
\Li_{f \chi } \dl & = & \dl \Li_{f \chi }.
\end{eqnarray}
Equation (\ref{liefx})
can be used to define Lie derivatives recursively on any form.
There does not seem to be a way to generalize (\ref{lixv}), {\em i.e.} to
introduce Lie derivatives of {\em vector fields} along {\em
arbitrary} elements of $\A\cross\U$ or $\A\cross{\cal T}_{q}$
in the quantum case. Exceptions are the right-invariant vector fields
$\widehat{x} \in \A\cross\U$, where
\begin{equation}
\Li_{\widehat{x}}(\phi ) = \widehat{x_{(1)}} \phi
\widehat{S^{^{-1}}x_{(2)}},\z \mbox{for }
\phi  \in \A\cross\U.
\end{equation}

\section{Universal Cartan Calculus}

The equations presented in this section were obtained in collaboration
with P. Watts starting directly from Hopf algebras without explicitly
referring to any bases.\\
As we have already mentioned in the section on quantum Lie algebras,
given (infinite) linear bases $\{ e_{i}\}$ and $\{f^{i}\}$  of the
Hopf algebras \U\ and
 of \A, we can always construct new counit-free elements
$\vec{e_{i}} := e_{i} - 1 \epsilon (e_{i})$  and $\vec{f^{i}} := f^{i} - 1
\epsilon (f^{i})$
that span (infinite) spaces $\tq^{u}$ and $R^{\perp u}$ respectively satisfying
properties $i)$ through $v)$; in fact $1 \oplus \tq^{u} = \U$ and
$1 \oplus R^{\perp u} = \A$ as vector spaces.
Using some Schmidt orthogonalization procedure one can rearrange
the infinite bases of \U\
and \A\ in such a way that $e_{0} = 1_{{\cal U}}$,
$f^{0} = 1^{{\cal A}}$ and $e_{i}$, $f^{i}$ with
$\epsilon (e_{i}) = \epsilon (f^{i}) = 0$  for $i = 1,\ldots,\infty $ span
$\tq^{u}$ and $R^{\perp u}$
respectively. Greek indices $\alpha ,\beta ,\ldots$ will run from
$0$ to $\infty $, whereas roman indices $i,j,k,\ldots$ will only
take on values from $1$ to $\infty $, unless otherwise stated. To avoid
confusion with the finite dimensional quantum Lie algebras we will
use the symbol $\dg$ instead of $\dl$  for the exterior derivative.

Given orthonormal  linear basis $\{e_{i}\}$
and $\{f^{i}\}$ of $\tq^{u}$ and $\R^{\perp u}$ we can now express \dg\ on
functions $a \in \A$ as
\begin{equation}
\dg (a) = - \om_{S^{-1}f^{i}} \Li_{e_{i}-1 \epsilon (e_{i})}(a);
\label{hopfd}
\end{equation}
note, however, that {\em all} of these
$\om_{S^{-1}f^{i}}$s are treated as linearly independent
and even in the classical limit stay linearly independent because (\ref{hopfd})
in conjunction with the Leibniz rule for \dg\ only gives trivial
commutation relations
($a \om_{b} =
\om_{b S^{-1}a_{(2)}} a_{(1)} - \epsilon (b) \om_{S^{-1}a_{(2)}} a_{(1)}$)
for forms with functions that do not permit
reorganization of the infinite set of $\om_{S^{-1}f^{i}}$s into a finite
basis of 1-forms. This is
the case for Connes' non-commutative geometry (\cite{Co} and
references therein) and is in contrast to the ordinary text book
treatment of differential calculi that has forms commuting with functions.

Here is a summary of basis-free commutation relations for
the Universal Cartan Calculus valid on any form. All of these
equations
are identical to the corresponding quantum Lie algebra relations
when written in terms of the bases $\{e_{\alpha }\}$ and $\{f^{\alpha }\}$.
$x,y \in \U$, $a \in \A$,
$\al$ is a $p$-form and $v \in \A\cross\U$ is a vector field.
\begin{eqnarray}
\Li_{x} a & = & a_{(1)} < x_{(1)}, a_{(2)} > \Li_{x_{(2)}}\\
\Li_{x} \dg(a) & = & \dg(a_{(1)}) < x_{(1)}, a_{(2)} > \Li_{x_{(2)}}\\
\Li_{x} \al & = & \Li_{x_{(1)}}(\al)\:\Li_{x_{(2)}}\\
\I_{x} a & = & a_{(1)} <x_{(1)},a_{(2)}> \I_{x_{(2)}}\\
\I_{x} \dg(a) & = & a_{(1)} < x -1 \epsilon (x), a_{(2)} > -
\dg(a_{(1)}) <x_{(1)},a_{(2)}> \I_{x_{(2)}}\\
\I_{x} \al & = & \I_{x}(\al) +(-1)^{p}
\Li_{x_{(1)}}(\al)\;\I_{x_{(2)}}\\
\dg \al & = & \dg(\al) + (-1)^{p} \al \dg\\
\dg \dg(\al) & = & - (-1)^{p} \dg(\al) \dg\\
\nonumber \\
\Li_{x}(v) & = & x_{(1)} v S(x_{(2)}) \label{lixv}\\
\nonumber \\
\dg^{2}&=&0\\
\dg \Li_{x} &=& \Li_{x} \dg\\
\Li_{x} &=& \dg \I_{x} + 1 \epsilon (x) + \I_{x} \dg\z\mbox{\it (generalized
Cartan
identity)}\\
\Li_{x} \Li_{y} &=& \Li_{y^{(1)}} <x_{(1)},y^{(2)'}> \Li_{x_{(2)}}\\
\Li_{x} \I_{y} &=& \I_{y^{(1)}} <x_{(1)},y^{(2)'}> \Li_{x_{(2)}}
\end{eqnarray}
The ``generalized Cartan identity'' is due to  P. Watts.

\chapter{Quantum Planes Revisited}

With the new tools that we have developed in the previous sections
we are now ready to take a second look at quantum planes.
The first two sections that follow will be devoted to the
realization and action of quantum Lie algebra generators
on a quantum plane. After introducing the basic equations we will
spend some time on the important question of their covariance.
The third section finally gives an introduction to the construction of a
Cartan calculus on quantum planes with the surprising result
--- first observed by Prof. B. Zumino \cite{pZ} in the example
of the 2-dimensional quantum plane --- that
the $\Li_{\partial }-x$ commutation must contain inner derivation terms in
order
to be consistent with a Lie derivative that commutes with $\dl$.
For simplicity we will however suppress these inner derivation terms
in the following two sections.

\section{Induced Calculus} \label{indcal}

In this section we wish to show how the calculus of the symmetry quantum
group induces a calculus on the plane. Originally, I was interested in this
topic trying to develop as general applicable a formalism for a
calculus on quantum planes as we have presented it in part I in the case
of quantum groups. As we have already mentioned, quantum planes
do not have a Hopf algebra structure --- at least not in the unbraided
theory --- and so we have to look for a different approach than the one
that we used to construct the cross product algebra. Later it turned
out that a better approach is based on \U-coactions leading
to the introduction of the generalized product rule in the first
section of this chapter. The material presented here is however of
interest in its own right: We will study realizations of
quantum group generators in terms of the calculus on a quantum
plane. This will also give an explanation for the appearance
of ``inner derivation terms'' in the generalized product rule.

The central idea of this section,
inspired by a comment of Prof. B. Zumino, is to
give the coordinate functions on the quantum plane functional
coefficients in \A, i.e. to make them variable with respect to the
action of vector fields in \U. Let $x_{0}^{i} \in \fum$ be the ``fixed''
coordinate functions and define new variable ones via
$x^{i} := (t^{-1})^{i}{}_{j} x_{0}^{j}$. Instead of the
differentials $\dl x_{0}^{i}$ we will
use $\dg x^{i} = -(\Omega  x)^{i}$ because
\begin{equation}
\dg x = \dg t^{-1}\cdot  x_{0} = t^{-1}\cdot t\cdot \dg(t^{-1})\cdot x_{0} =
-t^{-1}\cdot \dg(t)\cdot t^{-1}\cdot x_{0} = -\Omega \cdot x,
\end{equation}where $\dg$ is the exterior derivative on the
quantum group and $\Omega  = t^{-1} \dg t$ is the Maurer-Cartan Matrix.
By ``pullback'' the group derivative will become the derivative on
the plane, inducing a differential calculus there.
It then immediately follows that $\DA(\dl x^{i}) = \dl x^{j} \otimes
(t^{-1})^{j}{}_{i}$, which will ultimately give us the desired commutation
between
Lie derivatives and $\dl$.

Turn now to the quantum group. Reserving Latin indices $i,j,\ldots$ for
the plane coordinates, let us use Greek indices for the adjoint
representation of the quantum group. Let $\{v_{\alpha }\}$\footnote{We write
$v$
instead of $\chi $ here to avoid confusion with coordinate functions $x \in
\fum$.} be a basis of
bicovariant generators with coproduct
$\Delta  v_{\alpha } = v^{\alpha } \otimes 1 + O_{\alpha }{}^{\beta } \otimes
v_{\beta }$ spanning $\tq \subset \U$
and let $\{\omega _{\alpha }\}$ be the dual basis of 1-forms; $\I_{v_{\alpha
}}(\omega ^{\beta }) = \delta _{\alpha }^{\beta }$,\x
$\Omega ^{i}{}_{j} = \omega ^{\alpha } \I_{v_{\alpha }}(\Omega ^{i}{}_{j}) =
-\omega ^{\alpha } <v_{\alpha },(t^{-1})^{i}{}_{j}>$. Via the Cartan identity
$\Li_{v} = \I_{v} \dg + \dg \I_{v}$ one computes actions of $\tq$ on $\fum$:
\begin{equation}
v_{\alpha } \tr x^{i} = \I_{v_{\alpha }}(\dg x^{i}) = <v_{\alpha
},(t^{-1})^{i}{}_{j}> x^{j}.
\end{equation}
Now we can make an ansatz for a realization of the
group generators in terms of
functions and derivatives on the plane\footnote{$\doteq$ means: ``equal
when evaluated on $\fum$''}
\begin{equation}
v_{\alpha } \doteq J_{\alpha }^{i} \partial _{i},
\end{equation}
where $J_{\alpha }^{i} \in \fum$ is easily computed, using $\partial
_{i}(x^{j}) = \delta _{i}^{j}$ to be
\begin{equation}
J_{\alpha }^{i} = v_{\alpha }(x^{i}) = <v_{\alpha },(t^{-1})^{i}{}_{j}> x^{j}.
\end{equation}
In some lucky cases there is an inverse expression for the
partial derivatives on the plane in terms of the group generators.
With $\tilde{J_{i}^{\alpha }} \in \fum$
\begin{equation}
\partial _{i} = \tilde{J_{i}^{\alpha }} \otimes v_{\alpha }\tr,
\end{equation}
an expression that is classically only valid locally  and may exclude
some points unless we are dealing with an inhomogeneous group, but will
give explicit $\partial -x$ commutation relations if it exists:
\begin{equation}
\partial _{i} x^{j} = \tilde{J_{i}^{\alpha }} v_{\alpha } x^{j} = \partial
_{i}(x^{j}) +
\underbrace{\tilde{J_{i}^{\alpha }} O_{\alpha }{}^{\beta }(x^{j})
J_{\beta }{}^{k}}_{L_{i}{}^{k}(x^{j})} \partial _{k}. \label{nicedx}
\end{equation}

\subsubsection{Example: GL${}_{\frac {1}{q}}(2)$, Manin-Wess-Zumino Quantum
Plane}

The coordinate functions $x,y$ of the Manin plane satisfy commutation
relations $x y = q y x$ that are covariant under coactions of the
quantum matrix group GL${}_{\frac {1}{q}}(2)$. This quantum group has four
bicovariant
generators $v_{1},v_{2},v_{+},v_{-}$; we will focus on the last two for the
moment,
giving their fundamental $t^{-1}$ representations
\begin{equation}
<v_{+},t^{-1}> = \left(\begin{array}{cc}0 & q^{3}\\0 & 0\end{array}\right),\z
<v_{-},t^{-1}> = \left(\begin{array}{cc}0 & 0\\q & 0\end{array}\right)
\end{equation}
and the first tensor product representations
\begin{equation}
<\Delta  v_{+},t_{1}^{-1} \otimes t_{2}^{-1}> =
\left(\begin{array}{cccc}0&q^{4}&q^{3}&0\\
0&0&0&q^{5}\\0&0&0&q^{4}\\0&0&0&0\end{array}\right),\x
<\Delta  v_{-},t_{1}^{-1} \otimes t_{2}^{-1}> =
\left(\begin{array}{cccc}0&0&0&0\\
q^{2}&0&0&0\\q&0&0&0\\0&q^{3}&q^{2}&0\end{array}\right).
\end{equation}
All these were obtained from
\begin{equation}
r_{\frac {1}{q}} = \left(\begin{array}{cccc}\frac
{1}{q}&0&0&0\\0&1&0&0\\0&-\lambda &1&0\\
0&0&0&\frac {1}{q}\end{array}\right).
\end{equation}
We immediately find
\begin{equation}
\partial _{x} = \tilde{J}_{x}^{\alpha } v_{\alpha } =q^{-3} y^{-1} v_{+},\z
\partial _{y} = q^{-1} x^{-1} v_{-},
\end{equation}
which we only have to check on pairs of functions because of the
form of (\ref{nicedx}):
\begin{equation}
\partial _{x} \left(\begin{array}{c}x x\\x y\\y x\\y y\end{array}\right) =
\left(\begin{array}{c}(1+q^{2}) x\\q^{2} y\\q y\\0\end{array}\right),\z
\partial _{y} \left(\begin{array}{c}x x\\x y\\y x\\y y\end{array}\right) =
\left(\begin{array}{c}0\\q x\\x\\y+q^{2} y\end{array}\right).
\end{equation}
{}From this we read off the following $\partial -x$ commutation relations in
perfect agreement with the results given in \cite{WZ}
\begin{eqnarray}
\partial _{x} x & = & 1 + q^{2} x \partial _{x} + (q^{2}-1) y \partial _{y},\\
\partial _{x} y & = & q y \partial _{x},\\
\partial _{y} x & = & q x \partial _{y},\\
\partial _{y} y & = & 1 + q^{2} y \partial _{y}.
\end{eqnarray}
Using the other two generators $v_{1},v_{2}$ gives identical results.
This method works for all linear quantum planes \cite{pZ} and
can be formulated abstractly in terms of $r$-matrices.
If one does not want to extend the algebra by introducing inverses
$y^{-1},x^{-1}$ of the coordinate functions, it is also possible
to obtain the above commutation relations as a vanishing ideal
of $x y$ thereby also avoiding the questionable use of $\tilde{J}$.

\section{Covariance}

Let us collect some of the equations valid on a quantum plane.
Let $f,g \in \fum$ be functions and $\partial _{i}$ be  derivatives on the
quantum
plane, let $v_{a}$ be generators of the quantum Lie algebra
--- corresponding to the symmetry quantum group of the plane ---
with coproduct $\Delta  v_{a} = v_{a} \otimes 1 + O_{a}{}^{b} \otimes v_{b}$,
and let
$L_{i}{}^{j}$ be a linear automorphism of $\fum$:
\begin{eqnarray}
v_{a} f & = & v_{a}(f) + O_{a}{}^{b}(f) v_{b},\\
v_{a} & \doteq & J_{a}^{i} \partial _{i},\footnotemark\\
\partial _{i} f & = & \partial _{i}(f) + L_{i}{}^{j}(f) \partial _{j}.
\end{eqnarray}
\footnotetext{Careful: An expression linear in the partials may not
always exist, in particular for $e_{q}(2)$ we get a power series
instead. It {\em does} exist for Wess-Zumino type quantum planes
and then we have $J_{a}^{i} = <v_{a},(t^{-1})^{i}{}_{j}> x^{j}$.}
{}From this equations we can form a new one
\begin{equation}
J_{a}^{i} L_{i}^{k}(f) = O_{a}^{b}(f) J_{b}^{k},\label{newone}
\end{equation}
that can sometimes be rewritten as
\begin{equation}
L_{i}{}^{k}(f) = \tilde{J}_{i}^{a} O_{a}{}^{b}(f) J_{b}^{k}.
\end{equation}

\subsubsection{\underline{Examples}}

\paragraph{Quantum group as plane: $\fum := \A$.}
\begin{quote}
Left-Invariant Generators: $\partial _{i} := v_{i}
\Rightarrow$ $J_{i}^{j} = \delta _{i}^{j},\x L_{i}{}^{j} =
O_{i}{}^{j}.$
\end{quote}
\begin{quote}
Plane-Like Generators: $\partial _{(ij)}:= t^{i}{}_{k} X^{k}{}_{j} \Rightarrow$
$J_{(kl)}^{(ij)} = (t^{-1})^{k}{}_{i} \delta _{l}^{j},$ $\x
L_{(lj)}{}^{(nm)}(f) = t^{l}{}_{i} O_{(ij)}{}^{(km)}(f) (t^{-1})^{k}{}_{n}.$
\end{quote}
\paragraph{Linear quantum plane:} The algebra of functions on the
linear quantum plane is invariant under coactions
of GL${}_{q}(N)$; $\DA(x^{i}) = x^{j} \otimes S t^{i}{}_{j}$,
$J_{a}^{i} = <v_{a},S t^{i}{}_{j}> x^{j}$. Using (\ref{newone}) we find
$$x^{l}(<v_{a},S t^{i}{}_{l}> L_{i}{}^{k}(x^{j}) - <O_{a}{}^{b},S
t^{j}{}_{l}><v_{b},S t^{k}{}_{n}> x^{n}) = 0,$$
so that $L_{i}{}^{k}(x^{j})$ should be homogenous of first order in $x$, which
suggests
$$L_{i}{}^{k}(x^{j}) = <L_{i}{}^{k},S t^{j}{}_{l}> x^{l},\z L_{i}{}^{k} \in
\U.$$

\subsubsection{\underline{Covariance of: $v f = v_{(1)}(f) v_{(2)}$}}

Here: $v \in \U,\,f \in \fum$ and $v(f) = f^{(1)}<v,f^{(2)'}>$.
\paragraph{Covariance of $v(f)$ alone:}
\begin{equation}
\begin{array}{rcl}
\DA(v)\left(\DA f\right)
        & = & f^{(1)}<v^{(1)},f^{(2)'}> \otimes v^{(2)'}f^{(3)'}\\
        & = & f^{(1)}<v,f^{(3)'}> \otimes f^{(2)'}S(f^{(4)'})f^{(5)'}\\
        & = & f^{(1)} \otimes v(f^{(2)'})\\
        & = & \DA( v(f) ),\z\Box
\end{array}
\end{equation}
where we have used identity (\ref{usef}).
\paragraph{Covariance of the complete commutation relation:}
\begin{equation}
\begin{array}{rcl}
\DA v \cdot  \DA f
        & = & f^{(1)}<v^{(1)}{}_{(1)},f^{(2)'}>v^{(1)}{}_{(2)} \otimes
v^{(2)'}f^{(3)'}\\
        & = & f^{(1)}\underline{v^{(1)}{}_{(2)}} \otimes
        \underline{v^{(2)'}\,\widehat{v^{(1)}{}_{(1)}}(f^{(2)'})}\\
        & = & f^{(1)} v_{(2)}{}^{(1)} \otimes v_{(1)}(f^{(2)'})\,
v_{(2)}{}^{(2)'}\\
        & = & \DA(v_{(1)}(f) ) \DA(v_{(2)})\\
        & \stackrel{def}{=} & \DA(v f).\z\Box
\end{array}
\end{equation}
The underlined parts were rewritten using a compatability
relation between the right \A-coaction and the coproduct in \U:
\begin{equation}
v_{(2)}{}^{(1)} \otimes v_{(1)}(f^{(2)'})\, v_{(2)}{}^{(2)'} =
v^{(1)}{}_{(2)} \otimes v^{(2)'}\,\widehat{v^{(1)}{}_{(1)}}(f^{(2)'}).
\end{equation}
Please refer to section~\ref{S:RaLP} for the definition of the right
projector ``$\widehat{\x}$''.

\subsubsection{\underline{Covariance of:
$\partial _{i} f = \partial _{i}(f) + L_{i}{}^{j}(f) \partial _{j}$}}

See section~\ref{covdfl}.
The main result was the following condition on $L_{i}{}^{j}$:
\begin{equation}
\left( L_{i}{}^{j}(f^{(2)'}) S^{2}t^{k}{}_{j} - S^{2}t^{l}{}_{i}
\widehat{L_{l}{}^{k}}(f^{(2)}) \right)
\otimes f^{(1)} = 0.
\end{equation}

\subsubsection{\underline{Covariance of: $J_{a}^{i} L_{i}{}^{k}(f) =
O_{a}{}^{b}(f) J_{b}^{k}$}}

This proof is somewhat involved and we should keep in mind
that equation $v_{a} f = v_{a}(f) + O_{a}{}^{b}(f) v_{b}$ is already based on
$\DA$ being an algebra homomorphism; nevertheless, in several
steps:
\paragraph{$\DA$ is a homomorphism of}$\fum\cross\tqm$.
Proof on a function $f$:
\begin{equation}
\begin{array}{rcl}
\DA(J_{a}^{i} \partial _{i} f)
        & = & \DA(v_{a} f)\\
        & = & v_{a}^{(1)}f^{(1)} \otimes v_{a}^{(2)'}f^{(2)'}\\
        & = & <\underline{v_{a}^{(1)}, S t^{k}{}_{l}}> x^{l} \partial _{k}
f^{(1)}
                \otimes \underline{v_{a}^{(2)'}} f^{(2)'}\\
        & = & <v_{a}, S t^{s}{}_{r}> x^{l} \partial _{k} f^{(1)} \otimes S
t^{r}{}_{l} S^{2}t^{k}{}_{s} f^{(2)'}\\
        & = & \underline{x^{l}} \partial _{k} f^{(1)} \otimes \underline{S
t^{r}{}_{l}
                <v_{a},S t^{s}{}_{r}>} S^{2}t^{k}{}_{s} f^{(2)'}\\
        & = & \DA J_{a}{}^{s} \,\DA(\partial _{s} f),\z\Box
\end{array}
\end{equation}
and also
\begin{equation}
\begin{array}{rcl}
\DA(J_{a}^{i} \partial _{i}( f) )
        & = & x^{j} \partial _{i}(f^{(1)}) \otimes S t^{s}{}_{j}<v_{a},S
t^{r}{}_{s}> S^{2}t^{i}{}_{r} f^{(2)'}\\
        & = & \DA J_{a}{}^{r}\, \DA(\partial _{i})\left(\DA f\right)\\
        & = & \DA J_{a}^{r}\, \DA\left(\partial _{i}(f)\right).\z\Box
\end{array}
\end{equation}
\paragraph{A short aside,} checking consistency of $O_{a}{}^{b}(f)J_{b}^{k}$
with
$\DA$ being an algebra homomorphism of $\fum$.
\begin{equation}
\begin{array}{rcl}
\underline{\DA(O_{a}{}^{b}(f) J_{b}^{i})}\, \DA(\partial _{i})
        & \stackrel{def}{=} & \DA(O_{a}{}^{b}(f) J_{b}^{i} \partial _{i})\\
        & = & \DA(O_{a}{}^{b}(f) v_{b})\\
        & = & \DA(O_{a}^{b}(f))\,\DA(v_{b})\\
        & = & \DA(O_{a}{}^{b}(f))\,\DA(J_{b}^{i} \partial _{i})\\
        & \stackrel{def}{=} &
        \underline{\DA(O_{a}{}^{b}(f))\,\DA(J_{b}^{i})}\,\DA(\partial
_{i}).\z\Box
\end{array}
\end{equation}
\paragraph{Synthesis:} Comparing
$$v_{a} f = v_{a}(f) + O_{a}{}^{b}(f) v_{b}$$ and $$J_{a}^{i} \partial _{i} f
= J_{a}^{i} \partial _{i}(f) + J_{a}^{i} L_{i}{}^{j}(f) \partial _{j}$$ we
finally find:
\begin{equation}
\begin{array}{rcl}
\DA(J_{a}^{i} L_{i}{}^{k}(f) )
        & = & \DA(J_{a}^{i})\,\DA(L_{i}{}^{k}(f) )\\
        & = & \DA(O_{a}{}^{b}(f) J_{b}{}^{k} )\\
        & = & \DA(O_{a}{}^{b}(f) )\,\DA(J_{b}{}^{k}).\z\Box
\end{array}
\end{equation}
{\em Remark:} Given a linear operator $L_{i}{}^{j}:\fum \to  \fum$, satisfying
the appropriate consistency conditions, --- equation
\begin{equation}
J_{a}^{i} L_{i}{}^{k}(f) = O_{a}{}^{b}(f) J_{b}{}^{k}
\end{equation}
could very well be used to give explicit covariant $x-x$ commutation
relations.

\section{Cartan Calculus on Quantum Planes}

So far we have only dealt with functions and (partial) derivatives
that we combined into an algebra of differential operators on the
quantum plane via commutation relations
\begin{equation}
\partial _{i} f = \partial _{i}(f) + L_{i}{}^{j}(f)
\partial _{j},\z \partial _{i}\in\tqm, \, f\in\fum.
\end{equation}
Now we would like to construct  differential forms through an
exterior derivative $\dl:\fum \to  \Lambda ^{1}(\fum)$ that is nilpotent
and satisfies the usual graded Leibniz rule.
Lie derivatives are introduced next, recalling that
they {\em act} on functions like the ordinary derivatives, that they
correspond to $\Li_{\partial _{i}}(f) = \partial _{i}(f)$, and requiring that
they
commute with the exterior derivative $\Li_{\partial _{i}} \circ \dl = \dl \circ
\Li_{\partial _{i}}$. Just like it was the case for quantum Lie algebras, the
linear operator $L_{i}{}^{j}$  should also act like a Lie derivative, $i.e.$
we extend its definition from functions to forms by requiring that
it commute with $\dl$. Inner derivations $\I_{\partial _{i}}$ are defined as
graded linear
operators of degree -1 orthogonal to the natural basis $\xi ^{i} := \dl(x^{i})$
of 1-forms: $\I_{\partial _{i}}(\xi ^{j}) = \delta _{i}^{j}$ --- in consistency
with the Cartan identity
\begin{equation}
\Li_{\partial _{i}} = \I_{\partial _{i}} \dl + \dl \I_{\partial _{i}}
\end{equation}
that we want to postulate. For the exterior derivative of a function
we can choose between two expansions in terms of 1-forms
\begin{equation}
\dl(f) = \xi ^{i} a_{i} = b_{i} \xi ^{i}
\end{equation}
that we contract with $\I_{\partial _{j}}$ to find
\begin{equation}
\partial _{j}(f) = a_{j} = \I_{\partial _{j}}(b_{i} \xi ^{i})
\end{equation}
and
\begin{equation}
\dl(f) = \xi ^{i} \partial _{i}(f).
\end{equation}
The second expression has to wait while we
quickly  derive $x-\xi $-commutation
relations with the help of the first expression and the Leibniz
rule for $\dl$:
\begin{equation}
\begin{array}{rcl}
\dl f   & = & \xi ^{i}\partial _{i} f\\
        & = & \xi ^{i}\partial _{i}( f) + \xi ^{i} L_{i}{}^{j}(f) \partial
_{j}\\
        &= \dl(f) + f \dl & = \xi ^{i}\partial _{i}(f) + f \xi ^{j} \partial
_{j},
\end{array}
\end{equation}
valid on any function and hence
\begin{equation}
f \xi ^{j} = \xi ^{i} L_{i}{}^{j}(f),\label{fxico}
\end{equation}
so that the second expression takes the (not so pretty) form
\begin{equation}
\dl(f) = \left(S L_{i}{}^{j} \circ \partial _{j}\right)(f),
\end{equation}
which, unlike in the quantum group case, does not simplify any further.
Lie derivatives and inner derivations along arbitrary first order
differential operators $f^{i} \partial _{i},\x f^{i} \in \fum$ are introduced
by the
following set of consistent equations:
\begin{eqnarray}
\I_{f^{i} \partial _{i}} & = & f^{i} \I_{\partial _{i}},\\
\Li_{f^{i} \partial _{i}} & = & \dl  \I_{f^{i} \partial _{i}} + \I_{f^{i}
\partial _{i}} \dl,\\
\Li_{f^{i} \partial _{i}} & = & f^{i} \Li_{\partial _{i}} + \dl(f^{i})
\I_{\partial _{i}},\\
\Li_{f^{i} \partial _{i}} \dl & = & \dl \Li_{f^{i} \partial _{i}}.
\end{eqnarray}
We will not give a complete set of commutation relations here because
the reader can easily obtain most of them from the quantum group treatment
simply by replacing $\Lio{i}{j} \to  L_{i}{}^{j}$. The problem of defining a
Lie bracket of vector fields on the quantum {\em plane} has,
however, not found a satisfactory solution yet.

\section{Induced Cartan Calculus}

We would like to complete the program started in section~\ref{indcal},
where we induced a calculus on the plane from the calculus on the
symmetry quantum group of that plane using a realization
$v_{a} \doteq J_{a}^{i} \partial _{i}$ of the
bicovariant group generators in terms of functions and derivatives on the
plane. From this expression we get the following two relations for
the Cartan generators on the   plane:
\begin{eqnarray}
\I_{v_{a}}  & \doteq & \I_{J_{a}^{i} \partial _{i}}  =  J_{a}^{i} \I_{\partial
_{i}}\\
\Li_{v_{a}} & \doteq & \Li_{J_{a}^{i} \partial _{i}} = J_{a}^{i} \Li_{\partial
_{i}} + \dl(J_{a}^{i}) \I_{\partial _{i}}.
\end{eqnarray}
Commutation relations for the inner derivation with functions are
easily derived;
\begin{equation}
\I_{v_{a}} f = \Lio{a}{b}(f) \I_{v_{b}}
\end{equation}
and hence
\begin{equation}
J_{a}^{i} \I_{\partial _{i}} f = \Lio{a}{b}(f) J_{b}^{k} \I_{\partial _{k}}
\end{equation}
or, if a $\tilde{J}^{a}_{i}$ exists,
\begin{equation}
\I_{\partial _{i}} f = \tilde{J}^{a}_{i} \Lio{a}{b}(f) J_{b}^{k} \I_{\partial
_{k}},
\end{equation}
and finally
\begin{equation}
\I_{\partial _{i}} f = L_{i}{}^{k}(f) \I_{\partial _{k}}.
\end{equation}
Commutation relations for the Lie derivatives with functions can
now be calculated using the Cartan identity. We will present the
result of such a
computation for Wess-Zumino type linear planes (where $\tilde{J}^{a}_{i}$
exists):
\begin{equation}
\begin{array}{rcl}
\Li_{\partial _{i}} x^{l} & = & \delta _{i}^{l} + \underbrace{\tilde{J}^{a}_{i}
O_{a}{}^{b}(x^{l}) J_{b}^{k}}_{L_{i}{}^{k}(x^{l})} \Li_{\partial _{k}}\\
         &   & +\left( \dl(\tilde{J}^{a}_{i} O_{a}{}^{b}(x^{l}) J_{b}^{k}) -
                \tilde{J}^{a}_{i} \dl(O_{a}{}^{b}(x^{l})) J_{b}^{k} \right)
                \I_{\partial _{k}}.\label{innozero}
\end{array}
\end{equation}
Classically: $O_{a}{}^{b}(x^{l}) \to  \delta _{a}^{b} x^{l}$ and functions
commute with functions and
forms so that the last term in the above equation vanishes.
The quantum case has a little surprise for us:
As was first discovered by Prof. Zumino through purely
algebraic considerations in the case of the
GL${}_{q}(2)$-plane, an inner derivation term is necessary in the
$\Li_{\partial }-x$ commutation relations in order to get consistency with
the undeformed Cartan identity.
Let us illustrate this at our standard example.

\paragraph{Cartan Calculus for the 2-dimensional Quantum Plane.}
Using $x-\dl(x)$ commutation relations from (\ref{fxico})
\begin{eqnarray}
x \dl(x) & = & q^{2} \dl(x) x,\\
x \dl(y) & = & (q^{2}-1) \dl(x) y + q \dl(y) x,\\
y \dl(x) & = & q \dl(x) y,\\
y \dl(y) & = & q^{2} \dl(y) y,
\end{eqnarray}
we obtain
\begin{eqnarray}
\Li_{\partial _{x}} x & = & 1 + q^{2} x \Li_{\partial _{x}} + (q^{2}-1) y
\Li_{\partial _{y}}
                + q \lambda  \dl(x) \I_{\partial _{x}} + \lambda ^{2} \dl(y)
\I_{\partial _{y}},\\
\Li_{\partial _{x}} y & = & q y \Li_{\partial _{x}} + \lambda  \dl(y)
\I_{\partial _{x}}, \\
\Li_{\partial _{y}} x & = & q x \Li_{\partial _{y}} + \lambda  \dl(x)
\I_{\partial _{y}} ,\\
\Li_{\partial _{y}} y & = & 1 + q^{2} y \I_{\partial _{y}} + q \lambda  \dl(y)
\I_{\partial _{y}},
\end{eqnarray}
directly from (\ref{innozero}) after a lengthy computation.
Alternatively, we could have started with $\I_{\partial }-x$ commutation
relations
\begin{eqnarray}
\I_{\partial _{x}} x & = & q^{2} x \I_{\partial _{x}} + (q^{2}-1) y
\I_{\partial _{y}}, \\
\I_{\partial _{x}} y & = & q y \I_{\partial _{x}},\\
\I_{\partial _{y}} x & = & q x \I_{\partial _{y}} ,\\
\I_{\partial _{y}} y & = & q^{2} y \I_{\partial _{y}},
\end{eqnarray}
which have the great advantage that they have the exact same form
as the well-known $\partial -x$ relations. This also means that all of
our covariance considerations are still valid here.

\chapter{A Torsion-free Tangent Bundle for SU${}_{q}$(2)}

%

\section*{Introduction}

In the classical theory of Lie groups one can introduce a tangent bundle
over the group manifold. There are two natural choices for the connection:
Either one imposes the condition of zero curvature and then chooses a
vanishing connection in an appropriate gauge ---  such that the torsion
is given by the RHS of the Cartan-Maurer equation --- or one can attempt
to set the torsion equal to zero to obtain a (Riemannian or G-Structure
type) non-vanishing
curvature. The first scenario generalizes quite easily to the quantum group
case. In this chapter we will try to generalize the more interesting
case of vanishing torsion at the example of $SU_{q}(2)$.

To establish notation, a
review (including some additional relevant material) of the theory
of quantum Lie algebras is given in the next section, followed by
the description of a tangent bundle structure over a quantum group.
We then elaborate on the example of $SU_{q}(2)$ giving all R-matrices
and structure constants explicitly.

\section{Quantum Lie Algebras}

Quantum Lie Algebras are Hopf algebras \uqg\ that contain a finite-dimensional
sub vector space that closes under left and right coactions. Let
$\{ e_{i} \}$ be a linear basis of generators for this space\footnote{In
this chapter we will not consider a linear basis of the whole Hopf
algebra so there should not be any confusion from this notation.}
and $\{e^{j}\}$ a dual basis of 1-forms corresponding to a set of
functions $b^{j} \in \fun$ via $e^{j} \equiv S b^{j}_{(1)} \dl b^{j}_{(2)}$:
\begin{eqnarray}
\AD(e_{i}) &=& 1\otimes e_{i}, \nonumber \\
\DA(e_{i}) &=& e_{j} \otimes T^{j}{}_{i},\x T^{j}{}_{i}\in\fun,\\
\I_{e_{i}}(e^{j}) &  = & -<e_{i},S b^{j}> = \delta ^{j}_{i},\label{tebdual}\\
\AD(e^{i}) &=& 1\otimes e^{i},\\
\DA(e^{i}) &=& e^{j} \otimes S^{-1} T^{i}{}_{j}.
\end{eqnarray}
The exterior derivative on functions can be expressed in terms of these
bases as
\begin{equation}
\dl(a) = e^{i} (e_{i} \tr a) = e^{i} \Li_{e_{i}}(a).
\end{equation}
The Leibniz rule for $\dl$ requires that the generators
$\{ e_{i} \}$ have a coproduct of the form
\begin{equation}
\Delta (e_{i}) = e_{i} \otimes 1 + \theta _{i}{}^{j} \otimes e_{j}.
\end{equation}
A Cartan calculus can be introduced on these quantum Lie algebras
with equations like
\begin{eqnarray}
\Li_{e_{i}} \alpha
& = & \Li_{e_{i}}(\alpha ) + \Li_{\theta _{i}{}^{j}}(\alpha )\Li_{e_{j}}\\
\I_{e_{i}} \alpha & = & \I_{e_{i}}(\alpha ) + (-1)^{p}
\Li_{\theta _{i}{}^{j}}(\alpha ) \I_{e_{j}}\\
\Li_{e_{i}} & = & \dl \I_{e_{i}} + \I_{e_{i}} \dl\\
e^{i} & = & S b^{i}_{(1)} \dl b^{i}_{(2)} =: e_{b^{i}},
\end{eqnarray}
where $\alpha $ is a p-form, for a more complete list see
section~\ref{S:SoRCC}.
As in the classical case we make an ansatz for
the product of two forms in terms of tensor products
\begin{equation}
e^{i} \wedge e^{j} = e^{i} \otimes e^{j} - \hat{\sigma }^{ij}{}_{mn}
e^{m} \otimes e^{n},
\end{equation}
with as yet unknown numerical constants $\hat{\sigma }^{ij}{}_{mn} \in
k$ and define $\I_{e_{i}}$ to act on this product by contracting
in the first tensor product space. This leads to the following
explicit expression for $\hat{\sigma }^{ij}{}_{mn}$:
\begin{equation}
\hat{\sigma }^{ij}{}_{mn} = <S^{-1}\theta _{m}{}^{j},T^{i}{}_{n}>
\end{equation}
and, in a particular example that we will need later,
\begin{equation}
\begin{array}{rcl}
\dl e^{j} \equiv \dl e_{b^{j}}
& = & - e_{b^{j}_{(1)}} \wedge e_{b^{j}_{(2)}}\\
& = & -e_{S^{-1}(Sb^{j}_{(1)} b^{j}_{(3)})} \otimes e_{b^{j}_{(2)}}\\
& = & -e^{k} \otimes
e^{l} <-Se_{k},S^{-1}(Sb^{j}_{(1)} b^{j}_{(3)})><-Se_{l},b^{j}_{(2)}>\\
& = & -e^{k} \otimes e^{l} <\underbrace{(S^{-1}e_{k})_{(1)} e_{l}
S(S^{-1}e_{k})_{(2)}}_{S^{-1}e_{k} \ad e_{l}},Sb^{j}>.
\end{array}\label{de}
\end{equation}
In the previous equation we have introduced the adjoined action of
a left-invariant vector field on another vector field. A short
calculation gives
\begin{equation}
S^{-1}e_{k} \ad e_{l}
= e_{b} e_{c} (\delta ^{c}_{k}\delta ^{b}_{l} - \hat{\sigma }^{cb}{}_{kl})
= e_{a} <S^{-1}e_{k},T^{a}{}_{l}>=  e_{a} f'{}_{k}{}^{a}{}_{l}
\end{equation}
and similarly
\begin{equation}
e_{k} \ad e_{l}
\equiv \Li_{e_{k}}(e_{l}) = e_{b} e_{c} (\delta ^{c}_{k}\delta ^{b}_{l} -
\hat{R}^{cb}{}_{kl})
= e_{a} <e_{k},T^{a}{}_{l}> = e_{a} f_{k}{}^{a}{}_{l},
\end{equation}
where
\begin{equation}
\hat{R}^{cb}{}_{kl} = <\theta _{k}{}^{b},T^{c}{}_{l}>
\end{equation}
is the so-called ``big R-matrix'' related to $\sigma $\footnote{The Hat
``$\hat{\x}$'' denotes the action of the permutation matrix
$P^{ij}{}_{kl} =
\delta ^{i}_{l}\delta ^{j}_{k}$, {\it i.e.} $\hat{\sigma } \equiv P \sigma $.}
by
\begin{equation}
\sigma ^{ij}{}_{kl} R^{kl}{}_{mn} = \delta ^{i}_{m}\delta ^{j}_{n}.
\end{equation}
A little more work gives
\begin{equation}
f_{m}{}^{a}{}_{n} = -f'_{k}{}^{a}{}_{l} R^{kl}{}_{mn}.\label{ftofp}
\end{equation}
Were we to impose zero curvature now
and chose a vanishing connection, then the right
hand side of equation (\ref{de}) would give the torsion two form.

The calculus on quantum Lie algebras is by construction
covariant under left and right coactions. It has however
a closely related additional symmetry:
All equations that we have given are invariant under
linear changes of the bases ${e_{i}}$ and ${e^{j}}$:
\begin{equation}
e_{i} \rightarrow \chi _{i} = e_{l} M^{l}{}_{i},\z
e^{i} \rightarrow \tau ^{i} = (M^{-1})^{i}{}_{l} e^{l},\z M \in M_{N}(k).
\end{equation}
The adjoined matrix
representation $T$ and the braiding operator $\theta $ transform as expected
under this change of basis
\begin{equation}
T^{i}{}_{j} \rightarrow
\bar{T}^{i}{}_{j} = (M^{-1})^{i}{}_{l} T^{l}{}_{m} M^{m}{}_{j},
\end{equation}
\begin{equation}
\theta _{i}{}^{j} \rightarrow
\bar{\theta }_{i}{}^{j} = (M^{-1})^{i}{}_{l} \theta _{m}{}^{l} M^{m}{}_{j},
\end{equation}
such that now
\begin{equation}
\DA(\chi _{i}) = \chi _{j}
\otimes \bar{T}^{j}{}_{i}
\z \DA(\tau ^{i}) = \tau ^{j} \otimes S^{-1}\bar{T}^{j}{}_{i}
\end{equation}
and
\begin{equation}
\Delta \bar{T} =
\bar{T} \otimes \bar{T},\x \epsilon \bar{T} = I,\x S \bar{T} = \bar{T}^{-1},
\end{equation}
{\it i.e.} $\bar{T}$ (like $T$) satisfies the appropriate relations for
a matrix representation of \uqg.

\section{Quantum Tangent Bundle, Torsion, Curvature}

In this chapter we are  going to use a formulation \cite{SeUr} of the theory
of fiber bundles where all forms are pulled back to the base
manifold. This formulation is well suited for the generalization
to quantum groups because it makes it easier to keep track of
subtle distinctions between the calculi of base vs. fiber.

The base manifold in the problem under consideration is a quantum group,
implicitly defined by the Hopf algebra of functions \fun\ on it.
The typical fiber
of the tangent bundle is the invariant space span$\{e_{i}\}$,
i.e. the ``quantum
Lie algebra''.  We chose a basis $\{\chi _{i}\}$ of sections on the tangent
bundle and consider ``pointwise'' infinitesimal transformations
within the fiber along elements $A_{\mu }$
of \uqg
\begin{equation}
A_{\mu }\tr \chi _{i} = A_{\mu _{(1)}} \chi _{i} SA_{\mu _{(2)}}
= \chi _{j}<A_{\mu },\bar{T}^{i}{}_{j}>,\label{invtrans}
\end{equation}
where we have used $\DA\chi _{i} = \chi _{j} \otimes \bar{T}^{i}{}_{j}$.
In order to justify the word ``infinitesimal'' the $A_{\mu }$ should be
linear combinations of the $e_{i}$ and possibly
$S^{-1}e_{i}$\footnote{Higher powers of $S$ do  not result in
new generators in the example under consideration in the next section.}.
These heuristic considerations suggest that the connection
1-form should have the following form
\begin{equation}
\om = e^{\mu }A_{\mu },\z \om^{j}{}_{i} = e^{\mu }<A_{\mu },\bar{T}^{j}{}_{i}>
\end{equation}
which enters in the expression of the covariant derivative $\nabla $ on
the section basis:
\begin{equation}
\nabla \chi _{i} = \chi _{j} \otimes \om^{j}{}_{i}.\label{covder}
\end{equation}
This equation is basically a reformulation of (\ref{invtrans})
in differential form language and equation (\ref{covder}) replaces
the metricity condition on $\om$ in the sense of G-structures: In
the classical theory we construct classes of G-bases  fixing
one orthogonal basis $\{\chi _{i}\}$ and getting all other
orthogonal bases by transforming $\{\chi _{i}\}$ by a Lie
subgroup of the general linear group. For quantum groups
we choose transformation matrices of the form $<x,\bar{T}>$.
Later we will come back to the question which metric
--- if any --- is preserved by said transformations.
Using properties of $\nabla $
like
\begin{eqnarray}
\nabla (\chi + \psi ) &=& \nabla \chi + \nabla \psi ,\\
\nabla (f \psi ) &=& \dl f \otimes f \nabla \psi ,\z f \in \fun,\\
\nabla _{f u + v}\psi &=& f \nabla _{u} \psi +
\nabla _{v} \psi ,\z \nabla _{u}\psi \equiv \I_{u}(\psi ),
\end{eqnarray}
we can easily calculate the covariant
derivative of an arbitrary section $\psi = \chi _{i} \psi ^{i}$:
\begin{equation}
\nabla \psi = \chi _{i} \otimes (\nabla \psi )^{i}
= \chi _{i} \otimes (\dl \psi ^{i}
+ \om^{i}{}_{j} \psi ^{j}).
\end{equation}
For section-valued p-forms we introduce an exterior covariant
differentiation \DD:
\begin{equation}
\DD(\psi \otimes \al) := \nabla \psi \wedge \al + \psi \otimes \dl\al
\end{equation}
in accordance to the  {\em undeformed} Leibniz rule.

The last ingredient, enabling us to define torsion, is the
fusion form $\eta = \chi _{i} \otimes e^{i}$, viewed as a section valued
1-form.
It effectively identifies elements in the fibers of the
tangent bundle with the
tangent space over the points of the base manifold. One usually
takes the canonical element $e_{i} \otimes e^{i}$ as a natural choice for
the fusion form, but $\eta = \chi _{i} \otimes e^{i} =
e_{l} M^{l}{}_{i} \otimes e^{i}$, where $M^{l}{}_{i}$
is a constant numerical matrix that may however
differ from $\delta ^{l}{}_{i}$, is  also a mathematically acceptable
description and will in fact be quite important in the
quantum case as we shall see.
The torsion 2-form $\Theta $ is defined as the exterior covariant
derivative of the fusion form
\begin{equation}
\begin{array}{rcl}
\Theta = \DD \eta &  = &  \nabla \chi _{i} \wedge e^{i}
+ \chi _{i} \otimes \dl e^{i} \\
& = & \chi _{j} \otimes (\om^{j}{}_{i}\wedge e^{i} + \dl e^{j})\\
& =: & \chi _{j} \otimes \Theta ^{j}.
\end{array}
\end{equation}
We will later try to set $\Theta = 0$.
The curvature 2-form of a section $\psi $ is $\Omega
= \DD \nabla \psi $, i.e. the exterior covariant derivative
of the section valued 1-form $\nabla \psi $.
In terms of the section basis we find
\begin{equation}
\begin{array}{rcl}
\DD\nabla \chi _{i} & = & \DD(\chi _{j} \otimes \om^{j}{}_{i})\\
& = & \chi _{k} \otimes (\om^{k}{}_{j} \wedge \om^{j}{}_{i}
+ \dl \om^{k}{}_{i})\\
& =: & \chi _{k} \otimes \Omega ^{k}{}_{i}.
\end{array}
\end{equation}
The Ricci tensor can also be defined in this context:
\begin{equation}
e^{\mu }R_{\mu i} := \I_{e_{k}} \Omega ^{k}{}_{i}.
\end{equation}
For simple Lie algebras it has the particularly simple form of
the Killing metric times a constant.

Using tools from the previous section we can expand the torsion 2-form
in terms of tensor products
\begin{eqnarray}
\dl e^{j} & = & -e^{k} \otimes e^{l} f'_{k}{}^{j}{}_{l},\\
\om^{j}{}_{i}\wedge e^{i} & = & \om_{\mu }{}^{j}{}_{i} e^{\mu }\wedge e^{i} =
\om_{\mu }{}^{j}{}_{i} e^{k} \otimes e^{l}(\delta ^{\mu }_{k}\delta ^{i}_{l} -
\hat{\sigma }^{\mu i}{}_{kl})
\end{eqnarray}
and the condition of zero torsion becomes
\begin{equation}
\om_{\mu }{}^{j}{}_{i} (\delta ^{\mu }_{k}\delta ^{i}_{l} -
\hat{\sigma }^{\mu i}{}_{kl}) = f'_{k}{}^{j}{}_{l}.
\end{equation}
This is a set of linear equations for $\om_{\mu }{}^{j}{}_{i}$ with non-trivial
null space, i.e. we will get a solution $v_{\mu }{}^{j}{}_{i}$ and  vectors
$N_{\mu }{}^{a}{}_{i}$ with $N_{\mu }{}^{a}{}_{i}
(\delta ^{\mu }_{k}\delta ^{i}_{l} -
\hat{\sigma }^{\mu i}{}_{kl}) = 0$ such that
\begin{equation}
\om_{\mu }{}^{j}{}_{i} = v_{\mu }{}^{j}{}_{i} +
\sum _{a} n^{j}_{a} N_{\mu }{}^{a}{}_{i},\x n^{j}_{a}\in k. \label{lineq}
\end{equation}
To decide whether it is possible to find an
$\om_{\mu }{}^{j}{}_{i}$ that satisfies all conditions, in particular
\begin{equation}
\om_{\mu }{}^{j}{}_{i} \stackrel{?}{=} <A_{\mu }, \bar{T}^{j}{}_{i}>,
\end{equation}
it is now instructive to look  at the concrete example of $SU_{q}(2)$.

\section{Example SU${}_{q}(2)$}

\dots or $Sl_{q}(2)$ if one modifies the reality condition. Recall
\cite{RTF}, \cite{Z2} the commutation relations
for $SU_{q}(2)$, here written in compact matrix notation as
\begin{eqnarray}
r_{12}t_{1}t_{2} = t_{2}t_{1}r_{12}, & det_{q}t
= 1, & t^{\dagger } = t^{-1},
\nonumber \\
\Delta (t) = t \dot{\otimes} t, & \epsilon (t) = I, & S(t)=t^{-1},
\end{eqnarray}
where $t \in M_{n}(Fun(SU_{q}(2)))$ and $r$ is the ``small'' r-matrix
\begin{equation}
r = <\R,t_{1} \otimes t_{2}> = {1\over\sqrt {q}}\left(
\matrix{ q & 0 & 0 & 0 \cr 0 & 1 & 0 & 0 \cr 0
    &{q-{1\over q}}& 1& 0 \cr
    0 & 0 & 0 & q \cr  }\right).\end{equation}
The deformed universal enveloping algebra $U_{q}su(2)$,
dual to $Fun(SU_{q}(2))
$, is generated by operators $H$, $X_{+}$, $X_{-}$ satisfying
\begin{eqnarray}
\comm{H}{X_{\pm }}= \pm 2 X_{\pm }, & \comm{X_{+}}{X_{-}}
= \frac {q^{H}-q^{-H}}{ q-q^{-1}},
\nonumber \\
\Delta (H) = H \otimes 1 + 1 \otimes H, & \Delta (X_{\pm }) =
X_{\pm } \otimes q^{H/2} + q^{-H/2} \otimes X_{\pm }, \nonumber \\
\epsilon (H)= \epsilon (X_{\pm })=0, \\
S(H) = -H, & S(X_{\pm }) = -q^{\pm 1} X_{\pm }. \nonumber
\end{eqnarray}
Following \cite{RTF} these relations can be rewritten as
\begin{eqnarray}
r_{12}L^{\pm }_{2}L^{\pm }_{1}=L^{\pm }_{1}L^{\pm }_{2}r_{12},&
r_{12}L^{+}_{2}L^{-}_{1}=L^{-}_{1}L^{+}_{2}r_{12},\nonumber \\
\Delta (L^{\pm }) = L^{\pm } \dot{\otimes} L^{\pm }, & \epsilon (L^{\pm })=I,
\\
S(L^{\pm })=(L^{\pm })^{-1}, \nonumber
\end{eqnarray}
where $L^{\pm }$ are given by
\begin{equation}
L^{+} =  <\R,i\!d \otimes t> = \left( \begin{array}{lr}
q^{-H/2} &{1\over\sqrt {q}}\lambda X_{+} \\ 0 & q^{H/2}
\end{array} \right)
\end{equation}
and
\begin{equation}
 L^{-} = <\R,S t \otimes i\!d> =
\left( \begin{array}{lr} q^{H/2} & 0 \\ -\sqrt {q} \lambda X_{-} & q^{-H/2}
\end{array} \right),
\end{equation}
where $\lambda \equiv q - {1\over q}$.
Unitarity of $T$ implies $(L^{+})^{\dagger }=
(L^{-})^{-1}$, {\em i.e.} $\bar{H} = H$, $\overline{X_{\pm }}=X_{\mp }$.

Following the method described in section~\ref{Braid} we can construct a matrix
of bicovariant generators corresponding to an element $1\otimes 1 - \R^{21}\R$
of the ``pure braid group'':
\begin{equation}
\left(\begin{array}{cc} e_{1} & e_{2} \\ e_{3} & e_{4} \end{array}
\right) := {1\over \lambda }<1\otimes 1 - \R^{21}\R, t \otimes i\!d>
= {1\over \lambda } L^{+} SL^{-} =: X.
\end{equation}
The right coaction is then
\begin{equation}
\DA X^{i}{}_{l} = X^{j}{}_{k} \otimes S t^{i}{}_{j} t^{k}{}_{l},
\end{equation}
so that span$\{e_{i}\}$ forms an invariant subspace as required.
$c:=e_{1} + q^{-2} e_{4}$ by the way is the casimir.
The functions $b^{i} \in \fun$ see equation
(\ref{tebdual}) can be chosen as linear combinations
of the elements of $t$ \cite{SWZ2} because $t$
(and $S t$) form faithful (anti)representations of the $e_{i}$s.
Classical commutators become adjoint actions
$$e_{k} \ad e_{l} := e_{k_{(1)}} e_{l} S e_{k_{(2)}}
= e_{b} e_{c} (\delta ^{c}_{k}\delta ^{b}_{l} -
\hat{R}^{cb}{}_{kl})
= e_{a} f_{k}{}^{a}{}_{l},$$
where the $\hat{R}$ and $f$ can be calculated \cite{SWZ3} from $r$ (see
section~\ref{S:RMA})
\begin{equation}
\hat{R}^{(mn)(kl)}{}_{(ij)(pq)} = \left(({r_{31}}^{-1})^{T_{3}} r_{41} r_{24}
({r_{23}}^{T_{3}})^{-1}\right)^{ilmn}{}_{kjpq}
\end{equation}
and
\begin{equation}
f_{k}{}^{a}{}_{l} = {1\over \lambda }\left(I_{k} \delta ^{a}_{l} -
\sum _{i} \hat{R}^{a (ii)}{}_{k l}\right).
\end{equation}
Explicitly:\x $f^{a}{}_{(kl)}$
\begin{equation}
\matrix{ {{1 - {q^{2}}}\over {{q^{3}}}}& 0 & 0 & -{1\over q} + q & 0 & 0 &
   {1\over q} & 0 & 0 & -{1\over q} & 0 & 0 & 0 & 0 & 0 & 0 \cr 0 &
  {{1+q^{2}-q^{4}}\over{q^{3}}}& 0 & 0 & -{1\over q} & 0 & 0 & q & 0 & 0 & 0 &
0
    & 0 & -q & 0 & 0 \cr 0 & 0 & -q & 0 & 0 & 0 & 0 & 0 & {q^{-3}}
    & 0 & 0 &   -{1\over q} & 0 & 0 & {1\over q}
    & 0 \cr {{-1 + {q^{2}}}\over {{q^{3}}}} & 0 &
   0 & {1\over q} - q & 0 & 0 & -{1\over q} & 0 & 0 & {1\over q} & 0 & 0 & 0
& 0 & 0 & 0 \cr  }
\end{equation}
and $f'^{a}{}_{(kl)}$,
\begin{equation}
\matrix{ {{1 - {q^{2}}}\over {{q^{3}}}} & 0 & 0 & -{1\over q} + q & 0 & 0 &
   -{1\over q} & 0 & 0 & {1\over q} & 0 & 0 & 0 & 0 & 0 & 0 \cr 0 & -q & 0 & 0
    & {q^{-3}} & 0 & 0 & -{1\over q} & 0 & 0 & 0 & 0 & 0 & {1\over q} & 0 & 0
    \cr 0 & 0 & {{1+q^{2}-q^{4}}\over{q^{3}}}& 0
    & 0 & 0 & 0 & 0 & -{1\over q} &
   0 & 0 & q & 0 & 0 & -q & 0 \cr {{-1 + {q^{2}}}\over {{q^{3}}}} & 0 & 0 &
   {1\over q} - q & 0 & 0 & {1\over q} & 0 & 0 & -{1\over q} & 0 & 0 & 0 & 0
    & 0 & 0 \cr  }
\end{equation}
obtained by similar methods. In both matrices rows are labeled
by $a\in\{1,\dots,4\}$ and columns are labeled by
$(kl) \in \{(1,1),(1,2),\ldots,(4,4)\}$.

Using the explicit expressions for $\hat{\sigma }$
(see appendix) and $f'^{a}{}_{(kl)}$
we find the following particular solutions
$v^{j}{}_{(\mu i)}$ of (\ref{lineq}):
\begin{equation}
\matrix{ 0&0&0&0&0&0&{1\over q}&0&0&0&0&0&0&0&0&0\cr
   0&{1\over q}&0&0&0&0&0&q&0&0&0&0&0&0&0&0\cr
   0&0&-{q^{-3}}&0&0&0&0&0&0&0&0&-{1\over q}&0&0&0&0\cr
   0&0&0&0&0&0&-{1\over q}&0&0&0&0&0&0&0&0&0\cr}
\end{equation}
The null space of said linear equation, i.e.
of $\hat{\sigma } - I$, is spanned by $N^{a}_{(\mu i)},\x a = 1,\ldots,10$:
\begin{equation}
\matrix{ 0 & 0 & 0 & 0 & 0 & 0 & 0 & 0 & 0 & 0 & 1 & 0 & 0 & 0 & 0 & 0 \cr
0 & 0 & 0 & 0 & 0 & 0 & 1 & 0 & 0 & 1 & 0 & 0 & 0 & 0 & 0 & 0 \cr
0 & 0 & 0 & 0 & 0 & 0 & -1 + {q^{-2}} & 0 & 0 & 0 & 0 & 0 & 0 & 0 & 0 & 1 \cr
0 & 0 & 0 & 0 & 0 & 1 & 0 & 0 & 0 & 0 & 0 & 0 & 0 & 0 & 0 & 0 \cr
0 & 0 & 0 & 1 & 0 & 0 & 1 - {q^{-2}} & 0 & 0 & 0 & 0 & 0 & 1 & 0 & 0 & 0 \cr
0 & 0 & 1 & 0 & 0 & 0 & 0 & 0 & 1 & 0 & 0 & 0 & 0 & 0 & 0 & 0 \cr
0 & 0 & {q^{-4}} - {q^{-2}} & 0 & 0 & 0 & 0 & 0 & 0 & 0 & 0 & {q^{-2}} &
    0 & 0 & 1 & 0 \cr
0 & 1 & 0 & 0 & 1 & 0 & 0 & 0 & 0 & 0 & 0 & 0 & 0 & 0 & 0 & 0 \cr
0 & 1-{q^{-2}} & 0 & 0 & 0 & 0 & 0 & {q^{2}} & 0 & 0 & 0 & 0 & 0 & 1 & 0 & 0
\cr
1 & 0 & 0 & 0 & 0 & 0 & 0 & 0 & 0 & 0 & 0 & 0 & 0 & 0 & 0 & 0 \cr  }
\end{equation}
The fact that there are 10 null vectors shows
by the way that the number of independent 2-forms
is reduced from $4\times 4 = 16$ to $16 - 10 = 6 = 4\times 3/2$
as one would expect.

We will now investigate choices for $\om$ of gradually increasing
complexity starting with
a simple ansatz with $M = I$
\begin{equation}
\om_{\mu }{}^{j}{}_{i} = <A_{\mu }, T^{j}{}_{i}> =
A^{\nu }{}_{\mu }f_{\nu }{}^{j}{}_{i} +
A'^{\nu }{}_{\mu }f'_{\nu }{}^{j}{}_{i}
\end{equation}
corresponding to
\begin{equation}
A_{\mu }= A^{\nu }{}_{\mu }e_{\nu }+
A'^{\nu }{}_{\mu }S^{-1} e_{\nu }.\label{amu}
\end{equation}
In the classical case we would find
$A^{\nu }{}_{\mu }= - A'^{\nu }{}_{\mu }= \frac
{1}{4} \delta ^{\nu }_{\mu }$ as a
solution. Explicit computation shows however that there are no
solutions for $A^{\nu }{}_{\mu }$ and $A'^{\nu }{}_{\mu }$
in the quantum case. Next we try an
ansatz with trivial $A^{\nu }{}_{\mu }$ and $A'^{\nu }{}_{\mu }$ in analogy
to the classical solution, but we allow the basic generators
$e_{\nu }$ and $S^{-1} e_{\nu }$ in (\ref{amu}) to be multiplied by
elements $z, z' \in \uqg$
\begin{equation}
\om_{\mu }{}^{j}{}_{i} = < z e_{\mu }- z' S^{-1} e_{\mu }, T^{j}{}_{i}> =
Z^{j}{}_{k} f_{\mu }{}^{k}{}_{i} + Z'{}^{j}{}_{l} f'{}_{\mu }{}^{l}{}_{i}
\end{equation}
where $Z^{(')}{}^{j}{}_{k} = <z^{(')},T^{j}{}_{k}>$. Hence solving
\begin{equation}
v_{\mu }{}^{j}{}_{i} + n^{j}_{a} N_{\mu }{}^{a}{}_{i} =
Z'{}^{j}{}_{l} f'{}_{\mu }{}^{l}{}_{i} + Z^{j}{}_{k} f_{\mu }{}^{k}{}_{i}
\end{equation}
for $\{n^{j}_{a},Z'{}^{j}{}_{l},Z^{j}{}_{k}\}$ gives
$$
\matrix{
0 & {1\over {2\,q}} & 0 & 0 & 0 & 0 & 0 & 0 & 0 & 0\cr
0 & 0 & 0 & 0 & 0 & 0 & 0 & -{1\over {-q - {q^{3}}}} &
{q\over {1 + {q^{2}}}}
    & 0 \cr
0 & 0 & 0 & 0 & 0 & -{1\over {q + {q^{3}}}} &
-{q\over {1 + {q^{2}}}} & 0 & 0 & 0
    \cr
0 & {{-1}\over {2\,q}} & 0 & 0 & 0 & 0 & 0 & 0 & 0 & 0\cr}\hspace*{5cm}
$$
\begin{equation}
\hspace*{5cm} \matrix{ -{1\over 4} & 0 & 0 & 0 & {1\over 4} & 0 & 0 & 0 \cr
0 & -{{{q^{2}}}\over {{{\left( 1 + {q^{2}} \right) }^{2}}}} & 0 & 0 & 0 &
    {{{q^{2}}}\over {{{\left( 1 + {q^{2}} \right) }^{2}}}} & 0 & 0 \cr
0 & 0 & -{{{q^{2}}}\over {{{\left( 1 + {q^{2}} \right) }^{2}}}} & 0 & 0 & 0
    & {{{q^{2}}}\over {{{\left( 1 + {q^{2}} \right) }^{2}}}} & 0\cr
{1\over 4} & 0 & 0 & 0 & -{1\over 4} & 0 & 0 & 0 \cr
}
\end{equation}
as a particular solution and
\begin{equation}
\matrix{ 0 & 0 & 0 & 0 & 0 & 0 & 0 & 0 & 0 & 0 & 1 & 0 & 0 & 1 & 0 & 0 & 0
    & 0 \cr 0 & 0 & 0 & 0 & 0 & 0 & 0 & 0 & 0 & 0 & 0 & 0 & 0 & 0 & 1 & 0 & 0
    & 1 \cr  }
\end{equation}
as the corresponding null space. The first 10 columns in both
matrices are labeled by $a$, the next 4 columns
are labeled by $k$, and the last 4 are labeled by $l$. $j$ is the row index.
Two comments about the null space are in order: Note that the
first ten columns are zero. This means that $n^{j}_{a}$ and hence
$\om_{\mu }{}^{j}{}_{i}$ are in fact uniquely determined by our ansatz.
Note also that both $f$ and $f'$ and thereby $e_{\mu }$ {\em and}
$S^{-1}e_{\mu }$ were necessary to satisfy the equation.
All that remains is some arbitrariness in the definition of $K$
and $K'$. This actually comes from the existence
of an invariant form $e^{1}+e^{4}$. Being
invariant means $\dl(e^{1}+e^{4}) = 0$ or
$f'{}_{\mu }{}^{1}{}_{i} + f'{}_{\mu }{}^{4}{}_{i}
= 0$; by equation (\ref{ftofp})
the same is true for $f$.
We use this remaining freedom to diagonalize
\begin{equation}
K = -K' = \left(\matrix{ {1\over 4} & 0 & 0 & 0 \cr
0 & {{{q^{2}}}\over {{{\left( 1 + {q^{2}} \right) }^{2}}}} & 0 & 0 \cr
0 & 0 & {{{q^{2}}}\over {{{\left( 1 + {q^{2}} \right) }^{2}}}} & 0\cr
0 & 0 & 0 & {1\over 4}\cr }\right)
\end{equation}
corresponding for instance to
\begin{equation}
z = - z' = \frac {1}{4} + \frac {q(1-q^{2})}{4(1+q^{2})^{2}}(e_{4}
-   S^{-1}e_{4})
\end{equation}
and
\begin{equation}
A_{\mu }= z e_{\mu }.
\end{equation}
If $z$ and $z'$ had been invariant elements (casimirs) then $A_{\mu }$
would have had nice transformation properties. The way it is, the solution
is somewhat unsatisfying. Luckily it turns
out that $z$ can be eliminated without having to change our solution for
$\om$ if we allow for a non-trivial
$M$ matrix. As can be seen by inspection of the explicit forms of
$f$ and $f'$:
\begin{equation}
Z^{j}{}_{l} (f_{\mu }{}^{l}{}_{i} - f'{}_{\mu }{}^{l}{}_{i}) =
(M^{-1})^{j}{}_{k}
M^{\nu }{}_{\mu }M^{h}{}_{i} (f_{\nu }{}^{k}{}_{h} -
f'{}_{\nu }{}^{k}{}_{h}),
\end{equation}
where
\begin{equation}
M = \left(\matrix{ {{{q^{2}}}\over {{{\left( 1 + {q^{2}} \right) }^{2}}}}
& 0 & 0 & 0 \cr
0 & {{{q}}\over {{{2\left( 1 + {q^{2}} \right) }}}} & 0 & 0 \cr
0 & 0 & {{{q}}\over {{{2\left( 1 + {q^{2}} \right) }}}} & 0 \cr
0 & 0 & 0 & {{{q^{2}}}\over {{{\left( 1 + {q^{2}} \right) }^{2}}}}
}\right)
\end{equation}
such that now
\begin{equation}
\om_{\mu }{}^{j}{}_{i} = <\chi _{\mu }- S^{-1}\chi _{\mu }, \bar{T}^{j}{}_{i}>,
\z\mbox{\it i.e. } A_{\mu }= \chi _{\mu }- S^{-1}\chi _{\mu }.
\end{equation}

\section{Appendix}
$\hat{R}^{ij}{}_{kl}$: {\footnotesize
$$\left(
\matrix{
1 & 0 & 0 & 0 & 0 & 0  \cr
0 & 1 - {q^{-2}} & 0 & 0 & 1 & 0 \cr
0 & 0 & 1 - {q^{2}} & 0 & 0 & 0 \cr
1 + {q^{-4}} - {2\over {{q^{2}}}} & 0 & 0 & 2 - {q^{-2}} - {q^{2}} & 0 & 0 \cr
0 & {q^{-2}} & 0 & 0 & 0 & 0 \cr
0 & 0 & 0 & 0 & 0 & 1 \cr
1 - {q^{-2}} & 0 & 0 & 1 - {q^{2}} & 0 & 0 \cr
0 & -1 + {q^{-4}} - {q^{-2}} + {q^{2}} & 0 & 0 & 1 - {q^{-2}} & 0  \cr
0 & 0 & {q^{2}} & 0 & 0 & 0 \cr
-1 + {q^{-2}} & 0 & 0 & -1 + {q^{2}} & 0 & 0 \cr
0 & 0 & 0 & 0 & 0 & 0 \cr
0 & 0 & 0 & 0 & 0 & 0 \cr
0 & 0 & 0 & 1 & 0 & 0 \cr
0 & -1 + {q^{-2}} & 0 & 0 & 0 & 0 \cr
0 & 0 & -1 + {q^{2}} & 0 & 0 & 0 \cr
-1 - {q^{-4}} + {2\over {{q^{2}}}} & 0 & 0 &
   {{\left( -{1\over q} + q \right) }^{2}} & 0 & 0 \cr  }
\right.$$
\begin{equation}
\left.
\matrix{
0 & 0 & 0 & 0 & 0 & 0 & 0 & 0 & 0 & 0 \cr
0 & 0 &  0 & 0 & 0 & 0 & 0 & 0 & 0 & 0 \cr
0 & 0 & 1 & 0 & 0 & 0 & 0 & 0 & 0 & 0 \cr
-1 + q^{-2} & 0 &  0 & 1 - {q^{-2}} & 0 & 0 & 1 & 0 & 0 & 0 \cr
0 & 0 &  0 & 0 & 0 & 0 & 0 & 0 & 0 & 0 \cr
0 & 0 &  0 & 0 & 0 & 0 & 0 & 0 & 0 & 0 \cr
0 & 0 &  0 & 1 & 0 & 0 & 0 & 0 & 0 & 0 \cr
0 & 1 - q^{2} &  0 & 0 & 0 & 0 & 0 & {q^{2}} & 0 & 0 \cr
0 & 0 &  0 & 0 & 0 & 0 & 0 & 0 & 0 & 0 \cr
1 & 0 &  0 & 0 & 0 & 0 & 0 & 0 & 0 & 0 \cr
0 & 0 &  0 & 0 & 1 & 0 & 0 & 0 & 0 & 0 \cr
0 & 0 &  {q^{-4}} - {q^{-2}} & 0 & 0 & 1 - {q^{-2}} & 0 & 0 & {q^{-2}} &  0 \cr
0 & 0 &  0 & 0 & 0 & 0 & 0 & 0 & 0 & 0 \cr
0 & 1 &  0 & 0 & 0 & 0 & 0 & 0 & 0 & 0   \cr
0 & 0 &  0 & 0 & 0 & 1 & 0 & 0 & 0 & 0   \cr
1 - q^{-2} & 0 &  0 & -1 + {q^{-2}} & 0 & 0 & 0 & 0 & 0 & 1 \cr  } \right)
\end{equation}  } 
Here is the ``big'' $\hat{\sigma }$, which describes the
$\wedge$ product  {\footnotesize
$$ \left(
\matrix{ 1 & 0 & 0 & 0 & 0 & 0 \cr
0 & 1 - {q^{2}} & 0 & 0 & 1 & 0 \cr
0 & 0 & 1 - {q^{-2}} & 0 & 0 & 0 \cr
1 + {q^{-4}} - {2\over {{q^{2}}}} & 0 & 0 & 2 - {q^{-2}} - {q^{2}} & 0 & 0 \cr
0 & {q^{2}} & 0 & 0 & 0 & 0 \cr
0 & 0 & 0 & 0 & 0 & 1 \cr
-1 + {q^{-2}} & 0 & 0 & -1 + {q^{2}} & 0 & 0 \cr
0 & 0 & 0 & 0 & {q^{-4}} - {q^{-2}} & 0 \cr
0 & 0 & {q^{-2}} & 0 & 0 & 0 \cr
1 - {q^{-2}} & 0 & 0 & 1 - {q^{2}} & 0 & 0  \cr
0 & 0 & 0 & 0 & 0 & 0 \cr
0 & 0 & -1 + {q^{-4}} - {q^{-2}} + {q^{2}} & 0 & 0 & 0 \cr
0 & 0 & 0 & 1 & 0 & 0  \cr
0 & -1 + {q^{2}} & 0 & 0 & 0 & 0  \cr
0 & 0 & -1 + {q^{-2}} & 0 & 0 & 0 \cr
-1 - {q^{-4}} + {2\over {{q^{2}}}} & 0 & 0 & -2 + {q^{-2}} + {q^{2}} &
0 & 0  \cr
}\right.
$$
\begin{equation}\left.
\matrix{
 0 & 0 & 0 & 0 & 0 & 0 & 0 & 0 & 0 & 0 \cr
 0 & 0 & 0 & 0 & 0 & 0 & 0 & 0 & 0 & 0 \cr
 0 & 0 & 1 & 0 & 0 & 0 & 0 & 0 & 0 & 0    \cr
 1 - q^{-2} & 0 & 0 & -1 + {q^{-2}} & 0 & 0 & 1 & 0 & 0 & 0 \cr
 0 & 0 & 0 & 0 & 0 & 0 & 0 & 0 & 0 & 0 \cr
 0 & 0 & 0 & 0 & 0 & 0 & 0 & 0 & 0 & 0 \cr
 0 & 0 & 0 & 1 & 0 & 0 & 0 & 0 & 0 & 0 \cr
 0 & 1 - q^{-2} & 0 & 0 & 0 & 0 & 0 & {q^{-2}} & 0 & 0 \cr
 0 & 0 & 0 & 0 & 0 & 0 & 0 & 0 & 0 & 0 \cr
 1 & 0 & 0 & 0 & 0 & 0 & 0 & 0 & 0 & 0 \cr
 0 & 0 & 0 & 0 & 1 & 0 & 0 & 0 & 0 & 0 \cr
 0 & 0 & 1 - {q^{-2}} & 0 & 0 & 1 - {q^{2}} & 0 & 0 & {q^{2}} & 0 \cr
 0 & 0 & 0 & 0 & 0 & 0 & 0 & 0 & 0 & 0 \cr
 0 & 1 & 0 & 0 & 0 & 0 & 0 & 0 & 0 & 0 \cr
 0 & 0 & 0 & 0 & 0 & 1 & 0 & 0 & 0 & 0 \cr
  -1 + q^{-2} & 0 & 0 & 1 - {q^{-2}} & 0 & 0 & 0 & 0 & 0 & 1 \cr  }\right)
\end{equation} } 

\chapter{Toward a BRST Formulation of Quantum Group Gauge Theory}

In this chapter we will give a brief introduction to a BRS type formalism
for quantum gauge theories. All fields will live on the base manifold.
A BRS formulation has two main advantages
here: It can be formulated as a purely algebraic theory with abstract
operators $\dg, \dl, t, \ldots$ (see \cite{Z0} for a beautiful  example of
the use of this abstract algebra in the context of anomalies)
and it emphasizes the coalgebra aspect
of the quantum structure group --- which is undeformed in the case of
matrix pseudo groups. This will lead to equations that are of
virtually identical form as their classical counterparts; this was
the base of Isaevs \cite{Is} approach to quantum group gauge theory.
We will however go a step beyond this work in as we will give an
interpretation of objects like $\dl(t) \neq 0$, where $\dl$ is the exterior
derivative on the {\em base} manifold of a bundle with quantum group valued
fiber, even though $t \in M_{n}(\A)$ may not have any base-dependence,
thereby justifying the coexistence of such different objects
within one algebra. We will not attempt any further (physical)
interpretations of $e.g.$ the connection form here, because this subject
is still controversial at the moment. Nevertheless we hope to give
an easy-to-use formalism that could serve as a starting platform for
further investigations. Articles of related interest are \cite{Du};
see \cite{BrMa} for an abstract treatment of quantum group gauge theory and
many examples.

Let $\A = \fun$ be the algebra of functions on the quantum structure
group and $\B = $Fun(M) be the --- possibly non-commutative --- algebra
of functions on the base manifold; for instance space-time.
The symbol $\dg$ shall denote the exterior derivative of $\Lambda (\A)$
and $\dl$ ditto of $\Lambda (\B)$ --- classically: $\dl = \dl(x^{\mu }) \frac
{\partial }{\partial  x^{\mu }}$; we will
require them to anticommute
\begin{equation}
\dg \dl = - \dl \dg.
\end{equation}
The quantum matrix $\left(t^{i}{}_{j}\right)_{i,j = 1}^{n} \in M_{n}(\A)$ (in
the fundamental
representation) shall describe the gauge transformation of a column
vector $\psi _{0}$
of fields, $A_{0}$ is the quantum Lie algebra valued
matrix of connection 1-forms and $v$ finally, the ``ghost'',
is an abbreviation for the
Cartan-Maurer form $t^{-1} \dg(t)$.
As in the chapter on the induced calculus we make $\psi _{0}$  and $A_{0}$
``variable'' with the help of \A-coactions:
\begin{eqnarray}
\psi  & := & t^{-1} \psi _{0} = ``\DA(\psi _{0})",\\
A & := & t^{-1} A_{0} t + t^{-1} \dl(t) = ``\DA(A_{0})".
\end{eqnarray}
To justify the name ``coaction" for
\begin{equation}
\DA({A_{0}}^{i}{}_l) = {A_{0}}^{j}{}_{k} \otimes S(t^{i}{}_{j}) t^{k}{}_{l} + 1
\otimes S(t^{i}{}_{j})
\dl(t^{j}{}_{l}) \label{questable}
\end{equation}
we have to extend the notion of the Hopf algebra $\A$ to a graded Hopf
algebra $\A \oplus
\A\otimes\dl(\A)$
via
\begin{eqnarray}
\Delta  \circ \dl & := & (\dl \otimes i\!d + i\!d \otimes \dl) \circ \Delta ,\\
\epsilon  \circ \dl & := & \dl \circ \epsilon : \A \to  \{0\},\\
S \circ \dl & := & \dl \circ S.
\end{eqnarray}
Consider $e.g.$
\begin{equation}
\epsilon (\dl a) = \cdot (S \otimes i\!d) \Delta (\dl a) = S(\dl a_{(1)})
a_{(2)} +
S(a_{(1)}) \dl(a_{(2)}) = 0,\z e.t.c.\, .
\end{equation}
It is straightforward to show that (\ref{questable}) does indeed
satisfy the axioms of a coaction:
\begin{equation}
(\DA \otimes i\!d) \DA = (i\!d \otimes \Delta ) \DA,\z
(i\!d \otimes \epsilon ) \DA = i\!d.
\end{equation}
We are now ready to derive a set of BRS transformations
\begin{eqnarray}
\dg( \psi  ) & = & \dg(t^{-1} \psi _{0}) = t^{-1} t \dg(t^{-1}) \psi _{0}
                        = -t^{-1} \dg(t) t^{-1} \psi _{0}\nonumber\\
             & = & - v \psi ,\\
\dg(\dl \psi )   & = & \dl(v) \psi  - v \dl(\psi ),\\
\dg(v)       & = & - v^{2},\\
\dg(t)       & = & t v,\\
\dg(t^{-1})     & = & - v t^{-1},\\
\dg(t^{-1}\dl(t))&= & - v t^{-1}\dl(t) - t^{-1}\dl(t)v - \dl(v),\\
\dg(A)       & = & \dg(t^{-1}) A_{0} t - t^{-1} A_{0} \dl(t) +
                        \dg(t^{-1}\dl(t))\nonumber \\
             & = & -v A - A v - \dl(v),
\end{eqnarray}
simply by applying $\dg$ and working out the algebra; the first and last
lines should give a flavor of these computations.
All these equations correspond via the Cartan identity $\Lix{i}
= \dg \Ix{i} + \Ix{i} \dg$ to infinitesimal gauge transformations.
The $\dg \Ix{i}$ term is actually zero on functions and on
left-invariant 1-forms like $v$, so we only need the second term $\Ix{i}
\dg$, $i.e.$ {\em all} gauge transformation information is already
contained in the BRS $\dg$; $e.g.$
\begin{equation}
\begin{array}{rcl}
\Lix{i}( \psi ) & = & \Ix{i}(\dg \psi )\\
            & = & -\Ix{i}(v \psi )\\
            & = & <\chi _{i},S t> \psi  =: \lambda _{i} \psi
\end{array}
\end{equation}
and
\begin{equation}
\begin{array}{rcl}
\Lix{i}(v) & = & \underbrace{\dg(\Ix{i} v)}_{0} + \Ix{i}(\dg v)\\
           & = & -\Ix{i}(v^{2})\\
           & = & -\Ix{i}(v) v + \Lio{i}{j}(v) \Ix{j}(v)\\
           & = & \lambda _{i} v + \Lio{i}{j}(v) \lambda _{j} = \{\lambda _{i} ,
v \}_{q}\\
           & = & \lambda _{i} v + M_{i}{}^{l} v (M^{-1})_{l}{}^{j} \lambda
_{j},
\end{array}\label{lixiv}
\end{equation}
with $M_{i}{}^{l} \equiv <O_{i}{}^{l},S t>$.
Next we introduce a covariant derivative $D$
such that $D \psi $ transforms covariantly
\begin{equation}
\dg(D \psi ) = - v (D \psi )
\end{equation}
in analogy to $\dg(\psi ) = - v \psi $. This is not really an extension of the
algebra as $D = \dl + A$ --- in fact that is exactly what motivated $A$'s
transformation properties.
{}From $\dl$ and $A$ we can construct another covariant tensor
\begin{equation}
F := \dl(A) + A A,
\end{equation}
the ``field strength''. A short (purely algebraic) computation gives
\begin{equation}
\dg(F) = - v F + F v.
\end{equation}

It is now time to give an interpretation to objects like $\dl(t)$, where
$\dl$ is the exterior derivative on the {\em base} space so that
we have to give
$\B$-dependence to $t$ in some way:\\
$i)$\ It is always possible
to  construct  a new explicitly $\B$-dependent $t_{W} \in M_{n}(\B \otimes \A)$
\begin{equation}
t_{W} := W^{-1} t W,
\end{equation}
where $W \in M_{n}(\B)$ is a pointwise invertible Matrix of functions on the
base
space. Here we were careful not to destroy $t$'s Hopf algebra properties
that are reminiscent of a representation, $i.e$ $\Delta  t_{W} = t_{W} \otimes
t_{W}$,
$S t_{W} = t_{W}^{-1}$ and $\epsilon  t_{W} = I$. This type of $\B$-dependence
is essentially  classical  because it could be obtained from the
adjoint action on $t$ of an element $\gamma  \in \B \otimes \U$
that is $\B$-dependent and group-like $\Delta \gamma  = \gamma  \otimes \gamma
$:
$\gamma  \ad t = <\gamma ^{-1},t> t <\gamma ,t>$;
see also \cite{BrMa2}.
More important is:\\
$ii)$ Implicit $\B$-dependence. Say, we have a $\B$-dependent gauge
transformation $g$, $i.e.$ $t(g) \in M_{n}(\B)$,
we then define $(\dl t)$ {\em on} it by
\begin{equation}
(\dl t)\left(g\right) := \dl\left(t(g)\right);
\end{equation}
that can be classically expressed as:
\begin{equation}
(\dl t)\left(g(x)\right) := \dl(x^{\mu }) \frac {\partial }{\partial  x^{\mu
}}\left(t\left(g(x)\right)\right).
\z\mbox{\it(classical)}
\end{equation}
(It would be interesting to see whether one could actually rewrite
$(\dl t)$ as a matrix
$$\dl t \approx  \dl(x^{\mu }) \phi _{\mu }^{\alpha } (\chi _{\alpha } \tr t)
\z\in
M_{n}(\Lambda ^{1}(\B) \otimes \B \otimes \A)$$
for every given choice of gauge, parameterized by $\phi _{\mu }^{\alpha } \in
\B$.)\\
{\em Remark:} In our formulation we are actually more interested in actions
than contractions, but
remembering $\Delta \circ \dl = (\dl \otimes 1 + 1 \otimes \dl) \circ \Delta $,
this is easily accomplished:
\begin{equation}
x \tr \dl(a) = \dl(a_{(1)}) <x, a_{(2)}> + a_{(1)} \dl(<x , a_{(2)}>).
\end{equation}

If we contract with an element $x$  of \U\ a {\em product} of say two functions
in \A, we look at the coproduct of \U\ to
determine how to split up $x$ into parts, each contracting its respective
function: $<x,a b> = <x_{(1)},a><x_{(2)},b>$.
As soon as $x$ becomes also a function on the base, say $x = \sum \beta ^{a}
\chi _{a} \in \B \otimes \U$, as is the case for
local gauge transformations --- and we are trying
to contract things like $t^{-1} \dl(t)$, we have a problem: we need to give
rules for where
to put the $\B$-dependence in coproducts like
\begin{equation}
<x , t^{-1} \dl(t)> \stackrel{?}{=} <x_{(1)},t^{-1}>\dl(<x_{(2)},t>)
\end{equation}
because otherwise it might sneak past the $\dl$ and escape to the left \ldots\
{}.
There is an infinity of possible rules for $\Delta x$; $\beta ^{a} \chi _{a(1)}
\otimes \chi _{a(2)}$,
$\chi _{a(1)} \otimes \beta ^{a} \chi _{a(2)}$, $\beta ^{a} \chi _{a} \otimes 1
+ \chi _{a(1)} \otimes \beta ^{a} (\chi _{a(2)}
- 1 \epsilon (\chi _{a(2)}) )$, \ldots\  are examples. Luckily $\beta ^{a} \in
\B$ and {\em not} $ \in k$, so that it
need not commute with $\otimes$ and one has at least the opportunity to give
rules.
No matter what we choose, we must not violate the Leibniz rule, in particular
we must be in consistency with
$\dl(1) = \dl(1 1) = 2 \dl(1)$, which implies that only $x$ with zero counit
can have $\B$-dependence. In the
classical case that singles out one natural choice:$$``\Delta " \beta ^{a} \chi
_{a}  = \beta ^{a} \chi _{a} \otimes 1 +
1 \otimes \beta ^{a} \chi _{a}. $$
This riddle is solved by extending the Cartan calculus to include Lie
derivatives along elements of $\B \otimes \tq$
via
\begin{equation}
\Li_{\beta ^{a} \chi _{a}} = \beta ^{a} \Li_{\chi _{a}} + \dl(\beta ^{a})
\Il_{\chi _{a}}.
\end{equation}
(Note the appearance of the exterior derivative $\dl$ of the {\em base} and the
corresponding inner derivation $\Il$
in this equation.) Here is an example, showing how $t^{-1} \dl(t)$ transforms
under a gauge transformation along
$\beta ^{a} \chi _{a}$:
\begin{equation}
\begin{array}{rcl}
\Li_{\beta ^{a} \chi _{a}}(t^{-1} \dl t) & = & \beta ^{a} \Li_{\chi
_{a}}(t^{-1} \dl t) + \dl(\beta ^{a}) \Il_{\chi _{a}}(t^{-1} \dl t)\\
    & = & \beta ^{a} <\chi _{a(1)}, t^{-1}> t^{-1} ( \dl(t)<\chi _{a(2)},t> +
t\underbrace{\dl(<\chi _{a(2)},t>)}_{ = 0})\\
    &   &+ \dl(\beta ^{a}) \Il_{\chi _{a}}(t^{-1} \dl t)\\
    & = & \beta ^{a} (<\chi _{a},t^{-1}> t^{-1} \dl(t) + <O_{a}{}^{b},t^{-1}>
t^{-1} \dl(t) <\chi _{b},t>)\\
    &   &+ \dl(\beta ^{a}) <-\chi _{a},t^{-1}>\\
    & = & \beta ^{a} \left[\lambda _{a} , t^{-1} \dl(t) \right]_{q} - \dl(\beta
^{a}) \lambda _{a}.
\end{array}
\end{equation}
(Compare to (\ref{lixiv}).)
This calculation implicitly used further relations of the extended Cartan
calculus:
\begin{eqnarray}
\Lix{{}} \dl & = & \dl \Lix{{}}\\
\Ix{{}} \dl & = & - \dl \Ix{{}}.
\end{eqnarray}
Before we leave the subject let us make a short remark about ordering problems.
If our base space has more than 1+1 dimension we cannot define a physical
(local) ordering on it; only a lexicographic
ordering is possible. Does this lead to contradictions if we are dealing
with non-commutative functions? Not necessarily, as long as we are ordering
within the column vector of fields and
otherwise use global commutation relations and in particular
just one global copy of $t$. Consider
for instance the quantum structure group $SU_{q}(2)$ and two
column vectors $\psi $ and $\psi '$
at different points on the base space. They will satisfy the following four
mixed commutation relations\footnote{In a more conservative approach along
the lines of the previous chapter the $\psi _{0}$ would be merely the
(commuting)
coefficients of a section basis --- the ordering problem would then
presumably show up somewhere else.}
$$\psi _{1} \psi _{2} = q \psi _{2} \psi _{1},\x \psi _{1} \psi '_{2} = q \psi
'_{2} \psi _{1},\x\psi '_{1} \psi _{2} = q \psi _{2} \psi '_{1},
\x\psi '_{1} \psi '_{2} = q \psi '_{2} \psi '_{1},$$
and they will both transform according to the same copy of $t$:
$$\psi \mapsto S^{-1}t \psi ,\z \psi ' \mapsto S^{-1} \psi '.$$
(An interesting idea would be to try and give $\B$-dependence to the braiding
operator $O_{a}{}^{b}$, but
that will affect the multiplication in \A\ in a way that may lead to
inconsistencies.)
Are we dealing with a non-local theory because of the global commutation
relations? The commutation
relations of the fields contained in $\psi $ are obviously non-local, however,
the real physical
observables are gauge invariant objects like tr${}_{q}(F)$ (see \cite{Mm}
for a discussion of such a set of observables) and those could very well be
central in the algebra and
in that sense ``local".  This subject matter is quite controversial, so we want
to leave it at that for now ---
hoping that the new tools provided will be beneficial in future discussions.

\end{document}